\PassOptionsToPackage{
  colorlinks=true,
  linkcolor=blue,
  citecolor=blue,
  urlcolor=blue
}{hyperref}

\documentclass{SciPost}
\hypersetup{
    colorlinks,
    linkcolor={red!50!black},
    citecolor={blue!50!black},
    urlcolor={blue!80!black}
}
\usepackage[bitstream-charter]{mathdesign}

\usepackage{graphicx}% Include figure files
\usepackage{dcolumn}% Align table columns on decimal point
\newcommand{\bm}[1]{{\mbox{\boldmath $#1$}}}
\usepackage{amsmath}
\usepackage{braket}
\usepackage{enumerate}
\usepackage{multirow}
\usepackage{hhline}
\usepackage{makecell}
\usepackage{rotating}
\usepackage{makecell}
\usepackage{environ}
\DeclareSymbolFont{usualmathcal}{OMS}{cmsy}{m}{n}
\DeclareSymbolFontAlphabet{\mathcal}{usualmathcal}
\fancypagestyle{SPstyle}{
\fancyhf{}
\lhead{\colorbox{scipostblue}{\bf \color{white} ~SciPost Physics}}
\rhead{{\bf \color{scipostdeepblue} ~Submission }}

\fancyfoot[C]{\textbf{\thepage}}
}

\begin{document}

\pagestyle{SPstyle}

\begin{center}{\Large \textbf{\color{scipostdeepblue}{
%%%%%%%%%% TODO: Write your article's title here
%Sixteen-fold
Sixteenfold Classification of Many-Body Multipoles \\
from Rotation-Compatible Canonical Symmetries
}}}\end{center}

\begin{center}\textbf{
% Use (full) first name (+ middle name initials) + surname format.
% Separate subsequent authors by a comma, omit comma and use "and" for the last author.
% Mark the corresponding author(s) with a superscript symbol in this order
% \star, \dagger, \ddagger, \circ, \S, \P, \parallel, ...
Shingo Kuniyoshi$^\star$, Rikuto Oiwa$^\dagger$, and Satoru Hayami$^\ddagger$
}\end{center}

\begin{center}
Graduate School of Science, Hokkaido University, Sapporo 060-0810, Japan
% \\
% {\bf 2} Affiliation2
% \\
% {\bf 3} Affiliation3
%%%%%%%%%% END TODO: AFFILIATIONS
% Provide email address of corresponding author(s)
\\[\baselineskip]
\end{center}

\section*{\color{scipostdeepblue}{Abstract}}
%
% abstは， 8行以内
%
\textbf{\boldmath{%
The conventional multipole framework provides a standard symmetry-based language for one-body electronic degrees of freedom, but fails to distinguish physically distinct sectors in many-body operator space.
We show that canonical symmetries compatible with rotations provide this structure through two additional $\mathbb Z_2$ labels associated with the particle-number gauge transformation $\mathcal G_{\pi/2}$ and the particle-hole transformation $\mathcal C_A$. 
These labels separate operators with different body numbers and particle-number changes, leading to a sixteenfold classification that organizes selection rules for nonzero expectation values, induced multipoles, and symmetry-allowed couplings in many-body multipole space.
}}

\vspace{\baselineskip}

%%%%%%%%%% BLOCK: Copyright information
% This block will be filled during the proof stage, and finilized just before publication.
% It exists here only as a placeholder, and should not be modified by authors.
\noindent\textcolor{white!90!black}{%
\fbox{\parbox{0.975\linewidth}{%
\textcolor{white!40!black}{\begin{tabular}{lr}%
  \begin{minipage}{0.6\textwidth}%
    {\small Copyright attribution to authors. \newline
    This work is a submission to SciPost Physics Core. \newline
    License information to appear upon publication. \newline
    Publication information to appear upon publication.}
  \end{minipage} & \begin{minipage}{0.4\textwidth}
    {\small Received Date \newline Accepted Date \newline Published Date}%
  \end{minipage}
\end{tabular}}
}}
}
%%%%%%%%%% BLOCK: Copyright information

% For convenience during refereeing we turn on line numbers:
% \linenumbers

\vspace{10pt}
\noindent\rule{\textwidth}{1pt}
\tableofcontents
\noindent\rule{\textwidth}{1pt}
\vspace{10pt}

\section{Introduction}
A central goal in the study of ordered phases is to identify the low-energy degrees of freedom that characterize a phase and the symmetries that constrain their behaviors.
While local-moment magnetism provides a paradigmatic example of spontaneous ordering, higher-rank multipoles involving spin and orbital degrees of freedom often play essential roles in $f$-electron systems, giving rise to a wide variety of unconventional ordered states and responses beyond conventional dipolar magnetism. 
This recognition has motivated the development of multipole representation theory, which decomposes electronic degrees of freedom into irreducible %components
representations of the rotation group and provides a %standard language
systematic framework for symmetry-based analysis in condensed matter physics
~\cite{Kuramoto_JPSJ_2008,Kusunose_JPSJ_2008,Santini_RMP_2009,Kuramoto_JPSJ_2009}.

The conventional classification of multipoles is based on rotational symmetry.
In recent years, this classification has been refined by incorporating spatial inversion $\mathcal P$ and time reversal $\mathcal T$.  
In this framework, multipoles are classified into four types: electric (E) $Q$, magnetic (M) $M$, magnetic-toroidal (MT) $T$, and electric-toroidal (ET) $G$, according to their parities under $\mathcal P$ and $\mathcal T$~\cite{dubovik1990toroid,Hayami_JPSJ_2018,SH_MY_YY_HK_mul_2018,HW_YY_Mul_2018,Kusunose_JPSJ_2020,Kusunose_IOP_2022,Hayami_JPSJ_2024}.
Each multipole is characterized by an irreducible representation of the rotation group and is denoted by $X_{kq}$ $(X=Q,M,T,G)$, where $k$ is the rank and $q$ labels its component. 
The corresponding parities $(\sigma_{\mathcal P},\sigma_{\mathcal T})$ under $(\mathcal P,\mathcal T)$ %parities 
are
\begin{align}
  Q_{kq} &: \quad
  (\sigma_{\mathcal P},\sigma_{\mathcal T})
  =
  \left((-1)^k,+1\right),
  \notag\\
  M_{kq} &: \quad
  (\sigma_{\mathcal P},\sigma_{\mathcal T})
  =
  \left((-1)^{k+1},-1\right),
  \notag\\
  T_{kq} &: \quad
  (\sigma_{\mathcal P},\sigma_{\mathcal T})
  =
  \left((-1)^k,-1\right),
  \notag\\
  G_{kq} &: \quad
  (\sigma_{\mathcal P},\sigma_{\mathcal T})
  =
  \left((-1)^{k+1},+1\right).
  \label{eq:main_conventional_QMTG}
\end{align}

This classification provides a powerful framework for identifying unconventional order parameters and organizing symmetry-allowed responses~\cite{SH_MY_YY_HK_mul_2018}.
For example, the ET monopole $G_0$ has emerged as a descriptor of electronic chirality~\cite{Hayami_PhysRevLett.122.147602,Oiwa_PRL_2022,Kishine_IJC_2022,hayami2023chiral,Kusunose_APL_2024,Hoshino_PRL_2023,Oiwa_PRR_2025,Miki_PRL_2025,ishitobi2026purely} and its associated response phenomena~\cite{1979JETPL,B.Gohler_nat_2011_CISS,O.Ben_nat_2017_CISS,K.Michaeli_PNAS_2019_CISS,Suda_natcom_2019_CISS,A.Inui2020prl_CrNb3S6,R.Neeman_ACR_2020_CISS,shitade2020geometric,Waldeck2021aplmat_CISS,Bloom_2024_chemrev_CISS}, whereas the MT monopole $T_0$ has been used to characterize time-reversal-odd scalar magnetic orders~\cite{Hayami_T0_PRB_2023,Hayami_T0_magnetism_2023,hayashida2024electric,kato2025electricfieldinducedmagnetictoroidalmoment}.
More recently, this framework has also been extended to particle-number-nonconserving one-body sectors, including operators of the form $\{c^\dagger c^\dagger,\,cc\}$, enabling a systematic classification of superconducting order parameters in multi-orbital systems~\cite{Kirikoshi_PRB_2024}.
Importantly, these developments remain within the one-body operator space.
Within this space, multipole operators form a complete orthonormal basis and provide a systematic framework for describing arbitrary anisotropies of electronic distributions~\cite{Kusunose_PRB_2023}.
Furthermore, practical first-principles-based schemes have been developed, making it possible to identify multipole degrees of freedom quantitatively in real materials~\cite{Ederer_PRB_2007,Spaldin_PRB_2013,Florian_PRB_2016,suzuki2018first,Oiwa_PRL_2022,Hoshino_PRL_2023,Inda_JCP_2024,Oiwa_PRB_2025,Oiwa_PRR_2025,Miki_PRL_2025,Xie_arxiv_2025}.
Taken together, these developments establish the conventional multipole classification as a mature and successful framework for one-body operator space.

As we show in this work, however, a fundamental limitation emerges when the conventional classification is extended to many-body operator space.
The problem is that physically distinct many-body multipole operators can belong to the same conventional sector.
That is, they may share the same rotational rank and the same conventional multipole type, and are therefore indistinguishable within the conventional $(\mathrm{rank},\sigma_{\mathcal P},\sigma_{\mathcal T})$ classification.

This limitation is already visible in the minimal local Fock space of a spinful $s$ orbital.
The operators
\begin{align}
N &= n_{\uparrow} + n_{\downarrow},
\\
U &= n_{\uparrow} n_{\downarrow},
\\
\eta_+ &= c^\dagger_\uparrow c^\dagger_\downarrow + c_\downarrow c_\uparrow
  \label{eq:intro_U}
\end{align}
represent the particle number, the onsite interaction, and a Hermitian pairing operator, respectively.
In the conventional multipole classification, all three operators are rotational scalars with electric character and hence belong to the same electric-monopole sector $Q_{0}$.
Nevertheless, they have distinct many-body structures: $N$ is a particle-number-conserving one-body operator, $U$ is a particle-number-conserving two-body operator, and $\eta_+$ is a particle-number-nonconserving one-body pairing operator.
Thus, the conventional labels do not resolve distinctions in body-number $N_{\rm b}$ and particle-number change $\Delta N$, even in the smallest local Fock space.

The above example exposes one limitation of the conventional classification: it does not distinguish body-number or net particle-number change.
However, many-body operator tensors present an additional challenge.
Unlike one-body operators, a normal-ordered higher-body tensor is not automatically a symmetry-adapted multipole.
Under canonical transformations that exchange creation and annihilation operators, such as particle-hole or chiral transformations, a normal-ordered many-body operator generally mixes with lower-body operators when the transformed operator is re-expressed in normal-ordered form.
As a result, a many-body operator classified by the conventional multipole type $(\mathrm{rank},\sigma_{\mathcal P},\sigma_{\mathcal T})$ may still contain lower-body components that break particle-hole or chiral symmetry unless those components are explicitly removed.
A consistent classification of many-body multipoles therefore requires not only additional symmetry labels but also a symmetry-adapted construction of the operator basis.

In this work, we construct such a symmetry-adapted many-body multipole basis by recursively subtracting lower-body components and then classify the resulting Hermitian multipoles.
The additional labels are associated with two canonical transformations intrinsic to even-fermion many-body operator space.
One is the particle-hole transformation $\mathcal C_A$, which distinguishes the parity of the body-number $N_{\rm b}$.
The other is the discrete particle-number gauge transformation $\mathcal G_{\pi/2}$, which distinguishes the net particle-number change $\Delta N$ modulo four.
Together with the conventional spatial-inversion and time-reversal labels, these give the sixteenfold classification
\begin{align}
  SO(3)
  \times
  \mathbb Z_2^{\mathcal P}
  \times
  \mathbb Z_2^{\mathcal T}
  \times
  \mathbb Z_2^{\mathcal C_A}
  \times
  \mathbb Z_2^{\mathcal G_{\pi/2}} .
  \label{eq:intro_extended_classification}
\end{align}
Each $\mathbb Z_2$ factor is represented on a Hermitian multipole by its eigenvalue
\begin{align}
\sigma_{\mathcal P}, \qquad \sigma_{\mathcal T}, \qquad \sigma_{\mathcal C_A} = (-1)^{N_{\rm b}}, \qquad \sigma_{\mathcal G} = (-1)^{\Delta N/2}.
\end{align}
The resulting classification is not merely formal: it yields selection rules for multipole expectation values and clarifies which many-body multipoles can be induced when a given symmetry is broken.

This paper is organized as follows.
In Sect.~\ref{sec:classification}, we define many-body multipole tensors and summarize the sixteenfold classification.
In Sect.~\ref{sec:example}, we illustrate its implications in two systems: a spinless $s$-$p$ system in the fixed $N=2$ sector and a four-site square cluster, and show how the new classification organizes expectation values and induced multipoles through symmetry-based selection rules.
The Appendices provide the rotation-compatible canonical transformations, the treatment of hybrid and multiplicity spaces, the construction of symmetry-adapted Hermitian many-body multipoles, and the explicit matrix representations used in the examples.

\section{Sixteenfold classification of many-body multipoles}
\label{sec:classification}

This section summarizes the classification used throughout the main text, while the detailed derivation is given in Appendices~\ref{app:single_j}--\ref{app:operators}.
We first introduce the canonical transformations that are compatible with rotations and act on fermionic creation and annihilation operators.
We then show how they lead to the sixteenfold classification of many-body operators.
We also construct symmetry-adapted many-body multipoles based on this classification.
As an illustrative example, we explicitly derive the symmetry-adapted many-body multipoles activated in the local Fock space of a spinful $s$ orbital.

\subsection{Canonical transformations and the sixteenfold labels}
\label{subsec:canonical_labels}

We consider fermion-parity-even operators built from creation and annihilation operators $c^\dagger_{\mu,j,m}$ and $c_{\mu,j,m}$.
The labels $j$ and $m$ denote the total angular momentum and its $z$ component, respectively, while $\mu$ collectively represents orbital angular momentum $l$, radial, sublattice, and other internal degrees of freedom.
Conventional multipoles are classified according to their transformation properties under rotations $R$, spatial inversion $\mathcal{P}$, and time reversal $\mathcal{T}$.

Let $\pi_\mu=\pm1$ denote the spatial parity associated with the index $\mu$.
Spatial inversion acts on the fermionic operators as
\begin{align}
  \mathcal P c^\dagger_{\mu,j,m}\mathcal P^{-1}
  &=
  \pi_{\mu} c^\dagger_{\mu,j,m},
  &
  \mathcal P c_{\mu,j,m}\mathcal P^{-1}
  &=
  \pi_{\mu} c_{\mu,j,m} .
  \label{eq:main_P_action}
\end{align}
Time reversal $\mathcal{T}$ is antiunitary.
With the standard spherical-tensor convention, it acts as
\begin{align}
  \mathcal T c^\dagger_{\mu,j,m}\mathcal T^{-1}
  &=
  (-1)^{j-m}c^\dagger_{\mu,j,-m},
  &
  \mathcal T c_{\mu,j,m}\mathcal T^{-1}
  &=
  (-1)^{j-m}c_{\mu,j,-m} .
  \label{eq:main_T_action}
\end{align}
One-body multipoles activated in a normal space are classified according to their parities under $\mathcal P$ and $\mathcal T$~\cite{SH_MY_YY_HK_mul_2018}.
For a rank-$k$ multipole, this yields the conventional fourfold classification shown in Eq.~\eqref{eq:main_conventional_QMTG}.

In many-body operator space, however, the conventional $(\mathrm{rank},\mathcal P,\mathcal T)$ classification is not sufficient to distinguish many-body multipole operators.
As shown in Eq.~\eqref{eq:intro_U}, the spinful $s$-orbital space includes the operators $N$, $U$, and $\eta_+$.
Although $N$, $U$, and $\eta_+$ all belong to the same electric-monopole sector $Q_{0}$, they differ in many-body structure: $N$ is a particle-number-conserving one-body operator, $U$ is a particle-number-conserving two-body operator, and $\eta_+$ is a particle-number-nonconserving one-body pairing operator.
This minimal example shows that the conventional classification does not resolve differences in body-number $N_{\rm b}$ and particle-number change $\Delta N$.

To resolve this degeneracy, we consider canonical transformations that preserve the rotational tensor structure.
The first is the discrete gauge transformation
\begin{align}
  \mathcal G_{\pi/2}: \quad
  c^\dagger_{\mu, j, m} \mapsto -i c^\dagger_{\mu, j, m},
  \qquad
  c_{\mu, j, m} \mapsto i c_{\mu, j, m}.
\end{align}
For an $N_{\rm b}$-body symmetry-adapted multipole operator $O$ with net particle-number change $\Delta N$, this transformation gives (See Appendix~\ref{app:single_j})
\begin{align}
  \mathcal G_{\pi/2}
  O \mathcal G_{\pi/2}^{-1}
  =
  e^{-i\pi \Delta N/2}O .
\end{align}
Within the even-fermion-parity operator space, this phase reduces to a $\mathbb Z_2$ eigenvalue,
\begin{align}
\sigma_{\mathcal{G}} =
\begin{cases}
+1 & \Delta N=0 \!\!\pmod 4, \\
-1 & \Delta N=2 \!\!\pmod 4.
\end{cases}
\end{align}
The label $\sigma_{\mathcal{G}}$ therefore distinguishes operator sectors with different particle-number changes modulo four.
In particular, it separates particle-number-conserving operators ($\Delta N=0$) from pairing operators with $\Delta N=\pm2$.

The second transformation is the antiunitary particle-hole transformation
\begin{align}
  \mathcal C_A: \quad
  c^\dagger_{\mu, j, m} \mapsto c_{\mu, j, m},
  \qquad
  c_{\mu, j, m} \mapsto c^\dagger_{\mu, j, m}.
\end{align}
In the many-body Hilbert space, this transformation exchanges the sectors with $N=n$ and $N=\Omega-n$, where $\Omega$ denotes the particle number at full filling, thereby reflecting the particle-hole structure of the local Fock space.
Furthermore, for an $N_{\rm b}$-body symmetry-adapted multipole operator $O$, it acts as (See Appendix~\ref{app:operators})
\begin{align}
  \mathcal C_A O \mathcal C_A^{-1}
  =
  (-1)^{N_{\rm b}} O.
\end{align}
Thus, the particle-hole parity is defined by
\begin{align}
\sigma_{\mathcal{C}_A} =
\begin{cases}
+1 & N_{\rm b} \text{: even}, \\
-1 &  N_{\rm b} \text{: odd}.
\end{cases}
\end{align}
The label $\sigma_{\mathcal C_A}$ therefore distinguishes operator sectors with different body-number parity.
For example, the one-body operators $N$ and $\eta_+$ have $\sigma_{\mathcal C_A}=-1$, whereas the two-body operator $U$ has $\sigma_{\mathcal C_A}=+1$.

In spherical-tensor language, it is often convenient to combine $\mathcal C_A$ with time reversal $\mathcal{T}$ and define the composite unitary chiral transformation
\begin{align}
  \Gamma \equiv \mathcal T \mathcal C_A.
\end{align}
The operator $\Gamma$ is unitary and acts as a chiral symmetry, analogous to the chiral symmetry appearing in the Altland--Zirnbauer classification~\cite{Altland_PhysRevB.55.1142, Schnyder_PhysRevB.78.195125, kitaev2009periodic, ryu2010topological}.
The corresponding parity eigenvalue is
\begin{align}
  \sigma_\Gamma
  =
  \sigma_{\mathcal T}\sigma_{\mathcal C_A}
  =
  (-1)^{N_{\rm b}} \sigma_{\mathcal T}.
  \label{eq:main_Gamma_rule}
\end{align}
The $\Gamma$ label provides an equivalent representation of the particle-hole parity in the spherical-tensor basis and is often convenient in practical calculations.

Consequently, a symmetry-adapted Hermitian many-body multipole carries the following parities
\begin{align}
  \left(
    \sigma_{\mathcal P},
    \sigma_{\mathcal T},
    \sigma_{\mathcal C_A},
    \sigma_\Gamma,
    \sigma_{\mathcal G}
  \right)
  =
  \left(
    \sigma_{\mathcal P},
    \sigma_{\mathcal T},
    (-1)^{N_{\rm b}},
    (-1)^{N_{\rm b}}\sigma_{\mathcal T},
    (-1)^{\Delta N/2}
  \right).
  \label{eq:main_16_labels}
\end{align}
Only four of these labels are independent, because $\sigma_\Gamma=\sigma_{\mathcal T}\sigma_{\mathcal C_A}$.
One may therefore choose either $(\sigma_{\mathcal P},\sigma_{\mathcal T},\sigma_{\mathcal C_A},\sigma_{\mathcal G})$ or equivalently $(\sigma_{\mathcal P},\sigma_{\mathcal T},\sigma_\Gamma,\sigma_{\mathcal G})$ as a complete set of symmetry labels.

The conventional fourfold classification is therefore refined into the sixteenfold classification as
\begin{align}
\label{ExtendGroup}
SO(3)\times
\mathbb Z_2^{\mathcal P}\times
\mathbb Z_2^{\mathcal T}\times
\mathbb Z_2^{\mathcal C_A}\times
\mathbb Z_2^{\mathcal G_{\pi/2}},
\end{align}
or equivalently,
\begin{align}
\label{ExtendGroup2}
SO(3)\times
\mathbb Z_2^{\mathcal P}\times
\mathbb Z_2^{\mathcal T}\times
\mathbb Z_2^{\Gamma}\times
\mathbb Z_2^{\mathcal G_{\pi/2}} .
\end{align}

Table~\ref{tab:16class} summarizes the resulting symmetry sectors characterized by $\mathcal P$, $\mathcal T$, $\mathcal C_A$, $\Gamma = \mathcal T \mathcal C_A$ and $\mathcal G_{\pi/2}$ for even-fermion-parity operators.
Once operators are organized into eigenoperators of the particle-hole transformation, $\mathcal C_A$ distinguishes the parity of the body-number $N_{\rm b}$: odd-body operators are $\mathcal C_A$-odd, whereas even-body operators are $\mathcal C_A$-even.
In the next subsection, we construct the symmetry-adapted many-body multipole operators associated with these sixteen symmetry sectors.

\begin{table*}[t]
  \centering
  \caption{
  Sixteenfold classification of even-fermion many-body multipoles.
  The first column lists the rank-$k$ multipole notation used in this work.
  The conventional type $(Q,M,T,G)$ is specified by $(\sigma_{\mathcal P},\sigma_{\mathcal T})$ together with the rank $k$, as in Eq.~\eqref{eq:main_conventional_QMTG}, while the additional labels characterize the body-number parity %fixes 
  $\sigma_{\mathcal C_A}=(-1)^{N_{\rm b}}$ and the particle-number-change sector $\sigma_{\mathcal G}=(-1)^{\Delta N/2}$.
  The chirality label is given by $\Gamma=\mathcal T\mathcal C_A$, with $\sigma_\Gamma=\sigma_{\mathcal T}\sigma_{\mathcal C_A}$.
  }
  \label{tab:16class}
  \renewcommand{\arraystretch}{1.5}
  \setlength{\tabcolsep}{3pt}
  \begin{tabular}{ccccccccc}
    \hline\hline
    Multipole
    & \makecell{Representative \\ structure}
    & \makecell{body-number \\ $N_{\rm b}$}
    & \makecell{particle-number change \\ $\Delta N \bmod 4$}
    & $\sigma_{\mathcal P}$
    & $\sigma_{\mathcal T}$
    & $\sigma_{\mathcal C_A}$
    & $\sigma_{\Gamma}$
    & $\sigma_{\mathcal G}$
    \\
    \hline
    $Q_{\mu,k,q}^{[N_{\rm b},0]}$
    & $c^\dagger c$
    & odd & $0$ & $(-1)^k$ & $+$ & $-$ & $-$ & $+$
    \\
    $M_{\mu,k,q}^{[N_{\rm b},0]}$
    & $c^\dagger c$
    & odd & $0$ & $(-1)^{k+1}$ & $-$ & $-$ & $+$ & $+$
    \\
    $T_{\mu,k,q}^{[N_{\rm b},0]}$
    & $c^\dagger c$
    & odd & $0$ & $(-1)^k$ & $-$ & $-$ & $+$ & $+$
    \\
    $G_{\mu,k,q}^{[N_{\rm b},0]}$
    & $c^\dagger c$
    & odd & $0$ & $(-1)^{k+1}$ & $+$ & $-$ & $-$ & $+$
    \\
    \hline
    $Q_{\mu,k,q}^{[N_{\rm b},2]}$
    & $c^\dagger c^\dagger$ or $cc$
    & odd & $2$ & $(-1)^k$ & $+$ & $-$ & $-$ & $-$
    \\
    $M_{\mu,k,q}^{[N_{\rm b},2]}$
    & $c^\dagger c^\dagger$ or $cc$
    & odd & $2$ & $(-1)^{k+1}$ & $-$ & $-$ & $+$ & $-$
    \\
    $T_{\mu,k,q}^{[N_{\rm b},2]}$
    & $c^\dagger c^\dagger$ or $cc$
    & odd & $2$ & $(-1)^k$ & $-$ & $-$ & $+$ & $-$
    \\
    $G_{\mu,k,q}^{[N_{\rm b},2]}$
    & $c^\dagger c^\dagger$ or $cc$
    & odd & $2$ & $(-1)^{k+1}$ & $+$ & $-$ & $-$ & $-$
    \\
    \hline
    $Q_{\mu,k,q}^{[N_{\rm b},0]}$
    & $c^\dagger c^\dagger cc$
    & even & $0$ & $(-1)^k$ & $+$ & $+$ & $+$ & $+$
    \\
    $M_{\mu,k,q}^{[N_{\rm b},0]}$
    & $c^\dagger c^\dagger cc$
    & even & $0$ & $(-1)^{k+1}$ & $-$ & $+$ & $-$ & $+$
    \\
    $T_{\mu,k,q}^{[N_{\rm b},0]}$
    & $c^\dagger c^\dagger cc$
    & even & $0$ & $(-1)^k$ & $-$ & $+$ & $-$ & $+$
    \\
    $G_{\mu,k,q}^{[N_{\rm b},0]}$
    & $c^\dagger c^\dagger cc$
    & even & $0$ & $(-1)^{k+1}$ & $+$ & $+$ & $+$ & $+$
    \\
    \hline
    $Q_{\mu,k,q}^{[N_{\rm b},2]}$
    & $c^\dagger c^\dagger c^\dagger c$ or $c^\dagger ccc$
    & even & $2$ & $(-1)^k$ & $+$ & $+$ & $+$ & $-$
    \\
    $M_{\mu,k,q}^{[N_{\rm b},2]}$
    & $c^\dagger c^\dagger c^\dagger c$ or $c^\dagger ccc$
    & even & $2$ & $(-1)^{k+1}$ & $-$ & $+$ & $-$ & $-$
    \\
    $T_{\mu,k,q}^{[N_{\rm b},2]}$
    & $c^\dagger c^\dagger c^\dagger c$ or $c^\dagger ccc$
    & even & $2$ & $(-1)^k$ & $-$ & $+$ & $-$ & $-$
    \\
    $G_{\mu,k,q}^{[N_{\rm b},2]}$
    & $c^\dagger c^\dagger c^\dagger c$ or $c^\dagger ccc$
    & even & $2$ & $(-1)^{k+1}$ & $+$ & $+$ & $+$ & $-$
    \\
    \hline\hline
  \end{tabular}
\end{table*}

\subsection{Construction of symmetry-adapted many-body multipole}
\label{subsec:many_body_construction}

We now construct the symmetry-adapted many-body multipole operators carrying the canonical symmetry labels given by Eq.~\eqref{eq:main_16_labels}.
The construction proceeds in three steps.
First, creation and annihilation operators are organized into irreducible spherical tensors.
Second, creation and annihilation blocks are coupled to a final tensor rank $k \in \mathbb{Z}$.
Third, lower-body contributions generated under particle-hole or chiral transformations are recursively removed so that the resulting operators transform covariantly under the canonical symmetries.
The detailed procedure is shown in Appendices~\ref{app:single_j}--\ref{app:operators}.

We first define irreducible creation and annihilation tensors associated with a single-particle basis labeled by $(\mu,j,m)$:
\begin{align}
  C^{[1]}_{\mu,j,m}
  &:={}
  c^\dagger_{\mu,j,m},
  \label{eq:main_C1_def}
  \\
  A^{[1]}_{\mu,j,m}
  &:={}
  \tilde c_{\mu,j,m}
  =
  (-1)^{j-m}c_{\mu,j,-m}.
  \label{eq:main_A1_def}
\end{align}
The operator $\tilde c_{\mu,j,m}$ transforms as a rank-$j$ spherical tensor, just like $c^\dagger_{\mu,j,m}$.
This convention allows creation and annihilation blocks to be coupled by the same Clebsch--Gordan (CG) coefficients as shown later.
We also introduce the zero-particle blocks
\begin{align}
  C^{[0]}_{0,0}
  =
  A^{[0]}_{0,0}
  =
  I,
  \label{eq:main_C0_A0_def}
\end{align}
which are rank-zero tensors.

For $N\ge2$, irreducible $N$-particle creation tensors are constructed recursively as
\begin{align}
  C^{[N]}_{\mu,j,m}
  :=
  \sum_{m',m''}
  \braket{j'm';j''m''|jm}\,
  C^{[N-1]}_{\mu',j',m'}
  C^{[1]}_{\mu'',j'',m''}.
  \label{eq:main_CN_def}
\end{align}
Here, $\braket{j'm';j''m''|jm}$ denotes the CG coefficient.
Similarly, annihilation tensors are defined by the opposite ordering convention
\begin{align}
  A^{[N]}_{\mu,j,m}
  :=
  \sum_{m',m''}
  \braket{j'm';j''m''|jm}\,
  A^{[1]}_{\mu'',j'',m''}
  A^{[N-1]}_{\mu',j',m'}.
  \label{eq:main_AN_def}
\end{align}
The collective label $\mu$ specifies the internal coupling channel, including $\mu',\mu'',j',j''$ and possible multiplicity labels.
The opposite ordering in Eq.~\eqref{eq:main_AN_def} is chosen so that Hermitian conjugation of a creation block naturally gives the corresponding annihilation block.

A normal-ordered many-body tensor is then obtained by coupling an $N_+$-particle creation block and an $N_-$-particle annihilation block:
\begin{align}
  O^{[N_{\rm b},\Delta N]}_{\mu,k,q}
  :=
  \sum_{m_+,m_-}
  \braket{j_+m_+;j_-m_-|kq}\,
  C^{[N_+]}_{\mu_+,j_+,m_+}
  A^{[N_-]}_{\mu_-,j_-,m_-}.
  \label{eq:main_def_many_body_O}
\end{align}
The body number $N_{\rm b}$ and particle-number change $\Delta N$ are defined by
\begin{align}
  2N_{\rm b}=N_+ + N_-,
  \qquad
  \Delta N=N_+-N_- .
  \label{eq:main_body_delta_def}
\end{align}
Since we restrict ourselves to even-fermion-parity operators, $N_{\rm b}$ and $\Delta N$ are integers and $\Delta N$ is even.
Here $k$ and $q$ are the final tensor rank and component, which are obtained by the CG coupling of the creation and annihilation blocks, and their allowed values satisfy
\begin{align}
  |j_+-j_-|
  \le
  k
  \le
  j_+ + j_-,
  \qquad
  q=m_+ + m_- .
  \label{eq:main_allowed_k}
\end{align}
Although the intermediate ranks $j_+$ and $j_-$ may be integer or half-integer, the final rank $k$ is always an integer for even-fermion operators.
In anomalous sectors, one of the two blocks may be the zero-particle scalar block in Eq.~\eqref{eq:main_C0_A0_def}.
For instance, a pair-creation operator corresponds to $N_+=2$ and $N_-=0$.
In this case, the annihilation block is the scalar tensor $A^{[0]}_{0,0}=I$, so the final rank $k$ is identical to the rank of the two-particle creation tensor.

The bare tensor $O^{[N_{\rm b},\Delta N]}_{\mu,k,q}$ does not always transform covariantly under particle-hole transformation $\mathcal C_A$ or chiral transformation $\Gamma$.
The reason is that these transformations generate contractions, which produce lower-body operators.
As shown in Appendix~\ref{app:operators}, such lower-body contributions can be removed recursively.
We denote the resulting symmetry-adapted tensor by
\begin{align}
  \widehat O^{[N_{\rm b},\Delta N]}_{\mu,k,q} =
  O^{[N_{\rm b},\Delta N]}_{\mu,k,q}
  +
  \text{lower-body counterterms}.
\end{align}
The counterterms are uniquely fixed by requiring covariant transformation law under the canonical symmetries.
The label $N_{\rm b}$ attached to $\widehat O$ refers to its highest-body sector, rather than to every term appearing in the lower-body-subtracted expression.

A simple example is provided by a spinful $s$-orbital monopole.
In this sector, the bare number operator $N=n_\uparrow+n_\downarrow$ transforms under $\Gamma$ as
\begin{align}
  \Gamma N  \Gamma^{-1}
  =
  2I-N,
\end{align}
and is therefore not a chiral eigenoperator.
The corresponding one-body symmetry-adapted monopole is
\begin{align}
  \widehat N=N-I,
  \qquad
   \Gamma \widehat N \Gamma^{-1}
  =
  - \widehat N .
\end{align}
Thus $\widehat N$ is the chiral eigenoperator with $\sigma_{\Gamma} = (-1)^{1} = -1$.
Similarly, the bare two-body scalar $U=n_\uparrow n_\downarrow$ transforms as
\begin{align}
  \Gamma U \Gamma^{-1}
  =
  (1-n_\uparrow)(1-n_\downarrow)
  =
  U-N+I,
\end{align}
so that the symmetry-adapted two-body scalar becomes
\begin{align}
  \widehat U
  =
  \left(n_\uparrow-\frac{1}{2}\right)
  \left(n_\downarrow-\frac{1}{2}\right)
  =
  U - \frac{1}{2} N + \frac{1}{4} I,
\end{align}
which satisfies
\begin{align}
  \Gamma \widehat U\Gamma^{-1}
  =
  \widehat U .
\end{align}
Thus, $\widehat U$ is the chiral eigenoperator with $\sigma_{\Gamma} = (-1)^{2} = +1$.
These examples illustrate that symmetry-adapted multipoles are obtained by supplementing bare operators with lower-body counterterms.
Although the identity operator $I$ is a zero-body scalar, it naturally appears in higher-body symmetry-adapted multipoles through this recursive subtraction procedure.

In the general case, Appendix~\ref{app:operators} shows that the lower-body-subtracted tensor can be constructed so as to satisfy
\begin{align}
  \Gamma
  \widehat O^{[N_{\rm b},\Delta N]}_{\mu,k,q}
  \Gamma^{-1}
  =
  (-1)^{N_{\rm b}}
  (-1)^{k-q}
  \left(
    \widehat O^{[N_{\rm b},\Delta N]}_{\mu,k,-q}
  \right)^\dagger .
  \label{eq:main_Gamma_hatO}
\end{align}
Equation~\eqref{eq:main_Gamma_hatO} shows that the action of $\Gamma$ is completely determined by the tensor structure and the body-number parity $(-1)^{N_{\rm b}}$.
The latter gives rise to the $\Gamma$ parity label in Eq.~\eqref{eq:main_Gamma_rule}.

Finally, the symmetry-adapted many-body multipoles are constructed as
\begin{align}
  X^{[N_{\rm b},\Delta N]}_{\mu,k,q;+}
  &:=
  \widehat O^{[N_{\rm b},\Delta N]}_{\mu,k,q}
  +
  (-1)^{k-q}
  \left(
    \widehat O^{[N_{\rm b},\Delta N]}_{\mu,k,-q}
  \right)^\dagger ,
  \label{eq:main_def_Xplus}
  \\
  X^{[N_{\rm b},\Delta N]}_{\mu,k,q;-}
  &:=
  i\left[
  \widehat O^{[N_{\rm b},\Delta N]}_{\mu,k,q}
  -
  (-1)^{k-q}
  \left(
    \widehat O^{[N_{\rm b},\Delta N]}_{\mu,k,-q}
  \right)^\dagger
  \right].
  \label{eq:main_def_Xminus}
\end{align}
These operators satisfy the spherical-tensor Hermiticity condition
\begin{align}
  \left( X^{[N_{\rm b},\Delta N]}_{\mu,k,q;\pm} \right)^\dagger
  =
  (-1)^{k-q}
  X^{[N_{\rm b},\Delta N]}_{\mu,k,-q;\pm},
  \label{eq:main_X_Hermiticity}
\end{align}
and transform as
\begin{alignat}{5}
 \mathcal P
  X^{[N_{\rm b},\Delta N]}_{\mu,k,q;\pm}
  \mathcal P^{-1}
  &=
  \sigma_{\mathcal{P}}
  X^{[N_{\rm b},\Delta N]}_{\mu,k,q;\pm},
  &\qquad &
  \sigma_{\mathcal{P}} = \pm 1,
   \label{eq:main_X_P_symmetry}
  \\
  \mathcal T
  X^{[N_{\rm b},\Delta N]}_{\mu,k,q;\pm}
  \mathcal T^{-1}
  &=
  \sigma_{\mathcal{T}}
  (-1)^{k-q}
  X^{[N_{\rm b},\Delta N]}_{\mu,k,-q;\pm},
  &\qquad&
  \sigma_{\mathcal{T}} = \pm 1,
   \label{eq:main_X_T_symmetry}
  \\
  \mathcal C_A
  X^{[N_{\rm b},\Delta N]}_{\mu,k,q;\pm}
  \mathcal C_A^{-1}
  &=
  \sigma_{C_{A}}
  X^{[N_{\rm b},\Delta N]}_{\mu,k,q;\pm},
  &\qquad&
  \sigma_{C_{A}} = \sigma_{\mathcal{T}}\sigma_{\Gamma} = (-1)^{N_{\rm b}},
   \label{eq:main_X_CA_symmetry}
  \\
  \Gamma
  X^{[N_{\rm b},\Delta N]}_{\mu,k,q;\pm}
  \Gamma^{-1}
  &=
  \sigma_{\Gamma}
  (-1)^{k-q}
  X^{[N_{\rm b},\Delta N]}_{\mu,k,-q;\pm},
 & \qquad&
  \sigma_{\Gamma} = (-1)^{N_{\rm b}} \sigma_{\mathcal{T}},
   \label{eq:main_X_Gamma_symmetry}
  \\
  \mathcal G_{\pi/2}
  X^{[N_{\rm b},\Delta N]}_{\mu,k,q;\pm}
  \mathcal G_{\pi/2}^{-1}
  &=
 \sigma_{\mathcal{G}}
  X^{[N_{\rm b},\Delta N]}_{\mu,k,q;\pm},
  &\qquad&
  \sigma_{\mathcal{G}} =  (-1)^{\Delta N/2}.
   \label{eq:main_X_G_symmetry}
\end{alignat}
Therefore, the symmetry-adapted Hermitian $N_{\rm b}$-body rank-$k$ multipole $X^{[N_{\rm b},\Delta N]}_{\mu,k,q;\pm}$ is characterized by
\begin{align}
  \left(
    \sigma_{\mathcal P},
    \sigma_{\mathcal T},
    \sigma_{\mathcal C_A},
    \sigma_{\mathcal G}
  \right)
  =
  \left(
    \sigma_{\mathcal P},
    \sigma_{\mathcal T},
    (-1)^{N_{\rm b}},
    (-1)^{\Delta N/2}
  \right),
  \label{eq:main_X_summary_labels_CA}
\end{align}
or equivalently by
\begin{align}
  \left(
    \sigma_{\mathcal P},
    \sigma_{\mathcal T},
    \sigma_{\Gamma},
    \sigma_{\mathcal G}
  \right)
  =
  \left(
    \sigma_{\mathcal P},
    \sigma_{\mathcal T},
    (-1)^{N_{\rm b}}
    \sigma_{\mathcal T},
    (-1)^{\Delta N/2}
  \right).
  \label{eq:main_X_summary_labels_Gamma}
\end{align}
Indeed, when they are transformed into a real (tesseral) basis, the multipoles $X^{[N_{\rm b},\Delta N]}_{\mu,k,q;\pm}$ acquire definite parities under the four symmetry operations.
The inversion and time-reversal parities determine the conventional multipole type,
\begin{align}
Q_{\mu,k,q}^{[N_{\rm b},\Delta N]} &: \quad (\sigma_{\mathcal P},\sigma_{\mathcal T}) = \left((-1)^k,+1\right), \notag\\
M_{\mu,k,q}^{[N_{\rm b},\Delta N]} &: \quad (\sigma_{\mathcal P},\sigma_{\mathcal T}) = \left((-1)^{k+1},-1\right), \notag\\
T_{\mu,k,q}^{[N_{\rm b},\Delta N]} &: \quad (\sigma_{\mathcal P},\sigma_{\mathcal T}) = \left((-1)^k,-1\right), \notag\\
G_{\mu,k,q}^{[N_{\rm b},\Delta N]} &: \quad (\sigma_{\mathcal P},\sigma_{\mathcal T}) = \left((-1)^{k+1},+1\right).
\label{eq:C_conventional_QMTG}
\end{align}
The additional labels $(\sigma_{\mathcal C_A},  \sigma_{\mathcal G}) = \left((-1)^{N_{\rm b}},(-1)^{\Delta N/2}\right)$ or equivalently by $(\sigma_{\Gamma},\sigma_{\mathcal G})$ further resolve each conventional multipole sector into four many-body subclasses.
Consequently, the conventional fourfold classification is refined into the sixteenfold classification summarized in Table~\ref{tab:16class}.

\subsection{Representative sectors and local four-state example}

As the simplest nontrivial realization of the sixteenfold classification, we reconsider the local Hilbert space of a spinful $s$ orbital introduced in the Introduction.
Despite its minimal dimension, this example already contains representatives of several distinct many-body sectors and illustrates how the additional symmetry labels introduced in this work distinguish operators that are indistinguishable within the conventional multipole classification.

The four local Fock states are
\begin{align}
  \mathcal{H}_{s} = \left\{ \ket{0}, \ket{\uparrow}, \ket{\downarrow}, \ket{\uparrow\downarrow} \right\}.
\end{align}
The corresponding operator space consists of all linear operators acting on $\mathcal H_s$ and therefore has dimension $4^2=16$.
This space is decomposed according to fermion parity $(-1)^{N}$, where $N=n_\uparrow+n_\downarrow$ is the particle-number operator.
The even-parity sector is
\begin{align}
  \mathcal H_{\rm even} = \left\{ \ket{0},\ket{\uparrow\downarrow} \right\},
\end{align}
whereas the odd-parity sector is
\begin{align}
  \mathcal H_{\rm odd} = \left\{ \ket{\uparrow},\ket{\downarrow} \right\}.
\end{align}
Since all operators considered in the present work carry even fermion parity, they preserve the decomposition
$\mathcal H_{\rm even}\oplus\mathcal H_{\rm odd}$.
Consequently, the local even-fermion operator algebra consists of independent Hermitian operators acting within the even- and odd-parity sectors.
Its dimension is therefore
$\dim \mathcal O_{\rm even}=2^2+2^2=8$.

Following the tensor construction introduced in Sec.~\ref{subsec:many_body_construction}, the local operator basis is organized according to body number, particle-number-change sector, and total angular-momentum rank.
For a spinful $s$ orbital, the orbital degree of freedom is trivial, and the only angular momentum originates from spin, $j=1/2$.
The number-conserving one-body sector is therefore obtained from
\begin{align}
  \frac{1}{2}\otimes\frac{1}{2} = 0 \oplus 1 .
  \label{eq:main_spin_half_coupling}
\end{align}
In addition to the identity operator
\begin{align}
Q_0^{[0,0]} := I ,
\end{align}
which constitutes the unique zero-body E monopole, the scalar component with $k = 0$ gives the E monopole
\begin{align}
Q_{0}^{[1,0]} := \widehat N = N - I,
\label{eq:main_N}
\end{align}
where $I$ denotes the $4\times4$ identity operator.
The subtraction of the identity removes the trivial zero-body contribution and produces a symmetry-adapted one-body multipole.
On the other hand, the vector component with $k = 1$ gives the three spin M dipoles
\begin{align}
&M_{x}^{[1,0]} :=  \sigma_{x} =  c^{\dagger}_{\uparrow} c^{}_{\downarrow} + c^{\dagger}_{\downarrow} c^{}_{\uparrow},
\label{eq:main_Sx}\\
&M_{y}^{[1,0]} :=  \sigma_{y} = -i \left( c^{\dagger}_{\uparrow} c^{}_{\downarrow} - c^{\dagger}_{\downarrow} c^{}_{\uparrow} \right),
\label{eq:main_Sy}\\
&M_{z}^{[1,0]} :=  \sigma_{z} = n_{\uparrow} - n_{\downarrow}.
\label{eq:main_Sz}
\end{align}

We next consider the anomalous sector with particle-number change $\Delta N=\pm2$, with the pair-creation channel characterized by $N_+ = 2$ and $N_-=0$ and its Hermitian conjugate.
For a single $s$ orbital, fermionic antisymmetry leaves only the spin-singlet pair $c^{\dagger}_{\uparrow}c^{\dagger}_{\downarrow}$, which carries $j_+ = 0$.
Coupling this creation block to the vacuum block with $j_- = 0$ yields only a scalar multipole with $k=0$.
The corresponding Hermitian operators are E monopole and MT monopole, which is given by
\begin{align}
&Q_{0}^{[1,2]} := \eta_+ = c^{\dagger}_{\uparrow}c^{\dagger}_{\downarrow} + c^{}_{{\downarrow}}c^{}_{{\uparrow}},
\\
&T_{0}^{[1,2]} := \eta_- =  -i(c^{\dagger}_{\uparrow}c^{\dagger}_{\downarrow} - c^{}_{{\downarrow}}c^{}_{{\uparrow}}).
\end{align}
These operators act within the even-parity sector and mix the empty and doubly occupied states.

The remaining basis element is the two-body E monopole:
\begin{align}
  Q_{0}^{[2,0]} := \widehat U =
  \left(n_\uparrow-\frac{1}{2}\right)
  \left(n_\downarrow-\frac{1}{2}\right)
  =
  U-\frac{1}{2} N+\frac{1}{4} I .
\end{align}
This operator represents the local interaction channel and provides the simplest nontrivial example of a genuine two-body multipole.
Unlike $\widehat N$, it cannot be reduced to a one-body operator.
The lower-body terms are uniquely fixed by the symmetry-adapted construction.
Furthermore, $\Gamma\widehat U\Gamma^{-1}=\widehat U$, so that $\widehat U$ carries the chirality label $\sigma_\Gamma=+1$.

The resulting symmetry-adapted even-fermion basis is summarized in Table~\ref{tab:local_s_orbital_operators}.
In the present minimal $s$-orbital example, no ET multipole appears among the eight symmetry-adapted basis elements.
Their matrix representations in the ordered basis $\left\{ \ket{0}, \ket{\uparrow}, \ket{\downarrow}, \ket{\uparrow\downarrow} \right\}$ are
\begin{alignat}{4}
  I
  &=
  \begin{pmatrix}
  1&0&0&0\\
  0&1&0&0\\
  0&0&1&0\\
  0&0&0&1
  \end{pmatrix}, &\quad&
  \widehat N
  =
  \begin{pmatrix}
  -1&0&0&0\\
  0&0&0&0\\
  0&0&0&0\\
  0&0&0&1
  \end{pmatrix}, &\quad&
  \widehat U
  =
  \begin{pmatrix}
  \frac{1}{4}&0&0&0\\
  0&-\frac{1}{4}&0&0\\
  0&0&-\frac{1}{4}&0\\
  0&0&0&\frac{1}{4}
  \end{pmatrix},
    \cr
  \sigma_x
  &=
  \begin{pmatrix}
  0&0&0&0\\
  0&0&1&0\\
  0&1&0&0\\
  0&0&0&0
  \end{pmatrix}, &\quad&
  \sigma_y
  =
  \begin{pmatrix}
  0&0&0&0\\
  0&0&-i&0\\
  0&i&0&0\\
  0&0&0&0
  \end{pmatrix}, &\quad&
  \sigma_z
  =
  \begin{pmatrix}
  0&0&0&0\\
  0&1&0&0\\
  0&0&-1&0\\
  0&0&0&0
  \end{pmatrix},
  \cr
  \eta_+
  &=
  \begin{pmatrix}
  0&0&0&1\\
  0&0&0&0\\
  0&0&0&0\\
  1&0&0&0
  \end{pmatrix}, &\quad&
  \eta_-
  =
  \begin{pmatrix}
  0&0&0&i\\
  0&0&0&0\\
  0&0&0&0\\
  -i&0&0&0
  \end{pmatrix}\nonumber.
\end{alignat}

This simple local example already illustrates two central features of the many-body classification.
First, the conventional labels $(\mathrm{rank},\mathcal P,\mathcal T)$ do not distinguish all physically distinct local operators.
In particular, $\widehat N$, $\eta_+$, and $\widehat U$ are all electric monopoles with $(k,\sigma_{\mathcal P},\sigma_{\mathcal T}) =(0,+,+)$, yet they belong to three different sectors $(N_{\rm b},\Delta N) = (1,0),\, (1,2),\, (2,0)$.
The operator $\widehat N$ is a number-conserving one-body E monopole, $\eta_+$ is an anomalous one-body E monopole with $\Delta N=2$, and $\widehat U$ is a number-conserving two-body E monopole.
These operators are therefore indistinguishable within the conventional fourfold multipole classification but become distinct once the additional symmetry labels $\sigma_{\mathcal C_A}=(-1)^{N_{\rm b}}$ and $\sigma_{\mathcal G}=(-1)^{\Delta N/2}$ are included.
This provides the simplest explicit realization of the sixteenfold classification proposed in this work.

Second, symmetry-adapted many-body multipole operators need not be homogeneous polynomials of a fixed body number.
Rather, an $N_{\rm b}$-body multipole is defined by its highest-body component together with lower-body counterterms required by covariance under the canonical symmetries.
The operator $\widehat U$ provides the simplest example of this structure.
The many-body multipole classification is therefore fundamentally a classification of operator space rather than of angular-momentum rank alone.
The additional labels $\sigma_{\mathcal C_A}$ and $\sigma_{\mathcal G}$ encode information about body number and particle-number structure that is invisible in conventional multipole theory, thereby providing a complete symmetry-based organization of many-body operators.

\begin{table*}[t]
  \centering
  \caption{
  Hermitian even-fermion multipole operators in the four-dimensional local Fock space of a spinful $s$ orbital, $\{\ket{0},\ket{\uparrow},\ket{\downarrow},\ket{\uparrow\downarrow}\}$.
  The table illustrates how the conventional multipole classification is refined by the additional labels
$\sigma_{\mathcal C_A}=(-1)^{N_{\rm b}}$ and $\sigma_{\mathcal G}=(-1)^{\Delta N/2}$,
which distinguish body-number parity and particle-number-change sectors, respectively.
  The chiral label is defined by $\Gamma=\mathcal T\mathcal C_A$ and $\sigma_\Gamma=\sigma_{\mathcal T}\sigma_{\mathcal C_A}$.
  The symbols $+$ and $-$ in the symmetry columns denote the eigenvalues $+1$ and $-1$, respectively.
  }
  \label{tab:local_s_orbital_operators}
  \renewcommand{\arraystretch}{1.5}
  \setlength{\tabcolsep}{3pt}
  \footnotesize
  \begin{tabular}{cccccccccc}
    \hline\hline
    Multipole
    & Operator
    & Fermionic expression
    & \makecell{body-number \\ $N_{\rm b}$}
    & \makecell{particle-number change \\ $\Delta N \bmod 4$}
    & $\sigma_{\mathcal P}$
    & $\sigma_{\mathcal T}$
    & $\sigma_{\mathcal C_A}$
    & $\sigma_{\Gamma}$
    & $\sigma_{\mathcal G}$
    \\
    \hline
       $I$
    & $I$
    & $I$
    & $0$
    & $0$
    & $+$
    & $+$
    & $+$
    & $+$
    & $+$
    \\
    \hline
    $Q_{0}^{[1,0]}$
    & $\widehat N$
    & $n_\uparrow+n_\downarrow-I$
    & $1$
    & $0$
    & $+$
    & $+$
    & $-$
    & $-$
    & $+$
    \\
    $M_{1}^{[1,0]}$
    &  $\sigma_x$
    & $c^\dagger_\uparrow c_\downarrow
       +
       c^\dagger_\downarrow c_\uparrow$
    & $1$
    & $0$
    & $+$
    & $-$
    & $-$
    & $+$
    & $+$
    \\
    $M_{1}^{[1,0]}$
    & $\sigma_y$
    & $-i
       \left(
       c^\dagger_\uparrow c_\downarrow
       -
       c^\dagger_\downarrow c_\uparrow
       \right)$
    & $1$
    & $0$
    & $+$
    & $-$
    & $-$
    & $+$
    & $+$
    \\
    $M_{1}^{[1,0]}$
    & $\sigma_z$
    & $n_\uparrow-n_\downarrow$
    & $1$
    & $0$
    & $+$
    & $-$
    & $-$
    & $+$
    & $+$
    \\
    \hline
    $Q_{0}^{[1,2]}$
    & $\eta_+$
    & $c^\dagger_\uparrow c^\dagger_\downarrow + c_\downarrow c_\uparrow$
    & $1$
    & $2$
    & $+$
    & $+$
    & $-$
    & $-$
    & $-$
    \\
    $T_{0}^{[1,2]}$
    & $\eta_-$
    & $-i \left( c^\dagger_\uparrow c^\dagger_\downarrow - c_\downarrow c_\uparrow \right)$
    & $1$
    & $2$
    & $+$
    & $-$
    & $-$
    & $+$
    & $-$
    \\
    \hline
    $Q_{0}^{[2,0]}$
    & $\widehat U$
    &  \makecell{$\left(n_\uparrow-\frac{1}{2}\right) \left(n_\downarrow-\frac{1}{2}\right)$ \\
       $= n_\uparrow n_\downarrow - \frac{1}{2} (n_\uparrow+n_\downarrow) + \frac{1}{4} I$}
    & $2$
    & $0$
    & $+$
    & $+$
    & $+$
    & $+$
    & $+$
    \\
    \hline\hline
  \end{tabular}
\end{table*}

%%%%%%%%%%%%%%%%%%%%%%%%%%%%%%%%%%%%%%%%%%%%%%%%%%%%%%%%%%%%%%%%%% 

\section{Selection rules for multipole expectation values}
\label{sec:example}

The sixteenfold classification constructed in Sec.~\ref{sec:classification} is a classification of operator sectors.  
Its physical significance becomes apparent when one asks which multipole expectation values can be induced by a given Hamiltonian.  
In this section, we illustrate this point using two finite-dimensional toy models.  
The key observation is that expectation values allowed within the conventional $(\mathrm{rank},\mathcal P,\mathcal T)$ classification may still be forbidden, or may require additional induction conditions, once the additional symmetries are taken into account. 
We therefore focus on selection rules associated with $\mathcal C_A$, $\mathcal G_{\pi/2}$, and $\Gamma=\mathcal T\mathcal C_A$, as well as their interplay with conventional $(\mathcal P,\mathcal T)$ and point-group symmetries.

Throughout this section, we employ the following elementary symmetry argument.  
Let $U$ be a unitary or antiunitary symmetry of the Hamiltonian, and suppose that the equilibrium density matrix is invariant under $U$.  
If an operator $X$ is odd under $U$, 
\begin{align}
  U H U^{-1}=H,
  \qquad
  U X U^{-1}
  =
  \sigma_{U}(X) X,
  \qquad
  \sigma_{U}(X)=-1,
\end{align}
then its expectation value necessarily vanishes:
\begin{align}
  \langle X\rangle
  =
  \frac{1}{Z}\mathrm{Tr}\!\left(e^{-\beta H}X\right) =0,
  \qquad
  Z=
  \mathrm{Tr}\,e^{-\beta H}.
  \label{eq:selection_rule_general}
\end{align}
Accordingly, a nonzero expectation value of $X$ requires the corresponding symmetry to be absent or broken.
In the numerical examples below, we set $\beta=1$, or equivalently $k_{\rm B}T=1$. 
Our purpose is to identify which multipoles are symmetry-allowed; the magnitudes of the induced expectation values depend on the microscopic details of the model and are not discussed.

\subsection{$N=2$ sector of a spinless $s$-$p$ system}

We first consider the fixed-particle-number $N=2$ sector of a spinless $s$-$p$ system. 
This sector contains two $J=1$ triplets: one originating from the $sp$ configuration and the other from the $p^2$ configuration. 
We use the ordered basis
\begin{align}
  \{
  |p_1s\rangle,\,
  |p_0s\rangle,\,
  |p_0p_1\rangle,\,
  |p_{-1}s\rangle,\,
  |p_{-1}p_1\rangle,\,
  |p_{-1}p_0\rangle
  \},
  \label{eq:sp_N2_basis_main}
\end{align}
which spans a six-dimensional Hilbert space.
Operators acting within this sector are therefore represented by $6\times6$ matrices, and the corresponding Hermitian operator basis consists of 36 independent operators, including the identity.

Since all operators considered in this subsection conserve the particle number, $\Delta N=0$ and hence $\sigma_{\mathcal G}=+1$.
The gauge label is therefore trivial and will be omitted.
Instead, the relevant operators are classified by
\begin{align}
  (\sigma_{\mathcal P},\sigma_{\mathcal T},\sigma_{\Gamma}),
  \qquad
  \Gamma=\mathcal T\mathcal C_A.
\end{align}
Equivalently, since $\sigma_{\Gamma}=(-1)^{N_{\rm b}}\sigma_{\mathcal T}$, the $\Gamma$ label distinguishes one-body and two-body number-conserving operators that share the same conventional $(\mathcal P,\mathcal T)$ labels.

The active one-body multipoles in this sector include
\[
Q^{[1]}_0,\qquad M^{[1]}_1,\qquad Q^{[1]}_2,
\]
as well as hybrid $s$-$p$ multipoles
\[
Q^{[1]}_1,\qquad T^{[1]}_1 .
\]
The active two-body multipoles include
\[
Q^{[2]}_0,\qquad M^{[2]}_1,\qquad Q^{[2]}_2,
\]
together with the two-body hybrid multipoles
\[
G^{[2]}_0,\qquad M^{[2]}_0,\qquad G^{[2]}_2,\qquad M^{[2]}_2 .
\]
The explicit matrix representations of these multipoles are given in Appendix~\ref{app:matrix}.

\begin{table*}[htb]
  \centering
  \scriptsize
  \setlength{\tabcolsep}{2.5pt}
  \caption{
    Representative nonzero multipole expectation values in the fixed-particle-number $N=2$ sector of the spinless $s$-$p$ system.
    The triplets shown in parentheses denote the symmetry labels $(\sigma_{\mathcal P},\sigma_{\mathcal T},\sigma_{\Gamma})$.
    Checkmarks indicate multipoles whose expectation values can become nonzero for the corresponding Hamiltonian.
    Since all operators considered in this subsection conserve particle number, $\Delta N=0$ and the gauge label $\sigma_{\mathcal G}=+1$ is omitted.
 }
    \label{tab:exp}
  \resizebox{\textwidth}{!}{%
  \begin{tabular}{|c|*{12}{c|}} \hline&
        \shortstack{$\langle Q_0^{[1]} \rangle$ \\ {\scriptsize $(+,+,-)$}} & 
        \shortstack{$\langle M_1^{[1]} \rangle$\\ {\scriptsize $(+,-,+)$}} & 
        \shortstack{$\langle Q_2^{[1]} \rangle$\\ {\scriptsize $(+,+,-)$}} & 
        \shortstack{$\langle Q_1^{[1]} \rangle$\\ {\scriptsize $(-,+,-)$}} & 
        \shortstack{$\langle T_1^{[1]} \rangle$\\ {\scriptsize $(-,-,+)$}} & 
        \shortstack{$\langle Q_0^{[2]} \rangle$\\ {\scriptsize $(+,+,+)$}} & 
        \shortstack{$\langle M_1^{[2]} \rangle$\\ {\scriptsize $(+,-,-)$}} & 
        \shortstack{$\langle Q_2^{[2]} \rangle$\\ {\scriptsize $(+,+,+)$}}  & 
        \shortstack{$\langle G_0^{[2]} \rangle$\\ {\scriptsize $(-,+,+)$}} & 
        \shortstack{$\langle M_0^{[2]} \rangle$\\ {\scriptsize $(-,-,-)$}} & 
        \shortstack{$\langle G_2^{[2]} \rangle$\\ {\scriptsize $(-,+,+)$}} & 
        \shortstack{$\langle M_2^{[2]} \rangle$\\ {\scriptsize $(-,-,-)$}} \\ \hline
    \shortstack{$H_1 = M_1^{[1]}$\\{\scriptsize $(+,-,+)$}} 
      &  & $\checkmark$ &  &  &  & $\checkmark$ &  & $\checkmark$ &  &  &  &  \\ \hline
    \shortstack{$H_2 = T_1^{[1]}+M_1^{[1]}$\\{\scriptsize $(\times,-,+)$}} 
      &  & $\checkmark$ &  &  & $\checkmark$ & $\checkmark$ &  & $\checkmark$ & $\checkmark$ &  & $\checkmark$ &  \\ \hline   
    \shortstack{$H_3 = Q_1^{[1]}+Q_2^{[1]}$\\{\scriptsize $(\times,+,-)$}} 
      & $\checkmark$ &  & $\checkmark$ & $\checkmark$ &  & $\checkmark$ &  & $\checkmark$ &  &  & $\checkmark$ &  \\ \hline
    \shortstack{$H_4 = Q_1^{[1]}+Q_2^{[1]}+Q_2^{[2]}$\\{\scriptsize $(\times,+,\times)$}} 
      & $\checkmark$ &  & $\checkmark$ & $\checkmark$ &  & $\checkmark$ &  & $\checkmark$ & $\checkmark$ &  & $\checkmark$ &  \\ \hline           
  \end{tabular}%
  }
  % }
\end{table*}

To illustrate the resulting selection rules, we evaluate thermal expectation values for four representative Hamiltonians constructed from these multipoles.
The resulting expectation values are summarized in Table~\ref{tab:exp}.
The rows of Table~\ref{tab:exp} correspond to the four representative Hamiltonians, while the columns represent candidate multipole expectation values.
Checkmarks indicate multipoles whose thermal expectation values can become nonzero.
The table already reveals that multipoles sharing the same conventional $(\mathcal P,\mathcal T)$ labels do not necessarily exhibit the same expectation-value patterns.
As we show below, these differences originate from the additional chiral label $\Gamma$.
The allowed expectation values are determined by symmetry.
According to Eq.~\eqref{eq:selection_rule_general}, if a Hamiltonian $H$ is invariant under a symmetry transformation $U$, then the expectation value of any $U$-odd operator must vanish.
Therefore, only operators that are even under every preserved symmetry can acquire nonzero expectation values.

We now discuss the four representative Hamiltonians listed in Table~\ref{tab:exp}.
Throughout the following discussion, a symbol such as $M_1^{[1]}$ denotes a generic real linear combination of all components of the corresponding multipole.
For example,
\[
H=M_1^{[1]}
\]
stands for
\[
H
=
c_x M_x^{[1]}
+
c_y M_y^{[1]}
+
c_z M_z^{[1]},
\]
with generic real coefficients $c_x,c_y,c_z$.
This convention avoids accidental degeneracies and additional constraints associated with special coefficient choices or residual point-group symmetries, allowing us to focus on symmetry selection rules organized by
$(\mathcal P,\mathcal T,\Gamma)$.

We first consider
\[
 H_1=M_1^{[1]} .
\]
The corresponding symmetry labels are
\[
(\sigma_{\mathcal P},\sigma_{\mathcal T},\sigma_{\Gamma})
=
(+,-,+),
\]
so that $H_1$ preserves $\mathcal P$ and $\Gamma$ while breaking $\mathcal T$.
According to Eq.~\eqref{eq:selection_rule_general}, only operators that are even under both $\mathcal P$ and $\Gamma$ can acquire nonzero expectation values.
Table~\ref{tab:exp} therefore gives
\[
\langle M_1^{[1]}\rangle,\qquad
\langle Q_0^{[2]}\rangle,\qquad
\langle Q_2^{[2]}\rangle .
\]
This already demonstrates a selection rule beyond the conventional $(\mathcal P, \mathcal T)$ classification. 
The one-body E multipoles $Q_0^{[1]}$ and $Q_2^{[1]}$ have the same conventional $(\mathcal P,\mathcal T)=(+,+)$ labels as the two-body E multipoles $Q_0^{[2]}$ and $Q_2^{[2]}$, but they differ in $\Gamma$ parity. 
The one-body E multipoles are $\Gamma$-odd and therefore remain forbidden, whereas the two-body E multipoles are $\Gamma$-even and can be induced.

Next, consider
\[
H_2 = T_1^{[1]}+M_1^{[1]} .
\]
This Hamiltonian preserves $\Gamma$ and breaks $\mathcal T$, while inversion is no longer a good symmetry label:
\[
(\sigma_{\mathcal P},\sigma_{\mathcal T},\sigma_{\Gamma}) = (\times,-,+).
\]
Consequently, $\mathcal P$-odd and $\Gamma$-even channels, which were forbidden for $H=M_1^{[1]}$, become allowed. 
Table~\ref{tab:exp} shows that
\[
\langle G_0^{[2]}\rangle,\qquad
\langle G_2^{[2]}\rangle,\qquad
\langle T_1^{[1]}\rangle,
\]
become nonzero. 
In particular, the two-body ET monopole $G_0^{[2]}$ is induced once spatial inversion is broken while the relevant $\Gamma$-even sectors remain allowed.
This provides a simple example in which the spinless two-body ET monopole, absent in the onsite spinless one-body space, emerges through the hybridization of one-body multipoles~\cite{kuniyoshi2026theory}.

We next consider a time-reversal-even Hamiltonian,
\[
H_3=Q_1^{[1]}+Q_2^{[1]} .
\]
This Hamiltonian preserves $\mathcal T$ but breaks $\mathcal P$. 
From the conventional $(\mathcal P,\mathcal T)$ point of view, $\mathcal P$-odd and $\mathcal T$-even multipoles such as $G_0^{[2]}$ are not symmetry-forbidden. 
Table~\ref{tab:exp} indeed shows that several $\mathcal T$-even multipoles acquire expectation values, including
\[
\langle Q_0^{[1]}\rangle,\qquad
\langle Q_2^{[1]}\rangle,\qquad
\langle Q_0^{[2]}\rangle,\qquad
\langle Q_2^{[2]}\rangle,\qquad
\langle Q_1^{[1]}\rangle,\qquad
\langle G_2^{[2]}\rangle .
\]
However, the two-body ET monopole $G_0^{[2]}$ remains absent:
\[
\langle G_0^{[2]}\rangle=0 .
\]
This absence cannot be explained by the conventional $(\mathcal P, \mathcal T)$ labels alone. 
Instead, it originates from the chiral %structure
symmetry associated with the chiral transform $\Gamma=\mathcal T\mathcal C_A$.

For this Hamiltonian, one finds
\begin{align}
  \Gamma H_3\Gamma^{-1}=-H_3,
\label{eq:Gamma_anticommutes_H_Q1Q2}
\end{align}
so $\Gamma$ is not a conserved symmetry, but a chiral symmetry of $H$. 
For a nonzero-energy eigenstate satisfying
\[
H_3|\psi_E\rangle
=
E|\psi_E\rangle,
\qquad
E\neq0,
\]
one obtains
\[
E\langle \psi_E|\Gamma|\psi_E\rangle
=
\langle \psi_E|\Gamma H_3|\psi_E\rangle
=
-\langle \psi_E|H_3\Gamma|\psi_E\rangle
=
-E\langle \psi_E|\Gamma|\psi_E\rangle .
\]
Therefore,
\[
\langle \psi_E|\Gamma|\psi_E\rangle=0, \qquad (E \neq 0).
\]
This is a well-known consequence of the chiral antisymmetry.~\cite{Janos-topological-2016,Poli_2017}
Within the present $N=2$ sector, the unitary operator $\Gamma$ is represented, up to normalization and basis convention, by the two-body rank-zero ET multipole $G_0^{[2]}$:
\[
\Gamma \propto G_0^{[2]} .
\]
For generic coefficients, $H_3$ has no zero mode in this restricted sector.
This can be checked directly from the explicit matrix representations in Appendix~\ref{app:matrix}.
Therefore, the thermal expectation value also vanishes,
\[
\langle G_0^{[2]}\rangle
=
\frac{1}{Z}
\sum_E e^{-E}
\langle \psi_E|G_0^{(2)}|\psi_E\rangle
=0 .
\]
The absence of $G_0^{[2]}$ is therefore a genuine chiral selection rule.
It is invisible within the conventional $(\mathcal P,\mathcal T)$ classification but becomes transparent once the additional label $\Gamma$ is taken into account.

This interpretation is confirmed by the last row of Table~\ref{tab:exp}.
When the two-body E quadrupole $Q_2^{[2]}$ is added,
\[
H_4=Q_1^{[1]}+Q_2^{[1]}+Q_2^{[2]} .
\]
the chiral relation $\Gamma H\Gamma^{-1}=-H$ is no longer satisfied.
Consequently, the chiral selection rule discussed above ceases to apply, and Table~\ref{tab:exp} shows that
\[
\langle G_0^{[2]}\rangle\neq0 .
\]

The contrast between $H_3$ and $H_4$ therefore illustrates the central role of the additional label $\Gamma$.  
From the viewpoint of the conventional $(\mathcal P,\mathcal T)$ classification, the ET monopole $G_0^{[2]}$ is symmetry-allowed in both cases because $\mathcal P$ is broken while $\mathcal T$ is preserved.  
Nevertheless, $G_0^{[2]}$ is absent for $H_3$ and present for $H_4$.  
The distinction originates from the chiral %structure
symmetry associated with $\Gamma$: $H_3$ satisfies the anticommutation relation (\ref{eq:Gamma_anticommutes_H_Q1Q2}),
whereas $H_4$ does not.  
The appearance or absence of $G_0^{[2]}$ is therefore controlled not by the conventional $(\mathcal P,\mathcal T)$ labels alone, but by the additional chiral symmetry encoded in the sixteenfold classification.

\subsection{Many-body cluster multipoles in a four-site square cluster}
We next apply the sixteenfold classification to real-space cluster multipoles. 
As an example, we consider spinless $s$ electrons on a four-site square cluster, including particle-number-nonconserving operators. 
We label the four sites by A,B,C,D, with coordinates
\[
{\rm A}=(-1,-1),\quad
{\rm B}=(1,1),\quad
{\rm C}=(1,-1),\quad
{\rm D}=(-1,1).
\]
The four symmetry-adapted one-particle cluster orbitals are denoted by:
\begin{align}
\ket{0} &:\ (1,1,1,1),\\
\ket{x} &:\ (-1,1,1,-1),\\
\ket{y} &:\ (-1,1,-1,1),\\
\ket{xy} &:\ (1,1,-1,-1),
\end{align}
where the labels indicate their transformation properties under the point group symmetry of the square cluster. 
Correspondingly, from the site-basis creation operators ($c_{{\rm A}}^\dagger,c_{{\rm B}}^\dagger,c_{{\rm C}}^\dagger,c_{{\rm D}}^\dagger$), we define symmetry-adapted
 creation operators for cluster orbitals as
\begin{align}
\begin{pmatrix}
c_0^\dagger \\
c_x^\dagger \\
c_y^\dagger \\
c_{xy}^\dagger
\end{pmatrix}
=
W
\begin{pmatrix}
c_{{\rm A}}^\dagger \\
c_{{\rm B}}^\dagger \\
c_{{\rm C}}^\dagger \\
c_{{\rm D}}^\dagger
\end{pmatrix},
\qquad
W
=
\frac{1}{2}
\begin{pmatrix}
1 & 1 & 1 & 1\\
-1 & 1 & 1 & -1\\
-1 & 1 & -1 & 1\\
1 & 1 & -1 & -1
\end{pmatrix}.
\label{eq:cluster_mode_def}
\end{align}
Since the transformation matrix satisfies
$W W^\dagger = I$,
Eq.~\eqref{eq:cluster_mode_def} is a canonical transformation~\cite{Haule-2007,Walsh-2023}.
Thus one-electron operators can be represented equivalently in the symmetry-adapted cluster basis instead of the site basis. 

We work in the even-Fock subspace spanned by
\begin{align}
  \{
  |\mathrm{vac.}\rangle,\,
  |0,x\rangle,\,
  |0,y\rangle,\,
  |0,xy\rangle,\,
  |x,y\rangle,\,
  |y,xy\rangle,\,
  |x,xy\rangle,\,
  |\mathrm{full}\rangle
  \}.
  \label{eq:four_site_even_basis_main}
\end{align}
This Hilbert space is eight-dimensional
, and the full operator algebra acting on it has dimension $8\times8=64$.

For clarity, we restrict the following discussion to a $\mathcal{T}$-even Hermitian operator set composed of normal one-body, normal two-body, and anomalous pairing-type operators of the forms
\[
c_\mu^\dagger c_\nu + c_\nu^\dagger c_\mu,\qquad
c_\mu^\dagger c_\nu^\dagger + c_\nu c_\mu,\qquad
c_\mu^\dagger c_\nu^\dagger c_\rho c_\sigma .
  \label{eq:cluster_operator_forms}
\]
The explicit matrix representations are listed in Appendix~\ref{app:matrix}.  
Since all operators in this restricted set are $\mathcal T$ even, one has $\sigma_\Gamma=\sigma_{\mathcal C_A}$. 
We therefore use the three labels
\begin{align}
  (\sigma_{\mathcal P},\sigma_{\mathcal C_A},\sigma_{\mathcal G})
\end{align}
to display the selection rules.  
Normal one-body operators have $[N_{\rm b},\Delta N]=[1,0]$, normal two-body operators have $[2,0]$, and anomalous pairing-type operators have $[1,2]$.

Within this restricted $\mathcal T$-even set, the independent multipoles except for the identity include normal one-body multipoles:
\begin{align}
  &
  Q_0^{[1,0]},
  Q_0^{\prime[1,0]},
  Q_{xy}^{[1,0]},
  Q_v^{[1,0]},
  Q_{xy}^{\prime[1,0]},
  Q_x^{[1,0]},
  Q_x^{\prime[1,0]},
  Q_y^{[1,0]},
  Q_y^{\prime[1,0]},
\end{align}
normal two-body multipoles
\begin{align}
  &
  Q_0^{[2,0]},
  Q_0^{\prime[2,0]},
  Q_0^{\prime\prime[2,0]},
  Q_v^{[2,0]},
  Q_v^{\prime[2,0]},
  Q_v^{\prime\prime[2,0]},
  Q_{xy}^{[2,0]},
  G_z^{[2,0]},
  Q_x^{[2,0]},
  Q_x^{\prime[2,0]},
  Q_y^{[2,0]},
  Q_y^{\prime[2,0]},
\end{align}
and anomalous multipoles
\begin{align}
  &
  \tilde G_z^{[1,2]},
  \tilde Q_{xy}^{[1,2]},
  \tilde Q_x^{[1,2]},
  \tilde Q_x^{\prime[1,2]},
  \tilde Q_y^{[1,2]},
  \tilde Q_y^{\prime[1,2]} .
\end{align}
Here $v$ denotes the $x^2-y^2$-type channel.  
Primes distinguish multiple operators belonging to the same point-group representation and the same symmetry sector.

To avoid basis-dependent redundancy, operators with identical symmetry labels are grouped into generic linear combinations.
For example,
\begin{align}
\mathcal Q_0^{[1,0]}
&:= \alpha_1 Q_0^{[1,0]} + \alpha_2 Q_0^{\prime[1,0]},\\
\mathcal Q_{xy}^{[1,0]}
&:= \alpha_3 Q_{xy}^{[1,0]} + \alpha_4 Q_{xy}^{\prime[1,0]},\\
\mathcal Q_x^{[1,0]}
&:= \alpha_5 Q_x^{[1,0]} + \alpha_6 Q_x^{\prime[1,0]},\\
\mathcal Q_y^{[1,0]}
&:= \alpha_7 Q_y^{[1,0]} + \alpha_8 Q_y^{\prime[1,0]},\\
\mathcal Q_0^{[2,0]}
&:= \beta_1 Q_0^{[2,0]} + \beta_2 Q_0^{\prime[2,0]} + \beta_3 Q_0^{\prime\prime[2,0]},\\
\mathcal Q_v^{[2,0]}
&:= \beta_4 Q_v^{[2,0]} + \beta_5 Q_v^{\prime[2,0]} + \beta_6 Q_v^{\prime\prime[2,0]},\\
\mathcal Q_x^{[2,0]}
&:= \beta_7 Q_x^{[2,0]} + \beta_8 Q_x^{\prime[2,0]},\\
\mathcal Q_y^{[2,0]}
&:= \beta_9 Q_y^{[2,0]} + \beta_{10} Q_y^{\prime[2,0]},\\
\tilde{\mathcal Q}_x^{[1,2]}
&:= \gamma_1 \tilde Q_x^{[1,2]} + \gamma_2 \tilde Q_x^{\prime[1,2]},\\
\tilde{\mathcal Q}_y^{[1,2]}
&:= \gamma_3 \tilde Q_y^{[1,2]} + \gamma_4 \tilde Q_y^{\prime[1,2]} .
\end{align}
In the following tables and discussion, we suppress the calligraphic font and write these generic linear combinations simply as $Q$ or $\tilde Q$ when no confusion arises.  
Operators that appear only once in the present restricted set, such as $Q_v^{[1,0]}$, $Q_{xy}^{[2,0]}$, $G_z^{[2,0]}$, $\tilde G_z^{[1,2]}$, and $\tilde Q_{xy}^{[1,2]}$, are kept as individual basis operators.

The relevant multipoles are classified by $(\sigma_{\mathcal P},\sigma_{\mathcal C_A},\sigma_{\mathcal G})$ as follows:
\begin{align}
(+,-,+)&:\quad Q_0^{[1,0]},\ Q_{xy}^{[1,0]},\ Q_v^{[1,0]},\\
(-,-,+)&:\quad Q_x^{[1,0]},\ Q_y^{[1,0]},\\
(+,+,+)&:\quad Q_0^{[2,0]},\ Q_v^{[2,0]},\ Q_{xy}^{[2,0]},\ G_z^{[2,0]},\\
(-,+,+)&:\quad Q_x^{[2,0]},\ Q_y^{[2,0]},\\
(+,-,-)&:\quad \tilde G_z^{[1,2]},\ \tilde Q_{xy}^{[1,2]},\\
(-,-,-)&:\quad \tilde Q_x^{[1,2]},\ \tilde Q_y^{[1,2]}.
\label{eq:cluster_classification_PCG}
\end{align}
This classification explicitly shows how operators with the same spatial character are further resolved by the body-number label $\sigma_{\mathcal C_A}$ and the anomalous-sector label $\sigma_{\mathcal G}$.

We now evaluate expectation values for the following five representative Hamiltonians:
\begin{align}
H_1
&=
Q_x^{[1,0]} + Q_y^{[1,0]}
&\qquad&
\left(
\mathcal G_{\pi/2},\ 
\mathcal P\mathcal C_A=\mathrm{even}
\right),\\
H_2
&=
Q_0^{[2,0]} + Q_x^{[2,0]} + Q_y^{[2,0]}
&\qquad&
\left(
\mathcal G_{\pi/2},\ 
\mathcal C_A=\mathrm{even}
\right),\\
H_3
&=
Q_0^{[1,0]} + Q_{xy}^{[1,0]} + \tilde Q_x^{[1,2]}
&\qquad&
\left(
\mathcal P\mathcal G_{\pi/2}=\mathrm{even}
\right),\\
H_4
&=
Q_0^{[1,0]} + Q_v^{[1,0]} + \tilde Q_x^{[1,2]}
&\qquad&
\left(
\mathcal P\mathcal G_{\pi/2}=\mathrm{even},\
\mathcal G_{\pi/2}\mathcal M_x=\mathrm{even}
\right),\\
H_5
&=
Q_0^{[1,0]} + Q_v^{[1,0]} + \tilde Q_x^{[1,2]} + \tilde Q_{xy}^{[1,2]}
&\qquad&
\left(
\mathcal G_{\pi/2}\mathcal M_x=\mathrm{even}
\right).
\end{align}
The parentheses indicate representative symmetries preserved by each Hamiltonian. 
Here $\mathcal M_x$ denotes mirror reflection with respect to the plane $x=0$. 
The resulting nonzero multipole expectation values for each Hamiltonian are summarized in Table~\ref{tab:exp_H1_H5}.

\begin{table*}[t]
  \centering
  \scriptsize
  \setlength{\tabcolsep}{2.5pt}
  \caption{
  Nonzero multipole expectation values for the representative four-site cluster Hamiltonians $H_1$--$H_5$.
  Checkmarks indicate multipoles whose expectation values can become nonzero.
The symmetry labels shown in parentheses denote $(\sigma_{\mathcal P},\sigma_{\mathcal C_A},\sigma_{\mathcal G})$.
  A symbol $\times$ indicates that the corresponding Hamiltonian is not an eigenoperator of the
  associated symmetry
  transformation.  
  }
  \label{tab:exp_H1_H5}
  \resizebox{\textwidth}{!}{
  \begin{tabular}{|c|*{15}{c|}} \hline
    &
        \shortstack{$\langle Q_0^{[1,0]} \rangle$\\ {\scriptsize $(+,-,+)$}} &
        \shortstack{$\langle Q_x^{[1,0]} \rangle$\\ {\scriptsize $(-,-,+)$}} &
        \shortstack{$\langle Q_y^{[1,0]} \rangle$\\ {\scriptsize $(-,-,+)$}} &
        \shortstack{$\langle Q_{xy}^{[1,0]} \rangle$\\ {\scriptsize $(+,-,+)$}} &
        \shortstack{$\langle Q_v^{[1,0]} \rangle$\\ {\scriptsize $(+,-,+)$}} &
        \shortstack{$\langle Q_0^{[2,0]} \rangle$\\ {\scriptsize $(+,+,+)$}} &
        \shortstack{$\langle Q_v^{[2,0]} \rangle$\\ {\scriptsize $(+,+,+)$}} &
        \shortstack{$\langle Q_{xy}^{[2,0]} \rangle$\\ {\scriptsize $(+,+,+)$}} &
        \shortstack{$\langle G_z^{[2,0]} \rangle$\\ {\scriptsize $(+,+,+)$}} &
        \shortstack{$\langle Q_x^{[2,0]} \rangle$\\ {\scriptsize $(-,+,+)$}} &
        \shortstack{$\langle Q_y^{[2,0]} \rangle$\\ {\scriptsize $(-,+,+)$}} &
        \shortstack{$\langle \tilde Q_x^{[1,2]} \rangle$\\ {\scriptsize $(-,-,-)$}} &
        \shortstack{$\langle \tilde Q_y^{[1,2]} \rangle$\\ {\scriptsize $(-,-,-)$}} &
        \shortstack{$\langle \tilde G_z^{[1,2]} \rangle$\\ {\scriptsize $(+,-,-)$}} &
        \shortstack{$\langle \tilde Q_{xy}^{[1,2]} \rangle$\\ {\scriptsize $(+,-,-)$}}
        \\ \hline
    \shortstack{$H_1$\\{\scriptsize $(-,-,+)$}}
      &  & $\checkmark$ & $\checkmark$ &  &  & $\checkmark$ & $\checkmark$ & $\checkmark$ & $\checkmark$ &  &  &  &  &  &  \\ \hline
    \shortstack{$H_2$\\{\scriptsize $(-,+,+)$}}
      &  &  &  &  &  & $\checkmark$ & $\checkmark$ & $\checkmark$ & $\checkmark$ & $\checkmark$ & $\checkmark$ &  &  &  &  \\ \hline
    \shortstack{$H_3$\\{\scriptsize $(\times,-,\times)$}}
      & $\checkmark$ &  &  & $\checkmark$ & $\checkmark$ & $\checkmark$ & $\checkmark$ & $\checkmark$ & $\checkmark$ &  &  & $\checkmark$ & $\checkmark$ &  &  \\ \hline
    \shortstack{$H_4$\\{\scriptsize $(\times,-,\times)$}}
      & $\checkmark$ &  &  &  & $\checkmark$ & $\checkmark$ & $\checkmark$ &  &  &  &  & $\checkmark$ &  &  &  \\ \hline
    \shortstack{$H_5$\\{\scriptsize $(\times,-,\times)$}}
      & $\checkmark$ &  & $\checkmark$ &  & $\checkmark$ & $\checkmark$ & $\checkmark$ &  &  &  & $\checkmark$ & $\checkmark$ &  & $\checkmark$ & $\checkmark$ \\ \hline
  \end{tabular}
  }
\end{table*}

The first Hamiltonian $H_1$ is odd under both $\mathcal P$ and $\mathcal C_A$, but it preserves their product symmetry $\mathcal P\mathcal C_A$.
Therefore, expectation values that are odd under $\mathcal P\mathcal C_A$ are forbidden.
In particular, although the two-body operators $Q_x^{[2,0]}$ and $Q_y^{[2,0]}$ have the same spatial parity as $Q_x^{[1,0]}$ and $Q_y^{[1,0]}$, their $\mathcal C_A$ parity is opposite.  
Hence,
\[
\langle Q_x^{[2,0]} \rangle
=
\langle Q_y^{[2,0]} \rangle
=0.
\]
This is an example of a selection rule originating from $\mathcal C_A$ that is invisible from spatial-inversion parity alone.

For $H_2$, the Hamiltonian preserves $\mathcal C_A$.
Since normal one-body multipoles are $\mathcal C_A$-odd, all one-body expectation values vanish. 
This shows that one-body and two-body multipoles can be separated by $\mathcal C_A$ even when they belong to the same conventional electric-multipole label.

The Hamiltonian $H_3$ breaks $\mathcal P$ and $\mathcal G_{\pi/2}$ separately. 
 However, the normal terms in the Hamiltonian are $\mathcal P$-even and $\mathcal G_{\pi/2}$-even, while the anomalous term is $\mathcal P$-odd and $\mathcal G_{\pi/2}$-odd.  
 Thus, the composite symmetry
\[
\mathcal P\mathcal G_{\pi/2},
\]
is preserved.  
Consequently, the normal and anomalous channels are constrained differently:
a nonzero expectation value must be even under $\mathcal P\mathcal G_{\pi/2}$, which allows $\mathcal P$-even normal multipoles and $\mathcal P$-odd anomalous multipoles.

The Hamiltonian $H_4$ is obtained from $H_3$ by replacing the normal term $Q_{xy}^{[1,0]}$ with $Q_v^{[1,0]}$. 
It preserves not only
\[
\mathcal P\mathcal G_{\pi/2},
\]
but also the composite mirror-gauge symmetry
\[
\mathcal G_{\pi/2}\mathcal M_x.
\]
The allowed expectation values are therefore further restricted by mirror-gauge parity.
Comparing the $H_3$ and $H_4$ rows of Table~\ref{tab:exp_H1_H5}, one sees that several expectation values allowed by $\mathcal P\mathcal G_{\pi/2}$ alone are removed once $\mathcal G_{\pi/2}\mathcal M_x$ is also imposed.

Finally, in $H_5$, the additional term $\tilde Q_{xy}^{[1,2]}$ breaks the $\mathcal P\mathcal G_{\pi/2}$ symmetry preserved by $H_4$, while it still preserves $\mathcal G_{\pi/2}\mathcal M_x$. 
The induced channels change accordingly. 
In particular, expectation values forbidden by $\mathcal{P} \mathcal{G}_{\pi/2}$ can now appear, provided they remain compatible with $\mathcal G_{\pi/2}\mathcal M_x$.

These results demonstrate the role of the full sixteenfold structure in expectation-value selection rules.
Operators with the same conventional spatial multipole labels can differ by $N_{\rm b}$ and $\Delta N$, and hence by $\sigma_{\mathcal C_A}$ and $\sigma_{\mathcal G}$.
Thus, the sixteenfold classification developed in Sec.~\ref{sec:classification} provides a compact symmetry-based way to organize allowed and forbidden multipole expectation values in many-body operator space.

\section{Discussion}

The central result of this work is that the conventional multipole classification becomes incomplete once one moves from one-body observables to many-body or pairing operator space.
Operators that are indistinguishable within the conventional $(\mathrm{rank},\mathcal P,\mathcal T)$ framework can represent fundamentally different physical structures, including different body numbers, particle-number changes, and particle-hole characters.
We have shown that these hidden distinctions can be systematically resolved by canonical symmetries compatible with rotations.

For even-fermion-parity operator spaces, this resolution is achieved through two additional $\mathbb Z_2$ labels associated with the discrete gauge transformation $\mathcal G_{\pi/2}$ and the particle-hole transformation $\mathcal C_A$. 
Together with the conventional $(\mathcal P,\mathcal T)$ labels, they refine the familiar $Q/M/T/G$ fourfold classification into a sixteenfold classification in many-body multipole space. 
This refinement is not merely a finer labeling scheme.
Rather, it separates body-number, particle-number change, and particle-hole structure into distinct symmetry sectors while preserving rotational symmetry, thereby revealing structures that remain invisible in the conventional classification.

These additional labels are not introduced as model-dependent conveniences. 
As shown in the Appendix, $\mathcal G_{\pi/2}$ and $\mathcal C_A$ emerge from the classification of canonical transformations compatible with rotations. 
The discrete symmetries used in this work therefore originate from the intrinsic structure of many-body operator space itself. 
In this sense, the present framework is not an empirical refinement of conventional multipole theory but a symmetry-based classification derived from rotation-compatible canonical transformations.
It should be noted, however, that the additional symmetries considered in this work are restricted to canonical transformations acting on creation and annihilation operators. Although it is unclear whether this premise can be naturally generalized, such a generalization could further extend the classification by incorporating additional symmetries.

The examples discussed in this work illustrate complementary aspects of this framework. 
The local Fock space of a spinful $s$ orbital gives the minimal setting in which the coarse-grained nature of the conventional classification becomes apparent. 
The symmetry-adapted particle-number operator $\widetilde N$, the onsite-interaction operator $\widetilde U$, and the anomalous operator $\eta_+$ are all classified as electric monopoles within the conventional scheme. 
Nevertheless, they correspond respectively to a particle-number-conserving one-body operator, a particle-number-conserving two-body operator, and a particle-number-nonconserving one-body operator. 
Although they share the same conventional multipole label, they belong to different sectors of many-body operator space. 
The labels $\mathcal G_{\pi/2}$ and $\mathcal C_A$ resolve this distinction already at the level of the minimal local Hilbert space.

The $N=2$ sector of the spinless $s$-$p$ system demonstrates how the extended classification sharpens selection rules for multipole expectation values. 
A representative example is provided by the two-body electric-toroidal monopole $G_0^{[2]}$. 
Within the conventional $(\mathcal P,\mathcal T)$ classification alone, $G_0^{[2]}$ is not symmetry-forbidden for a Hamiltonian that preserves time reversal but breaks spatial inversion. 
Nevertheless, its expectation value vanishes when the relevant chiral symmetry is preserved. 
By including the derived label $\Gamma=\mathcal T\mathcal C_A$, this absence is understood as a symmetry selection rule associated with the chiral symmetry. 
Once a two-body electric channel is introduced, this chiral constraint is removed and $G_0^{[2]}$ can be induced. 
This example shows that expectation values allowed by conventional $(\mathcal P,\mathcal T)$ labels can still be constrained by the additional symmetry labels introduced in this work.
Note that, in general, a chiral transformation is not associated with a unique $G_0$.
As shown in Appendix~\ref{Representation_Canonical_Transformations}, chiral transformations are fundamentally described in terms of monopole degrees of freedom.
In spinless systems, each such transformation is characterized by one of the four monopole types, namely E, ET, M, or MT, and can be expressed as a linear combination of the corresponding monopoles.
However, when the state space is restricted, the chiral transformation can be described in terms of a single effective monopole, as in the case discussed in Sec.~\ref{sec:example}.

The four-site square cluster further demonstrates that the present framework is not restricted to onsite atomic multipoles but naturally extends to real-space cluster multipoles. 
In that example, one-body, two-body, and anomalous pairing-type multipoles were classified according to $(\mathcal P,\mathcal G_{\pi/2},\mathcal C_A)$ in a time-reversal-even setting. 
The resulting selection rules are controlled not only by the individual parities under these labels, but also by composite symmetries formed from $\mathcal G_{\pi/2}$, $\mathcal C_A$, and point-group operations. 
In particular, normal and anomalous channels are constrained simultaneously by such composite symmetries, leading to nontrivial conditions for inducing cluster-multipole expectation values. 
These results demonstrate that the classification developed here naturally extends from local orbital degrees of freedom to broader many-body multipole spaces involving multi-site, cluster, and anomalous degrees of freedom.

It is also useful to clarify the scope of the present framework.  
First, we have restricted our discussion to even-fermion-parity operator spaces, namely operators composed of an even number of fermionic operators. 
Operators with odd fermion parity, which can carry half-integer rank, therefore lie outside the direct scope of the present classification. 
Second, the classification identifies the symmetry sector of an operator but does not determine the magnitude of an expectation value.  
Whether a given expectation value becomes finite also depends additionally on the spectrum, matrix elements, and state under consideration.  
This limitation is intrinsic to any symmetry-based classification.
The essential point of the present work is that prohibitions, induced channels, and mixing structures that remain hidden within the conventional multipole framework can be systematically organized once $\mathcal C_A$, $\mathcal G_{\pi/2}$, and $\Gamma=\mathcal T\mathcal C_A$ are included.  
Third, additional basis-dependent relative signs can appear in multi-orbital or cluster bases. 
As discussed in Appendix~\ref{app:hybrid}, such signs can be useful in specific models. 
However, they depend on the choice of multiplicity basis and therefore do not constitute universal labels of the many-body multipole classification itself.

A natural next step is to incorporate crystalline symmetry into the present framework by combining point-group and space-group irreducible representations with the additional labels $\mathcal C_A$ and $\mathcal G_{\pi/2}$. 
The four-site square-cluster example discussed in this work already illustrates an intermediate step in this direction, showing that the classification naturally extends beyond local atomic multipoles to real-space cluster multipoles involving multi-site and anomalous degrees of freedom.
Extending the same framework to periodic crystals would provide a systematic classification of many-body multipoles in realistic materials.
Possible applications include multi-orbital and multi-site systems, interacting electron systems, electron-phonon coupled systems, and superconducting states. 
Beyond identifying order parameters, the framework may clarify which multipoles can be induced, mixed, or forbidden under a given symmetry environment.
Furthermore, combining the present classification with first-principles electronic-structure calculations may provide a practical route for identifying symmetry-adapted many-body multipoles and their associated response channels in realistic materials~\cite{nakamura2021respack, Oiwa_PRB_2025}.
In this sense, the present construction provides a foundation for extending multipole theory from a classification of observables to a symmetry-based organization of many-body operator space.

\section{Conclusion}

In this work, we have developed a symmetry-based classification of multipoles in many-body operator space from the viewpoint of rotation-compatible canonical symmetries.
For operator spaces composed of an even number of fermionic operators, we introduced two additional $\mathbb Z_2$ labels associated with the discrete gauge transformation $\mathcal G_{\pi/2}$ and the particle-hole transformation $\mathcal C_A$. 
These labels refine the conventional $Q/M/T/G$ classification based on $(\mathrm{rank},\mathcal P,\mathcal T)$ into a sixteenfold classification in many-body multipole space.

The resulting classification resolves physically distinct many-body operators that remain indistinguishable within the conventional classification. 
In particular, body number, particle-number change, and particle-hole structure emerge as independent symmetry characteristics encoded by $\mathcal C_A$ and $\mathcal G_{\pi/2}$. 
Through representative examples, we demonstrated that these additional symmetry labels systematically organize selection rules for multipole expectation values, induced multipoles, and symmetry-allowed mixing channels.

More broadly, the present framework extends multipole theory beyond the classification of one-body observables to a symmetry-based organization of many-body operator space itself. 
It provides a unified language for constructing symmetry-adapted many-body multipole operators, identifying order parameters, and analyzing emergent responses in interacting and cluster systems.
As such, the sixteenfold classification establishes a general foundation for the study of many-body multipoles and their symmetry-governed phenomena.

\section*{Acknowledgements}
% Acknowledgements should follow immediately after the conclusion.

\paragraph{Funding information}
This work was supported by JSPS KAKENHI Grants Numbers JP22H00101, JP23H04869, JP26K17075, JP26H00618 and by JST CREST (JPMJCR23O4) and JST FOREST (JPMJFR2366).

\appendix

\section{Rotation-compatible canonical transformations for a single-$j$ orbital}
\label{app:single_j}

In the conventional multipole classification, the rotational rank is supplemented by the parities under spatial inversion $\mathcal P$ and time reversal $\mathcal T$. 
Both transformations are canonical symmetries compatible with rotations: $\mathcal{P}$ is implemented by a unitary operator, whereas $\mathcal{T}$ is implemented by an antiunitary operator, and both preserve the canonical anticommutation relations of fermionic creation and annihilation operators.  
They therefore provide the canonical $\mathbb Z_2$ labels underlying the conventional multipole classification.
The main text extends this idea by introducing two additional $\mathbb Z_2$ labels associated with the particle-hole transformation $\mathcal C_A$ and the discrete gauge transformation $\mathcal G_{\pi/2}$.
The purpose of this Appendix is to show that these labels are not introduced ad hoc.
Rather, they arise naturally from a classification of canonical transformations compatible with rotations.

To this end, we classify canonical transformations acting on creation and annihilation operators that commute with rotations and preserve the properties relevant for multipole operator space: Fermionic anticommutation relations, Hermiticity, rotational rank, and time-reversal parity of operators composed of an even number of fermionic operators.
The result is that, within a single irreducible $j$ sector, the nontrivial canonical symmetries relevant to the multipole classification are generated by the particle-number gauge transformation $\mathcal G_\phi$, time reversal $\mathcal T$, the antiunitary particle-hole transformation $\mathcal C_A$, and the unitary composite transformation $\Gamma=\mathcal T\mathcal C_A$.  
Furthermore, requiring compatibility with Hermitian $\mathcal T$-even and $\mathcal T$-odd multipole sectors reduces the continuous gauge transformation to the discrete transformation $\mathcal G_{\pi/2}$, which becomes one of the additional $\mathbb Z_2$ labels used in the main text.
Throughout this Appendix, we restrict our attention to even-fermion-parity operator spaces, which contain local observables and local Hamiltonians. 

We first consider a single irreducible angular-momentum sector $\ket{jm}$. 
The label $j$ may represent either orbital angular momentum in a spinless problem or total angular momentum including spin. 
The magnetic quantum number takes values $m = -j, -j + 1, \ldots , j$.
We collect the corresponding annihilation and creation operators into column vectors as
\begin{align}
  {\bm c}_j := \left(c_{j,j},c_{j,j-1},\ldots,c_{j,-j}\right)^{T}, \\
  {\bm c}_j^\dagger := \left(c^\dagger_{j,j},c^\dagger_{j,j-1},\ldots,c^\dagger_{j,-j}\right)^{T},
\end{align}
where ${\bm c}_j^\dagger$ denotes a column vector of creation operators and should not be confused with the Hermitian conjugate of the vector ${\bm c}_j$.

Before proceeding to the classification of general canonical transformations compatible with rotations, we briefly review the conventional spatial-inversion and time-reversal transformations, which provide the standard $\mathbb Z_2$ labels in multipole theory.
We begin with spatial inversion $\mathcal{P}$. 
For a single orbital specified by orbital angular momentum $l$, $\mathcal{P}$ gives only a fixed sign determined by $l$,
\[
\pi_{l}=(-1)^l,
\]
and hence
\begin{align}
  \label{eq:inversion_cd}
  \mathcal P\, 
  {\bm c}_j^\dagger\,\mathcal P^{-1} = \pi_{l}\, 
  {\bm c}_j^\dagger, \\
  \label{eq:inversion_c}
  \mathcal P\, {\bm c}_j \,\mathcal P^{-1} = \pi_{l} {\bm c}_j.
\end{align}
Thus, within a single-orbital sector, $\mathcal{P}$ simply gives the conventional orbital parity: $\pi_{l} = +1$ for even-$l$ orbitals such as $s$ and $d$, and $\pi_{l} = -1$ for odd-$l$ orbitals such as $p$ and $f$.

For time reversal $\mathcal{T}$, we use the convention
\begin{align}
  \label{def_time_app}
  \mathcal T\ket{jm}=(-1)^{j-m}\ket{j,-m}.
\end{align}
The corresponding action on creation and annihilation operators is
\begin{align}
  \mathcal T\, c^\dagger_{j,m}\,\mathcal T^{-1}
  &=(-1)^{j-m} c^\dagger_{j,-m},
  \\
  \mathcal T\, c_{j,m}\,\mathcal T^{-1}
  &=(-1)^{j-m} c_{j,-m},
\end{align}
and since $\mathcal T$ is antiunitary,
\begin{align}
  \mathcal T\, i\, \mathcal T^{-1}=-i.
\end{align}
Although alternative phase conventions, $\mathcal T\ket{jm}=(-1)^{m}\ket{j,-m}$, are also commonly used in the literature, Eq.~\eqref{def_time_app} is convenient because the phase factor is always $\pm1$ for both integer and half-integer $j$.
With this convention, it is useful to introduce the metric tensor, also known as the Wigner $1jm$ symbol.~\cite{Edmonds1957,Varshalovich1988,Wigner1993}
% }
\begin{align}
\label{def_metric_g}
  g^{(j)}_{mn} := (-1)^{j-m}\delta_{m,-n}, \quad g^{(j)} g^{(j)\dagger} = I_{2j+1},
\end{align}
where $I_{2j+1}$ denotes the identity matrix in the $(2j+1)$-dimensional representation space.
Using this metric tensor, the time-reversed annihilation operators are defined by
\begin{align}
\label{gc}
  \tilde c_{j,m}
  &:=
  \sum_n g^{(j)}_{mn} c_{j,n}
  =
  (-1)^{j-m} c_{j,-m},
  \\
  \tilde{\bm c}_j &:= g^{(j)} {\bm c}_j .
\end{align}

\subsection{Canonical transformations preserving the one-body multipole space}

In this subsection, we identify the class of canonical transformations relevant to the multipole classification. 
A general Bogoliubov transformation mixes particle-number-conserving one-body operators $c^\dagger c$ with anomalous one-body operators $c^\dagger c^\dagger$ and $cc$. 
Such transformations do not preserve the conventional one-body multipole space as an invariant subspace. 
Since our purpose is to classify symmetries acting within the multipole operator space, we restrict attention to canonical transformations that map the normal one-body operator space onto itself.

A general linear canonical transformation $U$ acting on fermionic creation and annihilation operators can be represented as a Bogoliubov transformation
\cite{Bogoliubov1958,Valatin1958,Bravyi-2005,Kupsch-2014},
\begin{align}
\begin{split}
  U {\bm c}_j U^{-1} &= V^* {\bm c}_j + W^* {\bm c}_j^\dagger,
  \\
  U {\bm c}_j ^{\dagger} U^{-1}  &= V {\bm c}_j^\dagger + W {\bm c}_j,
\end{split}
  \label{def_Bogoliubov}
\end{align}
where the canonical anticommutation relations impose
\begin{align}
  VV^\dagger + WW^\dagger = I_{2j+1},
  \qquad
  VW^T + WV^T = 0.
\end{align}
The matrices $(V,W)$ describe the linear action on the operator space.
Whether $U$ is unitary or antiunitary will be imposed later through the corresponding covariance conditions.
For an antiunitary transformation, $c$-number coefficients are additionally complex conjugated.

Consider an arbitrary particle-number-conserving Hermitian one-body operator
\begin{align}
  \mathcal{O}_h := ({\bm c}_j^\dagger)^T h\, {\bm c}_j,  \qquad h = h^{\dagger}.
\end{align}
Under Eq.~(\ref{def_Bogoliubov}), it transforms as
\begin{align}
  U \mathcal{O}_h U^{-1} 
  =  ({\bm c}_j^\dagger)^T V^\dagger h V {\bm c}_j
  + ({\bm c}_j^\dagger)^T V^\dagger h W {\bm c}_j^\dagger
  + {\bm c}_j^T W^\dagger h V {\bm c}_j
  + {\bm c}_j^T W^\dagger h W {\bm c}_j^\dagger .
\end{align}
The second and third terms are anomalous terms of the $c^\dagger c^\dagger$ and $cc$ types, respectively.  
Requiring the transformed operator to remain within the normal one-body space for every Hermitian matrix $h$ gives
\begin{align}
  V^\dagger h W = 0,
  \qquad
  W^\dagger h V = 0
  \qquad
  (\forall\, h=h^\dagger).
\end{align}
Since this condition must hold for arbitrary $h$, either $V=0$ or $W=0$.
The canonical condition $VV^\dagger+WW^\dagger=I_{2j+1}$ then implies,
\begin{align}
 \label{VVd_WWd_I}
VV^\dagger=I_{2j+1} \qquad \text{or} \qquad WW^\dagger=I_{2j+1},
\end{align}
respectively.

Thus, a canonical transformation $U$ preserving the one-body multipole space can only take one of the following two forms:
\begin{alignat}{2}
  \label{W_condition_c2c}
&\text{unitary or antiunitary}, &\qquad& \bm{c}_{j}^{\dagger} \mapsto \bm{c}_{j}^{\dagger}
: \qquad   U\,{\bm c}_j^\dagger\,U^{-1} = W\,{\bm c}_j^\dagger, \\
  \label{W_condition_c2a}
&\text{unitary or antiunitary}, &\qquad& \bm{c}_{j}^{\dagger} \mapsto \bm{c}_{j}^{}
: \qquad     U\,{\bm c}_j^\dagger\,U^{-1} =W \,{\bm c}_j,
\end{alignat}
where $W$ is unitary.
These possibilities apply to both unitary and antiunitary implementations of $U$.  
The distinction between them enters only through the treatment of complex conjugation.

We next summarize the rotational transformation properties of creation and annihilation operators.  
Since half-integer angular momentum $j$ is included, the relevant symmetry group is the double cover $SU(2)$ of $SO(3)$.  
Under a rotation $R$, the creation operators transform according to the irreducible representation $D^{(j)}(R)$,
\begin{align}
  \label{UcU}
  R\, {\bm c}_j^\dagger\, 
  R^{-1}
  =
  D^{(j)}(R)\,{\bm c}_j^\dagger,
\end{align}
whereas the annihilation operators transform according to the complex-conjugate representation:
\begin{align}
  \label{UaU}
  R\, {\bm c}_j\, 
  R^{-1}
  =
  D^{(j)}(R)^*\,{\bm c}_j .
\end{align}
The metric tensor $g^{(j)}$ given by Eq.~\eqref{def_metric_g} relates these two representations:
\begin{align}
  \label{gd_dg}
  g^{(j)} D^{(j)}(R)^* = D^{(j)}(R)\, g^{(j)},
\end{align}
or equivalently
\begin{align}
  \label{gdg_d}
  g^{(j)} D^{(j)}(R)^* g^{(j)-1}  = D^{(j)}(R).
\end{align}
Therefore, $D^{(j)}$ and $D^{(j)*}$ are equivalent representations for $SU(2)$.
Defining the time-reversed annihilation multiplet
\begin{align}
\tilde{\bm c}_j
=
g^{(j)}
{\bm c}_j ,
\end{align}
one finds
\begin{align}
  R\, \tilde{\bm c}_j\, 
  R^{-1}
  =
  D^{(j)}(R)\,\tilde{\bm c}_j .
\end{align}
Hence $\tilde{\bm c}_j$ transforms according to the same irreducible representation as ${\bm c}_j^\dagger$.
This allows maps between creation and annihilation operators to be treated as intertwiners within the same irreducible $SU(2)$ representation, which will be the key ingredient in the Schur-lemma classification developed in the next subsection.

\subsection{Schur's lemma and rotation-compatible intertwiners}

As shown in the previous subsection, a canonical transformation that preserves the one-body multipole space is restricted to one of the two forms in Eqs.~(\ref{W_condition_c2c}) and (\ref{W_condition_c2a}).
With these preparations, using Schur's lemma~\cite{Weyl-1928,Fulton-Harris-1991}, 
we now classify the linear part $V$ or $W$ of rotation-compatible canonical transformations.

The key observation is that the rotation-covariance conditions reduce to the classification of intertwiners between irreducible representations of $SU(2)$.
In the present context, Schur's lemma has a simple interpretation: a matrix $W$ acting inside a single irreducible angular-momentum multiplet $\bm{c}_{j}^{\dagger}$ and compatible with all rotations cannot distinguish the different $m$ components.
Consequently, it must be proportional to the identity, $W = \lambda I_{2j+1}$ $(\lambda\in\mathbb C)$.

For example, consider a spinless $p$ orbital, i.e., the $j=1$ multiplet with $m=1,0,-1$.  
A matrix $W$ commuting with all rotations must in particular commute with the $z$-axis rotation
\begin{align}
  D_z^{(1)}(\theta)=
  \begin{pmatrix}
    e^{-i\theta}&0&0\\
    0&1&0\\
    0&0&e^{i\theta}
  \end{pmatrix} .
\end{align}
This already forbids mixing between different $m$ components, so $W$ must be diagonal.  
Furthermore, compatibility also with rotations about all axes requires identical treatment of the three $m$ components, forcing all diagonal entries to be equal.  
Therefore $W$ is proportional to the identity, $W = \lambda I_{3}$.
This is precisely the statement of Schur's lemma for this irreducible multiplet.  
By contrast, no nonzero rotation-compatible map exists between inequivalent irreducible representations.
For example, a map connecting an $s$ orbital ($j=0$) and a $p$ orbital ($j=1$) must vanish.

More generally, let $(\rho,V)$ and $(\rho',V')$ be irreducible representations of a group $G$ over $\mathbb C$.
A linear map $W:V\to V'$ is called an intertwiner, or a $G$-equivariant linear map, if it satisfies
\begin{align}
  W\rho(g)=\rho'(g)W
  \qquad
  (\forall g\in G).
\end{align}
In other words, the intertwiner $W$ is compatible with the group action. 
Schur's lemma states that an intertwiner between irreducible
representations is either zero or an isomorphism.
In particular,
\begin{align}
  \label{schur_i}
  W = 
  \begin{cases}
    0, & 
    (\rho,V)\not\simeq(\rho',V'),\\
    \lambda I, & 
    (\rho,V)\simeq(\rho',V'),
  \end{cases}
  \qquad
  \lambda\in\mathbb C,
\end{align}
where $\simeq$ denotes equivalence of representations.

We apply this result to the irreducible $SU(2)$ representation $D^{(j)}$ carried by the multiplet ${\bm c}_j^\dagger$.
For an intertwiner $W: V_j\to V_{j'}$, rotation covariance requires
\begin{align}
  W D^{(j)}(R) = D^{(j')}(R)W.
\end{align}
Schur's lemma therefore gives
\begin{align}
  \label{schur_i}
  W = 
  \begin{cases}
    0, & j\ne j',\\
    \lambda I_{2j+1}, & j=j',
  \end{cases}
  \qquad
  \lambda\in\mathbb C .
\end{align}

We also need the corresponding statement for the complex-conjugate representation $D^{(j)*}$.
As given by Eq.~\eqref{gdg_d}, for $SU(2)$, the irreducible representation $D^{(j)}$ is equivalent to its complex-conjugate representation $D^{(j)*}$. 
The invariant tensor $g^{(j)}$ satisfies Eq.~\eqref{gd_dg}, and hence gives an intertwiner $g^{(j)}$: $D^{(j)*}$ $\rightarrow$ $D^{(j)}$. 
Since $D^{(j)}$ and $D^{(j)*}$ are equivalent irreducible representations, $g^{(j)}$ is an isomorphism.
If $f: D^{(j)*} \to D^{(j)}$ is another intertwiner satisfying the same condition, then
\begin{align}
  g^{(j)-1}f: D^{(j)*} \to D^{(j)*}
\end{align}
is an intertwiner of the irreducible representation $D^{(j)*}$. 
Schur's lemma therefore gives
\begin{align}
  g^{(j)-1}f=\lambda I_{2j+1} ,  \qquad \lambda\in\mathbb C,
\end{align}
and hence
\begin{align}
  f=\lambda g^{(j)} .
\end{align}
Thus, the intertwiner between $D^{(j)*}$ and $D^{(j)}$ is unique up to an overall scalar factor $\lambda$.

Thus, up to an overall scalar factor, the identity matrix $I_{2j+1}$ is the unique rotation-compatible intertwiner from $V_j$ to itself, whereas $g^{(j)}$ is the unique rotation-compatible intertwiner from the conjugate representation $D^{(j)*}$ to $D^{(j)}$.
These two intertwiners form the basic building blocks of the classification developed below.

We now impose rotational covariance on canonical transformations. 
For unitary and antiunitary transformations satisfying Eqs.~\eqref{W_condition_c2c} and \eqref{W_condition_c2a}, we obtain
\begin{alignat}{4}
  \label{schur_1}
  &\text{unitary}, 
  &\qquad&  \bm{c}_{j}^{\dagger} \mapsto \bm{c}_{j}^{\dagger}: &\qquad& W D^{(j)}(R)=D^{(j)}(R) W, \\
  \label{schur_2}
  &\text{unitary}, 
  &\qquad& \bm{c}_{j}^{\dagger} \mapsto \bm{c}_{j}^{}:  &\qquad&  W D^{(j)}(R)=D^{(j)}(R)^* W, \\
  \label{schur_3}
  &\text{antiunitary}, 
  &\qquad& \bm{c}_{j}^{\dagger} \mapsto \bm{c}_{j}^{\dagger}: &\qquad& W D^{(j)}(R)^{*}=D^{(j)}(R) W, \\
  \label{schur_4}
  &\text{antiunitary}, 
 &\qquad& \bm{c}_{j}^{\dagger} \mapsto \bm{c}_{j}: &\qquad& W D^{(j)}(R)^*=D^{(j)}(R)^* W .
\end{alignat}
These four covariance conditions reduce the problem to classifying intertwiners $D^{(j)} \to D^{(j)}$ and $D^{(j)*} \to D^{(j)}$.
For antiunitary transformations, the rotation matrix is complex-conjugated when the transformation acts on it. 
Using Schur's lemma, these four conditions imply
\begin{alignat}{4}
\label{schur_unitary_c2c}
&\text{unitary},     &\qquad& \bm{c}_{j}^{\dagger} \mapsto \bm{c}_{j}^{\dagger}
&:\qquad& W = \lambda I_{2j+1},
&\qquad& \lambda\in\mathbb C,\\
\label{schur_unitary_c2a}
&\text{unitary},     &\qquad& \bm{c}_{j}^{\dagger} \mapsto \bm{c}_{j}
&:\qquad& W = \lambda g^{(j)},
&\qquad& \lambda\in\mathbb C,\\
\label{schur_antiunitary_c2c}
&\text{antiunitary}, &\qquad& \bm{c}_{j}^{\dagger} \mapsto \bm{c}_{j}^{\dagger}
&:\qquad& W = \lambda g^{(j)},
&\qquad& \lambda\in\mathbb C,\\
\label{schur_antiunitary_c2a}
&\text{antiunitary}, &\qquad& \bm{c}_{j}^{\dagger} \mapsto \bm{c}_{j}
&:\qquad& W = \lambda I_{2j+1},
&\qquad& \lambda\in\mathbb C .
\end{alignat}
Therefore, up to phase conventions, Schur's lemma leaves only four possible rotation-compatible canonical operations.

\subsection{Rotation-compatible canonical transformations}

We now identify the four canonical transformations obtained from the Schur-lemma classification in the previous subsection.
From Eq.~\eqref{schur_unitary_c2c}, the unitary $\bm{c}_{j}^{\dagger} \mapsto \bm{c}_{j}^{\dagger}$ case gives the particle-number $U(1)$ gauge transformation
\begin{align}
  \label{eq:g_pi}
  \mathcal G_\phi:
  \qquad
  {\bm c}_j^\dagger \mapsto e^{-i\phi}{\bm c}_j^\dagger,
  \qquad
  {\bm c}_j \mapsto e^{i\phi}{\bm c}_j .
\end{align}
From Eq.~\eqref{schur_unitary_c2a}, the unitary $\bm{c}_{j}^{\dagger} \mapsto \bm{c}_{j}^{}$ case gives the chiral transformation
\begin{align}
\label{eq:A_Gamma}
  \Gamma:
  \qquad
  {\bm c}_j^\dagger \mapsto \tilde{\bm c}_j =g^{(j)}{\bm c}_j ,
  \qquad
  {\bm c}_j \mapsto \tilde{\bm c}_j^\dagger =g^{(j)}{\bm c}_j^\dagger,
\end{align}
From Eq.~\eqref{schur_antiunitary_c2c}, the antiunitary $\bm{c}_{j}^{\dagger} \mapsto \bm{c}_{j}^{\dagger}$ case gives the time-reversal transformation used in this work
\begin{align}
  \label{eq:time}
  \mathcal T:
  \qquad
  {\bm c}_j^\dagger \mapsto \tilde{\bm c}_j^\dagger =g^{(j)}{\bm c}_j^\dagger,
  \qquad
  {\bm c}_j \mapsto \tilde{\bm c}_j =g^{(j)}{\bm c}_j.
\end{align}
From Eq.~\eqref{schur_antiunitary_c2a}, the antiunitary $\bm{c}_{j}^{\dagger} \mapsto \bm{c}_{j}^{}$ case gives the particle-hole transformation
\begin{align}
  \label{eq:CA}
  \mathcal C_A:
  \qquad
  {\bm c}_j^\dagger \mapsto {\bm c}_j,
  \qquad
  {\bm c}_j \mapsto {\bm c}_j^\dagger .
\end{align}
Thus, for a single irreducible $j$ sector, Schur's lemma reduces the possible rotation-compatible canonical transformations to the four operations $\mathcal G_\phi$, $\Gamma$, $\mathcal T$, and $\mathcal C_A$, up to phase conventions.
From their actions on the creation and annihilation operators, one finds that, for the present choice of phases, the composition of time reversal and the particle-hole transformation reproduces the chiral transformation:
\begin{align}
\Gamma = \mathcal T \mathcal C_A .
\end{align}
Therefore only two of the three operations $\Gamma$, $\mathcal T$, and $\mathcal C_A$ are independent.

\subsection{Reduction to discrete labels on the even multipole operator space}

The previous subsection identifies the rotation-compatible canonical transformations on fermionic operators.
We now examine the relations among $\mathcal G_\phi$, $\Gamma$, $\mathcal T$, and $\mathcal C_A$.
Schur's lemma guarantees the compatibility of their adjoint action on creation and annihilation operators with rotations, but the mutual relations among these transformations must still be examined.  
Since the multipole operators considered in this work are Hermitian, fermion-parity-even, and have definite time-reversal parity, we focus on their action within this restricted operator space.

We first examine the relation between $\Gamma$ and $\mathcal T$.
On the Nambu vector
\begin{align}
  \Psi_j :=
  \begin{pmatrix}
    {\bm c}_j^\dagger\\
    {\bm c}_j
  \end{pmatrix},
\end{align}
the transformations $\Gamma$ and $\mathcal T$ are represented as
\begin{align}
  \Gamma
  &=
  \begin{pmatrix}
    \bm 0 & g^{(j)} \\
    g^{(j)} & \bm 0
  \end{pmatrix},
  \\
  \mathcal T
  &=
  \begin{pmatrix}
    g^{(j)} & \bm 0 \\
    \bm 0 & g^{(j)}
  \end{pmatrix}\mathcal K ,
\end{align}
where $\mathcal K$ denotes complex conjugation.  
Using $g^{(j)*}=g^{(j)}$ and $g^{(j)}g^{(j)\dagger}=I_{2j+1}$, we obtain
\begin{align}
  \mathcal T \Gamma \mathcal T^{-1}
  &=
  \begin{pmatrix}
    g^{(j)} & \bm 0 \\
    \bm 0 & g^{(j)}
  \end{pmatrix}
  \begin{pmatrix}
    \bm 0 & g^{(j)} \\
    g^{(j)} & \bm 0
  \end{pmatrix}^{*}
  \begin{pmatrix}
    g^{(j)\dagger} & \bm 0 \\
    \bm 0 & g^{(j)\dagger}
  \end{pmatrix}
  =
    \begin{pmatrix}
    \bm 0 & g^{(j)} \\
    g^{(j)} & \bm 0
  \end{pmatrix}
  =
  \Gamma.
\end{align}
Thus, $\Gamma$ and $\mathcal T$ commute on the Nambu vector $\Psi_j$.

By contrast, on the Nambu vector $\Psi_j$, the continuous gauge transformation $\mathcal G_\phi$ is represented by
\begin{align}
  \mathcal G_\phi
  =
  \begin{pmatrix}
    e^{-i\phi} I & \bm 0 \\
    \bm 0 & e^{i\phi} I
  \end{pmatrix}.
\end{align}
Time reversal $\mathcal T$ sends $\mathcal G_\phi$ to its inverse $\mathcal G_{-\phi}$ as
\begin{align}
  \mathcal T \mathcal G_\phi \mathcal T^{-1}
  &=
  \begin{pmatrix}
    g^{(j)} & \bm 0 \\
    \bm 0 & g^{(j)}
  \end{pmatrix}
  \begin{pmatrix}
    e^{-i\phi} I & \bm 0 \\
    \bm 0 & e^{i\phi} I
  \end{pmatrix}^{*}
  \begin{pmatrix}
    g^{(j)\dagger} & \bm 0 \\
    \bm 0 & g^{(j)\dagger}
  \end{pmatrix}
  =
  \begin{pmatrix}
    e^{i\phi} I & \bm 0 \\
    \bm 0 & e^{-i\phi} I
  \end{pmatrix}
  =
  \mathcal G_{-\phi}.
\end{align}
Thus, a generic continuous gauge transformation $\mathcal G_\phi$ does not commute with $\mathcal T$ and it does not preserve the decomposition into $\mathcal{T}$-even and $\mathcal{T}$-odd Hermitian anomalous multipoles.
 
This can be seen explicitly from the pairing monopoles,
\begin{align}
  \eta_{+} := c^\dagger_{\uparrow}c^\dagger_{\downarrow} + c_{\downarrow}c_{\uparrow}
  , \qquad
  \eta_{-} := i(c^\dagger_\uparrow c^\dagger_\downarrow-c_\downarrow c_\uparrow).
\end{align}
Here $\eta_+$ is $\mathcal T$-even and $\eta_-$ is $\mathcal T$-odd.
Under $\mathcal G_\phi$,
\begin{align}
   \eta_+ \mapsto \cos(2\phi)\eta_+-\sin(2\phi)\eta_- . 
\end{align}
Therefore, $\mathcal G_\phi$ preserves the $\mathcal T$ parity of Hermitian multipoles only when $\sin(2\phi)=0$, i.e.,
\begin{align}
  \phi=\frac{n\pi}{2}, \qquad n\in\mathbb Z .
\end{align}
On even-fermion operators, $\phi=0$ and $\phi=\pi$ act trivially. 
The remaining nontrivial universal transformation is therefore
\begin{align}
  \label{g_pi_2}
  \mathcal G_{\pi/2}:
  \qquad
  {\bm c}_j^\dagger \mapsto -i{\bm c}_j^\dagger,
  \qquad
  {\bm c}_j \mapsto i{\bm c}_j .
\end{align}

For a Hermitian operator with net particle-number change $\Delta N$, this transformation gives
\begin{align}
  \mathcal G_{\pi/2} 
  \mathcal{O}_{h} \mathcal G_{\pi/2}^{-1} = e^{-i\pi\Delta N/2} 
  \mathcal{O}_{h}.
\end{align}
Since we restrict ourselves to even-fermion operators, $\Delta N$ is even and $e^{-i\pi\Delta N/2} = \pm1$.  
Hence $\mathcal G_{\pi/2}$ distinguishes $\Delta N=0\pmod 4$ from $\Delta N=2\pmod 4$.
For instance, $c^\dagger c$ is $\mathcal G_{\pi/2}$-even and $c^\dagger c^\dagger$ is $\mathcal G_{\pi/2}$-odd, whereas $c^\dagger c^\dagger c^\dagger c^\dagger$ is again $\mathcal G_{\pi/2}$-even because $\Delta N=4$. In all cases, the phase is $\pm1$, so the definite time-reversal character is preserved.

Finally, the relation between $\Gamma$ and $\mathcal G_{\pi/2}$ should be interpreted on the even-fermion operator space.  
On the single-fermion Nambu vector, one finds $\Gamma\mathcal G_{\pi/2}\Gamma^{-1}=\mathcal G_{-\pi/2}$.  
Although $\mathcal G_{-\pi/2}$ differs from $\mathcal G_{\pi/2}$ on a single fermion, the difference is a fermion-parity operation.
Since fermion parity acts trivially on even-fermion operators, $\Gamma$ and $\mathcal G_{\pi/2}$ commute on the multipole operator space considered here:
\begin{align}
  \Gamma \mathcal G_{\pi/2} \Gamma^{-1} = \mathcal G_{\pi/2} 
  .
\end{align}
Together with the commutation of $\Gamma$ and $\mathcal T$, this shows that
$\mathcal T$, $\Gamma$ or equivalently $\mathcal C_A$, and $\mathcal G_{\pi/2}$
provide mutually compatible discrete labels on the even Hermitian multipole operator space.

In summary, for a single irreducible $j$ sector, the independent rotation-compatible canonical transformations that preserve the even Hermitian multipole operator space yield the discrete labels associated with $\mathcal T$, $\mathcal C_A$ or equivalently $\Gamma=\mathcal T\mathcal C_A$, and $\mathcal G_{\pi/2}$.
Thus, the even multipole operator space in a single orbital can be classified by
\begin{align}
  SO(3)\times
  \mathbb Z_2^{\mathcal T}\times
  \mathbb Z_2^{\mathcal C_A}\times
  \mathbb Z_2^{\mathcal G_{\pi/2}} .
\end{align}
Equivalently, using $\Gamma=\mathcal T\mathcal C_A$, one may write
\begin{align}
  SO(3)\times
  \mathbb Z_2^{\mathcal T}\times
  \mathbb Z_2^{\Gamma}\times
  \mathbb Z_2^{\mathcal G_{\pi/2}} .
\end{align}
Including spatial inversion $\mathcal P$ gives the sixteenfold classification used in the main text.

\section{Hybrid systems and multiplicity spaces}
\label{app:hybrid}

In Appendix \ref{app:single_j}, we classified rotation-compatible canonical transformations within a single irreducible angular-momentum sector.  
We now generalize the discussion to systems with several orbital, spin-orbital, or site degrees of freedom.  
The essential result remains unchanged: Schur's lemma uniquely determines the action of a rotation-compatible transformation on each irreducible angular-momentum sector. 
The only additional degrees of freedom arise from multiplicity spaces with repeated irreducible representations, where nontrivial unitary mixing between equivalent copies is allowed.

\subsection{General multiplicity-space structure}
\label{app:general_multiplicity_structure}

Let the one-particle Hilbert space $\mathcal H$ be decomposed into irreducible representations of $SU(2)$ as
\begin{align}
  \mathcal H = \bigoplus_j \left( \mathcal M_j \otimes V_j \right).
\end{align}
Here $V_j$ is the irreducible representation space of angular momentum $j$, while $\mathcal M_j$ is the multiplicity space specifying how many times the same $j$ representation appears.  
We denote its dimension by $d_j=\dim\mathcal M_j$.
Creation operators in this sector can be written as
\begin{align}
\label{eq_cmujm}
  c^\dagger_{\mu, j, m},
  \qquad
  \mu=1,\ldots,d_j,
  \quad
  m=-j,\ldots,j .
\end{align}
The index $m$ belongs to the irreducible angular-momentum space $V_j$, whereas the index $\mu$ labels the copies of the same irreducible representation.
It is useful to collect the operators with the same $j$ into a vector
\begin{align}
  \bm{C}^\dagger_j
  :=
  \left(\bm c_{1,j}^\dagger, \bm c_{2,j}^\dagger,\ldots,
  \bm c_{d_j,j}^\dagger
  \right)^T,
\end{align}
where each $\bm{c}_{\mu, j}^{\dagger}$ is a $(2j+1)$-dimensional column vector in the $m$ space.
Under a rotation $R\in SU(2)$, the multiplicity index is invariant and the angular-momentum index transforms as
\begin{align}
\label{eq_rotation_action_creation}
  U_R\,
  \bm{C}_j^\dagger\,U_R^{-1} = \left( I_{d_j}\otimes D^{(j)}(R)  \right)
  \bm{C}_j^\dagger , \\
 \label{eq_rotation_action_annihilation}
  U_R\,
  \bm{C}_j\,U_R^{-1} = \left( I_{d_j}\otimes D^{(j)}(R)^*  \right)
  \bm{C}_j.
\end{align}

We now state how a rotation-compatible canonical transformation $U$ can act in such a multi-component space $\mathcal{M}_{j}$.  
As in Appendix~\ref{app:single_j}, there are four cases, depending on whether the transformation is unitary or antiunitary and whether it maps creation operators to creation or annihilation operators.  
Schur's lemma fixes the action on the irreducible space $V_j$ to be proportional to either the identity $I_{2j+1}$ or the metric tensor $g^{(j)}$.
Consequently, all nontrivial freedom is confined to the multiplicity space $\mathcal M_j$: 
\begin{alignat}{4}
  &{\rm unitary},
  &\qquad&
  \bm{C}_{j}^\dagger\to \bm{C}_{j}^\dagger, 
  &\qquad
  U\,{\bm C}_j^\dagger\,U^{-1}
  &=
  \left(
    u_j\otimes I_{2j+1}
  \right)
  {\bm C}_j^\dagger,
  &\qquad&
  u_j\in U(d_j),
  \label{eq:B_unitary_c_to_c}
  \\
  &{\rm unitary},
  &\qquad&
  \bm{C}_{j}^\dagger\to \bm{C}_{j}, 
  &\qquad
  U\,{\bm C}_j^\dagger\,U^{-1}
  &=
  \left(
    u_j\otimes  g^{(j)}
  \right)
  {\bm C}_j,
  &\qquad&
  u_j\in U(d_j),
  \label{eq:B_unitary_c_to_a}
  \\
  &{\rm antiunitary},
  &\qquad&
  \bm{C}_{j}^\dagger\to \bm{C}_{j}^\dagger, 
  &\qquad
  U\,{\bm C}_j^\dagger\,U^{-1}
  &=
  \left(
    u_j\otimes  g^{(j)}
  \right)
  {\bm C}_j^\dagger,
  &\qquad&
  u_j\in U(d_j),
  \label{eq:B_antiunitary_c_to_c}
  \\
  &{\rm antiunitary},
  &\qquad&
  \bm{C}_{j}^\dagger\to \bm{C}_{j}, 
  &\qquad
  U\,{\bm C}_j^\dagger\,U^{-1}
  &=
  \left(
    u_j\otimes I_{2j+1}
  \right)
  {\bm C}_j,
  &\qquad&
  u_j\in U(d_j),
  \label{eq:B_antiunitary_c_to_a}
\end{alignat}
The matrix $u_j$ acts only in the multiplicity space $\mathcal M_j$ and is constrained by the canonical condition. 
Any overall phase factor $\lambda \in \mathbb{C}$ is absorbed into $u_j$, so that the symbol $u_j$ should be regarded as an independent element of $U(d_j)$ in each line.
The factors $I_{2j+1}$ and $g^{(j)}$ are fixed by rotational compatibility in the irreducible space $V_j$.
The central result is therefore that rotational symmetry leaves no freedom inside an irreducible angular-momentum sector. 
All additional structure originates from multiplicity spaces associated with repeated irreducible representations.

Consequently, different $j$ sectors cannot be mixed by a rotation-compatible transformation, because they carry inequivalent irreducible representations of $SU(2)$.
By contrast, whenever the same irreducible representation appears multiple times, continuous unitary rotations within the multiplicity space are allowed.
These rotations depend on the detailed microscopic content of the one-particle Hilbert space and therefore do not constitute universal symmetry labels of multipole theory.
We next discuss how these possibilities appear in spinless and spinful systems.

\subsection{Spinless hybrid systems without repeated angular momenta}
\label{app:hybrid_spinless}

We first consider spinless hybrid systems composed of orbitals with different angular momenta such as $s$, $p$, $d$, and $f$.
In the absence of spin, the rotational quantum number is simply $j=l$. 
If only one radial copy of each orbital angular momentum is retained, each irreducible representation appears exactly once,
\begin{align}
  d_j=1,
\end{align}
and the one-particle Hilbert space contains no multiplicity degrees of freedom.
In this situation, a rotation-compatible transformation cannot mix different orbital sectors. 
For example, $s$ and $p$ orbitals transform according to inequivalent irreducible representations with $l=0$ and $l=1$, respectively, and Schur's lemma therefore forbids their mixing.
The only creation-preserving rotation-compatible transformation is restricted to independent phase transformations in each orbital block:
\begin{align}
  c_{l,m}^\dagger \mapsto e^{-i\phi_l}c_{l,m}^\dagger,
  \qquad
  c_{l,m} \mapsto e^{i\phi_l}c_{l,m}.
\end{align}

For a product of fermionic operators, an orbital-dependent phase is obtained by multiplying the phase factors carried by all creation and annihilation operators. 
For example, in the minimal case with one copy for each $l$,
\begin{align}
  c_{l_1,m_1}^\dagger c_{l_2,m_2}^\dagger
  c_{l_3,m_3} c_{l_4,m_4}
  \mapsto
  e^{-i\phi_{l_1}}
  e^{-i\phi_{l_2}}
  e^{i\phi_{l_3}}
  e^{i\phi_{l_4}}
  c_{l_1,m_1}^\dagger c_{l_2,m_2}^\dagger
  c_{l_3,m_3} c_{l_4,m_4}.
\end{align}
The common choice $\phi_l=\phi$ for all orbital blocks corresponds to the ordinary particle-number gauge transformation.  
As discussed in Appendix~\ref{app:single_j}, if one requires Hermitian even-fermion multipole operators to have definite time-reversal parity, this continuous gauge transformation yields a universal nontrivial $\mathbb Z_2$ label only at
\begin{align}
  \phi=\frac{\pi}{2},
\end{align}
which defines the discrete gauge transformation $\mathcal G_{\pi/2}$ used in the main text.
By contrast, assigning different $\pi/2$ phases to different orbitals generally gives a factor $\pm i$ to hybrid anomalous operators such as $c^\dagger_a c^\dagger_b$. 
Such transformations do not universally preserve Hermiticity or definite time-reversal parity and therefore do not define universal classification labels.  
Hence, $\phi=\pi/2$ has the meaning of a universal classification label only when it acts commonly on all orbitals.

The remaining physically meaningful transformations are relative sign changes between orbital blocks.
Let $\alpha$ label an orbital block with angular momentum $l_\alpha$. 
We define $\mathcal G_\alpha$ as the transformation that changes the sign of all fermion operators in block $\alpha$ and leaves the other blocks unchanged:
\begin{align}
  \mathcal G_\alpha :
  \qquad
  c^\dagger_{l_{\beta},m_{\beta}}
  \mapsto
  \begin{cases}
    -c^\dagger_{l_{\beta},m_{\beta}}, & \beta=\alpha,\\
    \ \ c^\dagger_{l_{\beta},m_{\beta}}, & \beta\neq\alpha .
  \end{cases}
\end{align}
The product over all $n_{\rm orb}$ orbital blocks gives $(-1)^{2n} = 1$ on any even operator of order $2n$, and therefore
\begin{align}
  \prod_{\alpha=1}^{n_{\rm orb}} \mathcal G_\alpha = I_{\rm n_{\rm orb}}.
\end{align}
Hence, only $n_{\rm orb}-1$ of them are independent.

Among these relative sign transformations, spatial inversion $\mathcal P$ is distinguished by its basis-independent physical meaning. 
If the parity of the orbital block $\alpha$ is
\begin{align}
  \pi_{l_\alpha}=(-1)^{l_\alpha},
\end{align}
then inversion acts as
\begin{align}
  \mathcal P:
  \qquad
  c^\dagger_{l_\alpha,m_{\alpha}}
  \mapsto
  \pi_{l_\alpha}
  c^\dagger_{l_\alpha,m_{\alpha}}
  , \qquad
  c_{l_\alpha,m_{\alpha}} \mapsto \pi_{l_{\alpha}} c_{l_\alpha,m_{\alpha}}.
\end{align}
Thus $\mathcal P$ is a specific relative sign transformation fixed by the physical parity of the orbitals.  
Unlike a generic $\mathcal G_\alpha$, spatial inversion has a basis-independent physical meaning once the orbitals are specified.  
For this reason, the universal classification adopted in the main text retains $\mathcal P$ as the standard spatial label, but does not include all possible relative signs as independent universal quantum numbers.

\subsection{Spinless orbitals with repeated radial copies}
\label{app:spinless_repeated_radial_copies}

The assumption $d_{l} = 1$ in the previous subsection is appropriate for a minimal hybrid model in which only one radial copy of each orbital angular momentum is retained.  
% This assumption, however, is not essential.  
% If several radial orbitals with the same angular momentum are included, for example two $s$-type orbitals distinguished by their radial quantum numbers, the same irreducible representation of the rotation group appears more than once.
% For example, 
% consider two $s$-type orbitals distinguished by their radial quantum numbers, such as $3s$ and $4s$. 
% 
This assumption, however, is not essential.  
If several radial orbitals with the same angular momentum are included,
the same irreducible representation of the rotation group appears multiple times. As a concrete example,
consider two $s$-type orbitals distinguished by
their radial quantum numbers, such as $3s$ and $4s$.
The $l=0$ sector is then
\begin{align}
  \mathcal H_{l=0}
  =
  \mathcal M_0\otimes V_0,
  \qquad
  \dim\mathcal M_0=2,
  \qquad
  \dim V_0=1,
  \label{eq:B_3s4s_decomp}
\end{align}
where $\mathcal M_0$ distinguishes the two radial copies.
Introducing
\begin{align}
  {\bm C}_{0}^\dagger
  =
  \begin{pmatrix}
    c_{3s}^\dagger\\
    c_{4s}^\dagger
  \end{pmatrix},
  \label{eq:B_3s4s_creation_vector}
\end{align}
a rotation-compatible creation-preserving canonical transformation may mix them as
\begin{align}
  U\,{\bm C}_{0}^\dagger\, U^{-1}
  =
  u_0\,{\bm C}_{0}^\dagger,
  \qquad
  u_0\in U(2).
  \label{eq:B_3s4s_U2}
\end{align}
This mixing is allowed by rotational symmetry because both orbitals transform according to the same irreducible representation $l=0$.  
Equivalently, Schur's lemma fixes the action in the angular-momentum space $V_0$ to be the identity, while the matrix $u_0$ acts only in the multiplicity space $\mathcal M_0$.

The same statement holds for any set of $d_l$ radial copies with the same orbital angular momentum $l$.  
If
\begin{align}
  {\bm C}_{l}^\dagger
  =
  \left(
    {\bm c}_{1,l}^\dagger,
    {\bm c}_{2,l}^\dagger,
    \ldots,
    {\bm c}_{d_l,l}^\dagger
  \right)^T ,
  \label{eq:B_spinless_repeated_l_vector}
\end{align}
then a rotation-compatible creation-preserving unitary transformation has the form
\begin{align}
 U\,{\bm C}_{l}^\dagger\, U^{-1}
  =
  \left(
    u_l\otimes I_{2l+1}
  \right)
  {\bm C}_{l}^\dagger,
  \qquad
  u_l\in U(d_l).
  \label{eq:B_spinless_repeated_l_Udl}
\end{align}
Thus, repeated copies of the same irreducible representation generate a $U(d_l)$ freedom acting solely in the multiplicity space.

Such multiplicity-space rotations should be distinguished from universal symmetry labels.  
Although the transformations above are compatible with rotational symmetry, they are generally not symmetries of a specific Hamiltonian. 
For example, $3s$ and $4s$ orbitals generally have different radial energies and different matrix elements, so an arbitrary $U(2)$ rotation between them is not physically realized as a symmetry.  
Therefore, multiplicity-space rotations represent model-dependent basis freedoms or additional accidental symmetries that may emerge in special situations. 
Here, the multiplicity labels need not correspond to radial degrees of freedom. 
They may instead represent internal labels distinguishing equivalent rotational irreps, such as pseudospin or flavor degrees of freedom.
As explained above, these multiplicity-space rotations 
do not constitute universal $\mathbb Z_2$ labels and are not included in the sixteenfold classification in the main text.

\subsection{Repeated irreducible representations in spinful hybrid systems}
\label{app:hybrid_spinful}

The multiplicity-space structure discussed in the previous subsection is not restricted to repeated radial copies in spinless systems.
In spinful systems with spin-orbit coupling, states originating from different orbitals can have the same total angular momentum $j$.  
In that case, the same irreducible representation $V_j$ appears multiple times in the one-particle Hilbert space, giving rise to a nontrivial multiplicity space $\mathcal M_j$.

As the simplest example, consider the $j=1/2$ doublet originating from
an $s$ orbital and the $j=1/2$ doublet originating from a $p$
orbital. 
In this case, the corresponding one-particle space is
\begin{align}
  \mathcal H_{j=1/2}
    =
  \mathcal M_{1/2}\otimes V_{1/2},
  \qquad
  \dim \mathcal M_{1/2}=2,
\end{align}
where $\mathcal M_{1/2}$ is the two-dimensional multiplicity space that distinguishes the orbital origin of the two copies of the same rotational irreducible representation.
Writing the creation-operator vector as
\begin{align}
  \bm{C}_{1/2}^\dagger
  :=
  \begin{pmatrix}
    {\bm c}_{s:j=1/2}^\dagger\\
    {\bm c}_{p:j=1/2}^\dagger
  \end{pmatrix},
\end{align}
a rotation-compatible creation-preserving unitary transformation takes
the form
\begin{align}
  U\,
  \bm{C}_{1/2}^\dagger\,U^{-1}
  =
  \left( u_{1/2}\otimes I_2 \right)
  \bm{C}_{1/2}^\dagger,
  \qquad u_{1/2} \in U(2).
\end{align}
Here $I_2$ is the $2\times2$ identity matrix acting on the irreducible angular-momentum space $V_{1/2}$,
and $u_{1/2}$ is a unitary matrix acting on the multiplicity space
$\mathcal M_{1/2}$.
More explicitly,
\begin{align}
  U\,
  \bm{C}_{1/2}^\dagger\,U^{-1}
  &=
  W\,
  \bm{C}_{1/2}^\dagger
  =
   \left( u_{1/2}\otimes I_2 \right) \bm{C}_{1/2}^\dagger =
  \begin{pmatrix}
     \alpha I_2 & \gamma I_2 \\
     \delta I_2 & \beta I_2
  \end{pmatrix}
  \begin{pmatrix}
    {\bm c}_{s:j=1/2}^\dagger\\
    {\bm c}_{p:j=1/2}^\dagger
  \end{pmatrix}
  =
  \left(
  \begin{pmatrix}
     \alpha & \gamma \\
     \delta & \beta
  \end{pmatrix}
  \otimes I_2
  \right)
  \begin{pmatrix}
    {\bm c}_{s:j=1/2}^\dagger\\
    {\bm c}_{p:j=1/2}^\dagger
  \end{pmatrix}.
\end{align}
The canonical condition $WW^\dagger=I$ implies
\begin{align}
  u_{1/2}
  :=
  \begin{pmatrix}
    \alpha & \gamma\\
    \delta & \beta
  \end{pmatrix}
  \in U(2).
\end{align}
The $U(1)$ component corresponding to the overall phase is the gauge freedom already discussed above.  Removing this overall phase leaves the nontrivial part characterized by the unimodular condition
\begin{align}
  \det u_{1/2}=1,
\end{align}
which is $SU(2)$.  
In other words, there exists a pseudospin $SU(2)$ rotation between the $s$- and $p$-derived $j=1/2$ doublets that commutes with rotations.  
In multipole language, the corresponding scalar operators include the diagonal component $n_s-n_p$ and off-diagonal hybridization components, such as electric-toroidal and magnetic monopoles; these form an $SU(2)$ algebra commuting with rotations.

The same conclusion holds generally. 
Whenever an irreducible representation $V_j$ appears with multiplicity $d_j$, rotation-compatible transformations contain an internal $U(d_j)$ freedom on $\mathcal M_j$.
At the level of multipole operators, this freedom is generated by operators that act within $\mathcal M_j$ while leaving the irreducible angular-momentum space $V_j$ unchanged.
For $d_j=2$, the nontrivial part forms an $SU(2)$ algebra and can be interpreted as a pseudospin-like rotation between the two equivalent copies of $V_j$.
Thus, repeated irreducible representations give rise to additional internal structure beyond the single-orbital case.

These continuous transformations are physically meaningful in a given model, but they are not universal symmetry labels.
They arise only when the same irreducible representation appears with nontrivial multiplicity, and therefore depend on the microscopic composition of the one-particle Hilbert space.
Since the multiplicity structure generally changes when the orbital content of the model is modified, the associated $U(d_j)$ freedom is model dependent.
Multiplicity-space rotations should therefore be regarded as additional internal structure rather than as part of the universal classification developed in this work.

\subsection{Consequence for the classification}

To summarize, the additional degrees of freedom arising in multi-orbital systems are of two types:
\begin{enumerate}
  \item orbital-dependent relative gauge phase in spinless hybrid systems;
  \item continuous rotations in the multiplicity space in spinless and spinful systems
  with repeated irreducible representations.
\end{enumerate}
The first category contains the physically distinguished case of spatial inversion $\mathcal P$, which acts as a fixed relative sign determined by orbital parity.  
The second corresponds to multiplicity-space rotations and appears only when the same irreducible representation occurs more than once.

To construct a universal classification applicable to both spinless and spinful systems, it is therefore natural to retain only those labels that follow directly from rotational compatibility and the canonical structure of fermionic operators.
For a single irreducible sector, these are
\begin{align}
  \mathcal T,\qquad
  \mathcal C_A \ \ (\text{or } \Gamma),\qquad
  \mathcal G_{\pi/2}.
\end{align}
These labels are independent of the microscopic orbital content and remain well defined in arbitrary hybrid systems.

Multi-orbital systems add the spatial inversion $\mathcal P$.
Thus the universal classification used in this work is then
\begin{align}
\label{eq:B_universal_classification_CA}
  SO(3)\times
  \mathbb Z_2^{\mathcal P}\times
  \mathbb Z_2^{\mathcal T}\times
  \mathbb Z_2^{\mathcal C_A}\times
  \mathbb Z_2^{\mathcal G_{\pi/2}},
\end{align}
or equivalently
\begin{align}
  \label{eq:B_universal_classification_Gamma}
  SO(3)\times
  \mathbb Z_2^{\mathcal P}\times
  \mathbb Z_2^{\mathcal T}\times
  \mathbb Z_2^{\Gamma}\times
  \mathbb Z_2^{\mathcal G_{\pi/2}} .
\end{align}

Other relative sign transformations in hybrid systems and continuous $U(d_j)$ rotations in multiplicity spaces may provide useful additional organization in a specific model.  
However, they depend on the detailed structure of the one-particle Hilbert space and are not universal consequences of rotational symmetry and canonical structure alone.
The sixteenfold classification used in the main text should therefore be understood as the universal part of the symmetry refinement. 
Additional multiplicity-space structures can be incorporated separately when required for a specific physical system.

\section{Symmetry-adapted Hermitian many-body multipole operators}
\label{app:operators}

In this Appendix, we explain how to construct symmetry-adapted Hermitian $N_{\rm b}$-body multipole operators.
We then show their transformation properties under the canonical symmetries introduced in Appendix~\ref{app:single_j}.
Our goal is to identify the universal discrete quantum numbers associated with many-body multipoles and to clarify their relation to the sixteenfold classification used in the main text. 

For an $N_{\rm b}$-body multipole operator with net particle-number change $\Delta N$, we show that the corresponding symmetry parities satisfy
\begin{align}
  \sigma_{\Gamma}
  =
  (-1)^{N_{\rm b}} \sigma_{\mathcal T},
  \qquad
  \sigma_{\mathcal C_A}
  =
  (-1)^{N_{\rm b}},
  \qquad
  \sigma_{\mathcal G}
  =
  (-1)^{\Delta N/2},
  \label{eq:C_final_labels}
\end{align}
where $\sigma_{\mathcal{T}} = \pm 1$ is the time-reversal parity.
These relations show that, once the body number $N_{\rm b}$, the particle-number transfer $\Delta N$, and the time-reversal parity are specified, the remaining canonical symmetry labels are fixed.
Together with the conventional spatial inversion parity $\sigma_{\mathcal{P}} = \pm 1$ and time-reversal parity $\sigma_{\mathcal{T}} = \pm 1$, they provide the foundation of the sixteenfold classification,
\begin{align}
  \mathbb Z_2^{\mathcal P}
  \times
  \mathbb Z_2^{\mathcal T}
  \times
  \mathbb Z_2^{\mathcal C_A}
  \times
  \mathbb Z_2^{\mathcal G_{\pi/2}},
  \label{eq:C_classification_CA}
\end{align}
or equivalently
\begin{align}
  \mathbb Z_2^{\mathcal P}
  \times
  \mathbb Z_2^{\mathcal T}
  \times
  \mathbb Z_2^{\Gamma}
  \times
  \mathbb Z_2^{\mathcal G_{\pi/2}}.
  \label{eq:C_classification_Gamma}
\end{align}

The remainder of this Appendix is devoted to deriving Eq.~\eqref{eq:C_final_labels} from the algebraic structure of fermionic operators and the symmetry-adapted multipole construction.

\subsection{Irreducible $N$-particle creation and annihilation operator tensors}

We first construct irreducible tensor operators composed of $N$ fermionic creation or annihilation operators.
The construction follows the standard angular-momentum coupling scheme and will serve as the building block for many-body multipole operators.

For $N=1$, the elementary creation operator tensor is
\begin{align}
\label{eq:C_C1_def}
  C^{[1]}_{\mu,j,m}:=c^\dagger_{\mu,j,m},
\end{align}
where $\mu$ denotes a multiplicity label in the sense of Eq.~\eqref{eq_cmujm}.
It collects all one-particle labels other than the angular-momentum labels $j,m$, such as orbital $l$, radial-copy, and so on.  

Higher-rank creation operator tensors are obtained recursively by angular-momentum coupling.
If $C^{[N-1]}$ is an antisymmetrized $(N-1)$-creation tensor, then
\begin{align}
  C^{[N]}_{\mu,j,m}
  :=
  \sum_{m',m''}
  \braket{j' m'; j'' m'' | j m}\,
  C^{[N-1]}_{\mu',j',m'}\,
  C^{[1]}_{\mu'', j'', m''}.
  \label{eq:C_def_Cn}
\end{align}
Here $\braket{j' m'; j'' m'' | j m}$ is the %Clebsch--Gordan (CG)
CG coefficient, and $\mu$ collectively denotes $\mu'$, $\mu''$, the intermediate angular momenta $j'$, $j''$, and so on.
We adopt the Condon--Shortley convention for CG coefficients.

For annihilation operators, we use the time-reversed $\tilde{c}_{\mu, j, m} = (-1)^{j-m}c_{\mu, j,-m}$ introduced in Eq.~\eqref{gc}, which transforms as a rank-$j$ tensor in the same way as $c^\dagger_{\mu, j, m}$.  
We therefore define the elementary annihilation tensor by
\begin{align}
  \label{eq:C_A1_def}
  A^{[1]}_{\mu, j, m}
  :=
  \tilde{c}_{\mu, j, m}.
\end{align}
The corresponding $N$-particle annihilation operator tensors are constructed recursively as
\begin{align}
  A^{[N]}_{\mu,j,m}
  :=
  \sum_{m',m''}
  \braket{j' m'; j'' m'' | j m}\,
  A^{[1]}_{\mu'', j'', m''}\,
  A^{[N-1]}_{\mu',j',m'},
  \label{eq:C_def_An}
\end{align}
where the ordering is intentionally opposite to that of Eq.~\eqref{eq:C_def_Cn}.

Schematically,
\begin{align}
  C^{[N]}
  &\sim
  c^\dagger_1 c^\dagger_2\cdots c^\dagger_N,
  \label{eq:C_order_schematic}
  \\
  A^{[N]}
  &\sim
  \tilde c_N\cdots \tilde c_2\tilde c_1 .
  \label{eq:A_order_schematic}
\end{align}
Thus $A^{[N]}$ is defined in the ordering naturally obtained from the Hermitian conjugating $C^{[N]}$. 
With this convention, the Hermitian conjugation relation between creation and annihilation tensors takes a simple block form and no additional fermionic ordering sign appears.

\subsection{Hermitian conjugation of $N$-particle creation and annihilation operator tensors}
\label{app:C_block_dagger}

We next establish the Hermitian-conjugation relation between the creation and annihilation tensor operators introduced above.

For $N=1$, the Hermitian conjugate of the elementary creation tensor is
\begin{align}
  \left(C^{[1]}_{\mu,j,m}\right)^\dagger
  =
  c_{\mu,j,m}
  =
  (-1)^{j+m}
  A^{[1]}_{\mu,j,-m},
  \label{eq:C1_dagger_A1}
\end{align}
where we used the definition $A^{[1]}_{\mu,j,m}=(-1)^{j-m}c_{\mu,j,-m}$.
For $N=2$, this relation can be checked explicitly.  From Eq.~\eqref{eq:C_def_Cn},
\begin{align}
  C^{[2]}_{\mu,j,m} 
  =
  \sum_{m_1,m_2}
  \braket{j_1m_1;j_2m_2|jm}\,
  C^{[1]}_{\mu_1,j_1,m_1}
  C^{[1]}_{\mu_2,j_2,m_2}.
  \label{eq:C_C2_def_explicit}
\end{align}
Taking the Hermitian conjugate, and using Eq.~\eqref{eq:C1_dagger_A1} together with the CG selection rule $m=m_1+m_2$, we obtain
\begin{align}
  \left( C^{[2]}_{\mu,j,m} \right)^\dagger
  &=
  \sum_{m_1,m_2}
  \braket{j_1m_1;j_2m_2|jm}\,
  \left( C^{[1]}_{\mu_2,j_2,m_2} \right)^\dagger
  \left( C^{[1]}_{\mu_1,j_1,m_1} \right)^\dagger
  \notag\\
  &=
  (-1)^{j_1+j_2+m}
  \sum_{m_1,m_2}
  \braket{j_1m_1;j_2m_2|jm}\,
  A^{[1]}_{\mu_2,j_2,-m_2}
  A^{[1]}_{\mu_1,j_1,-m_1}.
  \label{eq:C_C2_dagger_step2}
\end{align}
On the other hand, in the Condon--Shortley convention the CG coefficients satisfy
\begin{align}
  \braket{j_1m_1;j_2m_2|jm}
  =
  (-1)^{j_1+j_2-j}
  \braket{j_1,-m_1;j_2,-m_2|j,-m}.
  \label{eq:C_CG_reversal}
\end{align}
Therefore, using the definition of $A^{[2]}$ in Eq.~\eqref{eq:C_def_An}, we find
\begin{align}
  \left(
    C^{[2]}_{\mu,j,m}
  \right)^\dagger
  &=
  (-1)^{j_1+j_2+m}
  (-1)^{j_1+j_2-j}
  A^{[2]}_{\mu,j,-m}
  =
  (-1)^{j+m}
  A^{[2]}_{\mu,j,-m}.
  \label{eq:C_C2_dagger_final}
\end{align}
In the last step, we used the selection rule of the CG coefficient.  
If $\braket{j_1m_1;j_2m_2|jm}\neq0$, then $j_1+j_2-j\in\mathbb Z$, and $j_1+j_2+m+j_1+j_2-j = j+m+2(j_1+j_2-j)$.
Thus the two exponents differ by an even integer, and the remaining phase is $(-1)^{j+m}$.

The same argument extends recursively to arbitrary $N$. 
With the ordering convention in Eqs.~\eqref{eq:C_def_Cn} and \eqref{eq:C_def_An}, one obtains
\begin{align}
  \left(
    C^{[N]}_{\mu,j,m}
  \right)^\dagger
  =
  (-1)^{j+m}
  A^{[N]}_{\mu,j,-m}.
  \label{eq:C_block_dagger_Cn}
\end{align}
The Hermitian conjugate of the annihilation tensor is obtained similarly.  For $N=1$,
\begin{align}
  \left(A^{[1]}_{\mu,j,m}\right)^\dagger
  =
  \left[(-1)^{j-m}c_{\mu,j,-m}\right]^\dagger
  =
  (-1)^{j-m}
  C^{[1]}_{\mu,j,-m}.
  \label{eq:A1_dagger_C1}
\end{align}
The recursive construction then gives
\begin{align}
  \left(
    A^{[N]}_{\mu,j,m}
  \right)^\dagger
  =
  (-1)^{j-m}
  C^{[N]}_{\mu,j,-m}.
  \label{eq:C_block_dagger_An}
\end{align}

Thus the creation and annihilation tensor operators form mutually conjugate irreducible tensor multiplets. 
No additional fermionic ordering sign appears in Eqs.~\eqref{eq:C_block_dagger_Cn} and \eqref{eq:C_block_dagger_An}, precisely because the annihilation tensors were defined with the ordering opposite to that of the corresponding creation tensors.

\subsection{Symmetry properties of $N$-particle creation and annihilation operator tensors}
\label{app:C_symmetry_C_A_blocks}

\subsubsection*{Rotation $R$}

By construction, both $C^{[N]}_{\mu, j, m}$ and $A^{[N]}_{\mu, j, m}$ transform as irreducible rank-$j$ tensors under rotations:
\begin{align}
\label{eq:C_rotation_Cn}
  & R C^{[N]}_{\mu, j, m} R^{-1}
  =
  \sum_{m'}
  D^{(j)}_{m'm}(R)\,
  C^{[N]}_{\mu, j, m'},
\\
\label{eq:C_rotation_An}
  & R A^{[N]}_{\mu, j, m} R^{-1}
  =
  \sum_{m'}
  D^{(j)}_{m'm}(R)\,
  A^{[N]}_{\mu, j, m'}.
\end{align}

\subsubsection*{Spatial inversion $\mathcal{P}$}

For an elementary one-particle tensor, we write
\begin{align}
  \mathcal P c^\dagger_{\mu,j,m}\mathcal P^{-1}
  =
  \pi_\mu c^\dagger_{\mu,j,m},
  \qquad
  \mathcal P \tilde c_{\mu,j,m}\mathcal P^{-1}
  =
  \pi_\mu \tilde c_{\mu,j,m},
  \qquad
  \pi_\mu=\pm1 .
  \label{eq:C_P_elementary}
\end{align}
For an atomic orbital block $l$, $\pi_\mu=(-1)^{l}$.  
Therefore the inversion parity of an $N$-particle tensor is the product of the one-particle parities contained in the tensor:
\begin{align}
  \mathcal P C^{[N]}_{\mu,j,m}\mathcal P^{-1}
  &=
  \pi_\mu\,
  C^{[N]}_{\mu,j,m},
  \label{eq:C_P_Cn}
  \\
  \mathcal P A^{[N]}_{\mu,j,m}\mathcal P^{-1}
  &=
  \pi_\mu\,
  A^{[N]}_{\mu,j,m}.
  \label{eq:C_P_An}
\end{align}
The symbol $\pi_\mu$ denotes this product parity associated with the $N$-particle tensor label $\mu$.

\subsubsection*{Time reversal $\mathcal{T}$}

The time-reversal properties of $C^{[N]}_{\mu, j, m}$ and $A^{[N]}_{\mu, j, m}$ are
\begin{align}
 \mathcal T C^{[N]}_{\mu, j, m} \mathcal T^{-1}
 &=(-1)^{j-m}C^{[N]}_{\mu, j, -m},
 \label{eq:T_on_PNdag_jp}
 \\
 \mathcal T A^{[N]}_{\mu, j, m} \mathcal T^{-1}
 &=(-1)^{j-m} A^{[N]}_{\mu, j, -m}.
 \label{eq:T_on_tildePN_jp}
\end{align}
These relations follow recursively from Eqs.~\eqref{eq:C_def_Cn} and \eqref{eq:C_def_An}, together with the standard time-reversal property of CG coefficients.  
Thus, the action of $ \mathcal T $ is determined only by the irreducible tensor character, namely by the rank $j$ and component $m$, and not by the body order or particle-number structure of the operator.  
In this sense, $ \mathcal T $ acts universally at the level of irreducible tensor representations, independently of the microscopic construction of the many-body operator.

\subsubsection*{Chiral transformation $\Gamma=\mathcal T\mathcal C_A$}

% The particle-hole transformation $\mathcal C_A$ exchanges creation and annihilation operators.  
% After applying it to a normal-ordered many-body operator, one must normal order the result again.  
% This process generates contractions and hence gives the fermionic ordering sign given by Eq.~\eqref{eq:C_sN_def}. 
% It is therefore convenient to analyze the highest-body transformation using the unitary chiral operation $\Gamma=\mathcal T\mathcal C_A$ instead of the antiunitary $\mathcal C_A$ itself.
The particle-hole transformation $\mathcal C_A$ exchanges creation operators $c^\dagger$ and bare annihilation operators $c$, as defined in Eq.~\eqref{eq:CA}. Since the $N$-particle creation and annihilation operator tensors are expressed in terms of $c^\dagger$ and $\tilde c$, it is more transparent to analyze the transformation properties of higher-body operators using the chiral transformation $\Gamma=\mathcal T\mathcal C_A$ rather than the particle-hole transformation $\mathcal C_A$ itself.

In the convention of Appendix~\ref{app:single_j} given by Eq.~\eqref{eq:A_Gamma}, the elementary creation tensor is exchanged by $\Gamma$:
\begin{align}
  \Gamma C^{[1]}_{\mu,j,m}\Gamma^{-1}
  &=
  A^{[1]}_{\mu,j,m}.
  \label{eq:C_Gamma_C1}
\end{align}
Since $\Gamma$ acts on a product without reversing the order of the operators,
\begin{align}
  \Gamma
  \left(
    c^\dagger_1 c^\dagger_2\cdots c^\dagger_N
  \right)
  \Gamma^{-1}
  =
  \tilde c_1\tilde c_2\cdots \tilde c_N .
  \label{eq:Gamma_C_order}
\end{align}
However, the standard annihilation-tensor order in Eq.~\eqref{eq:A_order_schematic} is
$\tilde c_N\cdots\tilde c_2\tilde c_1$.  
Therefore, to express $\Gamma C^{[N]}\Gamma^{-1}$ in the standard $A^{[N]}$ convention, one must reverse the order of $N$ fermion operators.  
This requires $N(N-1)/2$ pairwise exchanges and gives the fermionic ordering sign
\begin{align}
  s_N
  =
  (-1)^{N(N-1)/2}.
  \label{eq:C_sN_def}
\end{align}
For $N=1$, $s_{1} = +1$, while for $N=2$,
\begin{align}
  \Gamma
  \left(
    c^\dagger_1c^\dagger_2
  \right)
  \Gamma^{-1}
  =
  \tilde c_1\tilde c_2
  =
  -
  \tilde c_2\tilde c_1,
  \label{eq:Gamma_C2_example}
\end{align}
so that $s_2=-1$.
Therefore, with the present ordering convention,
\begin{align}
  \label{eq:C_Gamma_Cn}
  \Gamma C^{[N]}_{\mu,j,m}\Gamma^{-1}
  &=
  s_N A^{[N]}_{\mu,j,m}.
\end{align}
It is a fermionic ordering sign fixed by the chosen convention for the relative ordering of creation and annihilation tensors.

We next derive the corresponding transformation of $A^{[N]}$.  
% 
% 
% From Eq.~\eqref{eq:C_Gamma_C1}, we have
From the definition in Eq.~\eqref{eq:A_Gamma}, we have
% \begin{align}
%   \Gamma c^\dagger_{\mu,j,m}\Gamma^{-1}
%   =
%   \tilde c_{\mu,j,m}
%   =
%   (-1)^{j-m}c_{\mu,j,-m}.
%   \label{eq:C_Gamma_cdagger_to_c}
% \end{align}
\begin{align}
  \label{eq:C_Gamma_cdagger_to_c}
  \Gamma c^\dagger_{\mu,j,m}\Gamma^{-1}
  &=
  \tilde c_{\mu,j,m},\\
  \label{eq:C_Gamma_c_minusm_to_cdagger}
  \Gamma \tilde c_{\mu,j,m} \Gamma^{-1}
  &=
  (-1)^{2j}c^\dagger_{\mu,j,m}.
\end{align}
% 
% Taking the Hermitian conjugate of Eq.~\eqref{eq:C_Gamma_cdagger_to_c} and replacing $m$ by $-m$, we obtain
% \begin{align}
%   \Gamma c_{\mu,j,-m}\Gamma^{-1}
%   =
%   (-1)^{j+m}c^\dagger_{\mu,j,m}.
%   \label{eq:C_Gamma_c_minusm_to_cdagger}
% \end{align}
% 
Therefore
% \begin{align}
%   \Gamma A^{[1]}_{\mu,j,m}\Gamma^{-1}
%   &=
%   \Gamma\left[
%     (-1)^{j-m}c_{\mu,j,-m}
%   \right]\Gamma^{-1}
%   =
%   (-1)^{j-m}
%   (-1)^{j+m}
%   c^\dagger_{\mu,j,m}
%   =
%   (-1)^{2j}
%   C^{[1]}_{\mu,j,m}.
%   \label{eq:C_Gamma_A1}
% \end{align}
\begin{align}
  \Gamma A^{[1]}_{\mu,j,m}\Gamma^{-1}
  &=
  (-1)^{2j}
  C^{[1]}_{\mu,j,m}.
  \label{eq:C_Gamma_A1}
\end{align}
% }
It is $+1$ for integer $j$ and $-1$ for half-integer $j$.

For a general $N$-particle annihilation tensor, acting with $\Gamma$ on $A^{[N]}_{\mu,j,m}$ exchanges each elementary annihilation tensor with a creation tensor.  
As in the creation-tensor case, restoring the standard ordering gives the fermionic sign $s_N$.  
In addition, the tensor-convention factor $(-1)^{2j}$ appears, as already seen in Eq.~\eqref{eq:C_Gamma_A1}. 
Hence
\begin{align}
  \Gamma A^{[N]}_{\mu,j,m}\Gamma^{-1}
  =
  s_N(-1)^{2j}
  C^{[N]}_{\mu,j,m}.
  \label{eq:C_Gamma_An}
\end{align}
The factor $(-1)^{2j}$ must be retained because the intermediate tensor rank $j$ of $C^{[N]}_{\mu,j,m}$ or $A^{[N]}_{\mu,j,m}$ can be half-integer, even though the final even-fermion many-body operator tensor has integer rank, as in Eq.~\eqref{eq:C_def_general_O}.

\subsubsection*{Particle-hole transformation $\mathcal C_A$}

The antiunitary particle-hole transformation $\mathcal C_A$ is related to the unitary chiral transformation by
\begin{align}
  \Gamma=\mathcal T\mathcal C_A,
  \qquad
  \mathcal C_A=\mathcal T^{-1}\Gamma .
  \label{eq:C_CA_from_Gamma}
\end{align}
Using Eq.~\eqref{eq:C_Gamma_Cn}, the particle-hole transformation of the creation tensor is
\begin{align}
  \mathcal C_A
  C^{[N]}_{\mu,j,m}
  \mathcal C_A^{-1}
  &=
  \mathcal T^{-1}
  \left(
    \Gamma
    C^{[N]}_{\mu,j,m}
    \Gamma^{-1}
  \right)
  \mathcal T
  \notag\\
  &=
  s_N
  \mathcal T^{-1}
  A^{[N]}_{\mu,j,m}
  \mathcal T
  \notag\\
  &=
  s_N
  (-1)^{j+m}
  A^{[N]}_{\mu,j,-m}.
  \label{eq:C_CA_Cn}
\end{align}

Similarly, using Eq.~\eqref{eq:C_Gamma_An}, the particle-hole transformation of the annihilation tensor is
\begin{align}
  \mathcal C_A
  A^{[N]}_{\mu,j,m}
  \mathcal C_A^{-1}
  &=
  \mathcal T^{-1}
  \left(
    \Gamma
    A^{[N]}_{\mu,j,m}
    \Gamma^{-1}
  \right)
  \mathcal T
  \notag\\
  &=
  s_N(-1)^{2j}
  \mathcal T^{-1}
  C^{[N]}_{\mu,j,m}
  \mathcal T
  \notag\\
  &=
  s_N(-1)^{2j}(-1)^{j+m}
  C^{[N]}_{\mu,j,-m}
  \notag\\
  &=
  s_N(-1)^{j-m}
  C^{[N]}_{\mu,j,-m}.
  \label{eq:C_CA_An}
\end{align}

\subsubsection*{Gauge transformation $\mathcal G_{\pi/2}$}

Under the discrete particle-number gauge transformation
\begin{align}
  \mathcal G_{\pi/2}:
  \qquad
  c^\dagger_{\mu,j,m}\mapsto -i c^\dagger_{\mu,j,m},
  \qquad
  c_{\mu,j,m}\mapsto i c_{\mu,j,m},
  \label{eq:C_G_elementary}
\end{align}
we also have $\tilde c_{\mu,j,m}\mapsto i\tilde c_{\mu,j,m}$.  
Hence
\begin{align}
  \mathcal G_{\pi/2}
  C^{[N]}_{\mu,j,m}
  \mathcal G_{\pi/2}^{-1}
  &=
  (-i)^N
  C^{[N]}_{\mu,j,m},
  \label{eq:C_G_Cn}
  \\
  \mathcal G_{\pi/2}
  A^{[N]}_{\mu,j,m}
  \mathcal G_{\pi/2}^{-1}
  &=
  i^N
  A^{[N]}_{\mu,j,m}.
  \label{eq:C_G_An}
\end{align}

\subsection{General many-body operator tensors}

A normal-ordered many-body operator tensor with $N_+$ creation operators and $N_-$ annihilation operators is obtained by coupling the creation operator tensors and the annihilation operator tensors:
\begin{align}
  O^{[N_{\rm b},\Delta N]}_{\mu,k,q}
  :=
  \sum_{m_+,m_-}
  \braket{j_+m_+;j_-m_-|kq}\,
  C^{[N_+]}_{\mu_+,j_+,m_+}\,
  A^{[N_-]}_{\mu_-,j_-,m_-}, 
  \qquad
  k, q \in \mathbb{Z}.
  \label{eq:C_def_general_O}
\end{align}
Here the body-number $N_{\rm b}$ and net particle-number change $\Delta N$ are
\begin{align}
  2N_{\rm b}=N_+ + N_-,
  \qquad
  \Delta N=N_+-N_-.
  \label{eq:C_body_delta_def}
\end{align}
We restrict ourselves to even-fermion operators, so $N_+ + N_-$ is even and $N_{\rm b}$ is an integer.  
The final tensor rank $k$ and component $q$ are therefore integers in the even-fermion operator space considered here.  
This should be distinguished from the intermediate tensor ranks $j_+$ and $j_-$, which can be integer or half-integer.  

For example,
\begin{align}
  N_{\rm b} = 1, \quad \Delta N = 0, \qquad (N_+,N_-)=(1,1) &\quad \Rightarrow\quad c^\dagger c \ \text{type},\\
  N_{\rm b} = 1, \quad \Delta N = 2, \qquad (N_+,N_-)=(2,0) &\quad \Rightarrow\quad c^\dagger c^\dagger \ \text{type},\\
  N_{\rm b} = 2, \quad \Delta N = 0, \qquad (N_+,N_-)=(2,2) &\quad \Rightarrow\quad c^\dagger c^\dagger c c \ \text{type},\\
  N_{\rm b} = 2, \quad \Delta N = 2, \qquad (N_+,N_-)=(3,1) &\quad \Rightarrow\quad c^\dagger c^\dagger c^\dagger c \ \text{type}.
\end{align}
By construction, $O^{[N_{\rm b},\Delta N]}_{\mu,k,q}$ transforms as a rank-$k$ irreducible tensor under rotations.

We next determine how this tensor is related to its Hermitian-conjugate channel.
Starting from Eq.~\eqref{eq:C_def_general_O} with $q$ replaced by $-q$, we obtain
\begin{align}
  \left(
    O^{[N_{\rm b},\Delta N]}_{\mu,k,-q}
  \right)^\dagger
  &=
  \sum_{m_+,m_-}
  \braket{j_+m_+;j_-m_-|k,-q}\,
  \left(
    A^{[N_-]}_{\mu_-,j_-,m_-}
  \right)^\dagger
  \left(
    C^{[N_+]}_{\mu_+,j_+,m_+}
  \right)^\dagger
  \notag\\
  &=
  \sum_{m_+,m_-}
  \braket{j_+m_+;j_-m_-|k,-q}\,
  (-1)^{j_- - m_-}
  (-1)^{j_+ + m_+}
  C^{[N_-]}_{\mu_-,j_-,-m_-}
  A^{[N_+]}_{\mu_+,j_+,-m_+},
  \label{eq:C_O_dagger_step1}
  \\ &=
  \sum_{m_+,m_-}
  (-1)^{j_-+m_-}
  (-1)^{j_+-m_+}
  \braket{j_+,-m_+;j_-,-m_-|k,-q}
  C^{[N_-]}_{\mu_-,j_-,m_-}
  A^{[N_+]}_{\mu_+,j_+,m_+}.
  \label{eq:C_O_dagger_step2}
\end{align}
In the second line, we used Eqs.~\eqref{eq:C_block_dagger_Cn} and \eqref{eq:C_block_dagger_An}.
In the third line, we relabeled $m_+\to -m_+$ and $m_-\to -m_-$.
Using the CG reversal identity
\begin{align}
  \braket{j_+,-m_+;j_-,-m_-|k,-q}
  =
  (-1)^{j_+ + j_- - k}
  \braket{j_+m_+;j_-m_-|kq},
  \label{eq:C_CG_reversal_for_general_O}
\end{align}
we obtain
\begin{align}
  \left(
    O^{[N_{\rm b},\Delta N]}_{\mu,k,-q}
  \right)^\dagger
  &=
  \sum_{m_+,m_-}
  (-1)^{
    j_+ + j_- - k
    +
    j_- + m_-
    +
    j_+ - m_+
  }
  \braket{j_+m_+;j_-m_-|kq}\,
  C^{[N_-]}_{\mu_-,j_-,m_-}
  A^{[N_+]}_{\mu_+,j_+,m_+}.
  \label{eq:C_O_dagger_step3}
\end{align}
Since the CG coefficient in Eq.~\eqref{eq:C_O_dagger_step3} is nonzero only when
$q=m_+ + m_-$, the exponent in Eq.~\eqref{eq:C_O_dagger_step3} satisfies
\begin{align}
  &j_+ + j_- - k
  +
  j_- + m_-
  +
  j_+ - m_+
  =
  2j_+ +2j_- -k +m_- -m_+
  =
  k-q
  +
  2(j_+ + j_- - k)
  +
  2m_- .
  \label{eq:C_phase_reduction_for_general_O}
\end{align}
As $m_-$ and $j_-$ have the same integer or half-integer character, $(-1)^{2m_-}=(-1)^{2j_-}$.  
In addition, since $j_+ +j_- -k\in\mathbb Z$, 
\begin{align}
  (-1)^{
    j_+ + j_- - k
    +
    j_- + m_-
    +
    j_+ - m_+
  }
  =
  (-1)^{k-q}
  (-1)^{2j_-}.
  \label{eq:C_phase_reduction_result}
\end{align}
Substituting this into Eq.~\eqref{eq:C_O_dagger_step3}, we find
\begin{align}
  \left(
    O^{[N_{\rm b},\Delta N]}_{\mu,k,-q}
  \right)^\dagger
  =
  (-1)^{k-q}
  (-1)^{2j_-}
  \sum_{m_+,m_-}
  \braket{j_+m_+;j_-m_-|kq}\,
  C^{[N_-]}_{\mu_-,j_-,m_-}
  A^{[N_+]}_{\mu_+,j_+,m_+}.
  \label{eq:C_O_dagger_final_explicit}
\end{align}

On the other hand, using Eqs.~\eqref{eq:C_Gamma_Cn} and \eqref{eq:C_Gamma_An}, $O^{[N_{\rm b},\Delta N]}_{\mu,k,q}$ transforms under $\Gamma$ as
\begin{align}
  \Gamma
  O^{[N_{\rm b},\Delta N]}_{\mu,k,q}
  \Gamma^{-1}
  =
  s_{N_+}s_{N_-}
  (-1)^{2j_-}
  \sum_{m_+,m_-}
  \braket{j_+m_+;j_-m_-|kq}\,
  A^{[N_+]}_{\mu_+,j_+,m_+}\,
  C^{[N_-]}_{\mu_-,j_-,m_-}.
  \label{eq:C_Gamma_general_O_before_normal_order}
\end{align}
The creation and annihilation blocks are exchanged.  
To restore normal order, the $N_+$ annihilation operators in $A^{[N_+]}$ must be moved past the $N_-$ creation operators in $C^{[N_-]}$, producing the fermionic sign $(-1)^{N_+N_-}$.  
The total highest-body sign is
\begin{align}
  s_{N_+}s_{N_-}(-1)^{N_+N_-}
  =
  (-1)^{
    \frac{N_+(N_+-1)}{2}
    +
    \frac{N_-(N_--1)}{2}
    +
    N_+N_-
  }
  =
  (-1)^{\frac{(N_+ + N_-)}{2}(N_+ + N_- - 1)}
=
  (-1)^{N_{\rm b} (2N_{\rm b} - 1)}
  =
  (-1)^{N_{\rm b}}.
  \label{eq:C_Gamma_highest_sign}
\end{align}
Normal ordering may also generate contraction terms with lower body number. 
Keeping the highest-body contribution explicitly, we obtain 
\begin{align}
\Gamma O^{[N_{\rm b},\Delta N]}_{\mu,k,q} \Gamma^{-1}
  = 
  (-1)^{N_{\rm b}}(-1)^{2j_-} 
  \sum_{m_+,m_-} 
  \braket{j_+m_+;j_-m_-|kq}\, 
  C^{[N_-]}_{\mu_-,j_-,m_-}\,
  A^{[N_+]}_{\mu_+,j_+,m_+}
 +
 \Delta^{(<N_{\rm b})}_{\mu,k,-q}
 \label{eq:C_Gamma_general_O_highest}
\end{align}
Here $\Delta^{(<N_{\rm b})}_{\mu,k,-q}$ denotes lower-body tensors generated by contractions during normal ordering. 

Comparing Eq.~\eqref{eq:C_Gamma_general_O_highest} with Eq.~\eqref{eq:C_O_dagger_final_explicit}, 
% we finally obtain
we obtain
\begin{align}
  \Gamma
  O^{[N_{\rm b},\Delta N]}_{\mu,k,q}
  \Gamma^{-1}
  =
  (-1)^{N_{\rm b}}
  (-1)^{k-q}
  \left(
    O^{[N_{\rm b},\Delta N]}_{\mu,k,-q}
  \right)^\dagger
  +
  \Delta^{(<N_{\rm b})}_{\mu,k,-q}.
  \label{eq:C_on_general_O}
\end{align}
This equation is the key highest-body transformation law.  
It shows that, up to lower-body contraction terms, the action of $\Gamma$ on an $N_{\rm b}$-body tensor differs from Hermitian conjugation and the standard tensor reversal only by the body-number sign $(-1)^{N_{\rm b}}$.
Applying the transformation $\Gamma$ once more, we obtain
  \begin{align}
    \label{Gamma_2_O}
  \Gamma^2
  O^{[N_{\rm b},\Delta N]}_{\mu,k,q}
  \Gamma^{-2}
  &=
  s_{N_+}^2s_{N_-}^2
  (-1)^{2j_+}(-1)^{2j_-}
  \sum_{m_+,m_-}
  \braket{j_+m_+;j_-m_-|kq}\,
  C^{[N_+]}_{\mu_+,j_+,m_+}\,
  A^{[N_-]}_{\mu_-,j_-,m_-}\nonumber\\
&=
  \sum_{m_+,m_-}
  \braket{j_+m_+;j_-m_-|kq}\,
  C^{[N_+]}_{\mu_+,j_+,m_+}\,
  A^{[N_-]}_{\mu_-,j_-,m_-}\nonumber\\
  &=
  O^{[N_{\rm b},\Delta N]}_{\mu,k,q}.
\end{align}

\subsection{Symmetry-adapted lower-body subtraction}
\label{app:C_lower_body_subtraction}

The term $\Delta^{(<N_{\rm b})}_{\mu,k,-q}$ in Eq.~\eqref{eq:C_on_general_O} arises from lower-body contributions generated by contractions during normal ordering.  
For $N_{\rm b}=1$ and $k=0$, it corresponds to the constant shift of the monopole.  
For $N_{\rm b}=2$, one-body tensors with the same rank $k$ can be mixed in, and in the scalar channel $k=0$ a constant term can also appear.  
In general, $\Delta^{(<N_{\rm b})}_{\mu,k,-q}$ is a linear combination of lower-body tensors with the appropriate rank-$k$ component.

We define a symmetry-adapted basis by subtracting lower-body components recursively.  
For one-body tensors, there is no lower-body operator with which they can mix, except in the electric-monopole sector.  
The one-body electric monopole contains an identity component: for a single fermionic mode, $n=c^\dagger c$ transforms as $n\mapsto 1-n$, so the covariant operator is $n-\frac{1}{2} I$.  
After this identity component is removed, all symmetry-adapted one-body tensors satisfy
\begin{align}
  \Gamma
  \widehat O^{[1,\Delta N]}_{\mu,k,q}
  \Gamma^{-1}
  =
  -
  (-1)^{k-q}
  \left(
    \widehat O^{[1,\Delta N]}_{\mu,k,-q}
  \right)^\dagger .
  \label{eq:C_Gamma_one_body_hat}
\end{align}

Next consider two-body operators.  
Acting with $\Gamma$ on a bare two-body tensor can generate one-body tensors with the same external rank and component, and in the scalar channel it can also generate the identity operator.  
Schematically, one may write
\begin{align}
  \Gamma
  O^{[2,\Delta N]}_{\mu,k,q}
  \Gamma^{-1}
  =
  (-1)^{k-q}
  \left[
    \left(
      O^{[2,\Delta N]}_{\mu,k,-q}
    \right)^\dagger
    +
    \sum_i c_i
    \left(
      \widehat O^{[1,\Delta N_i],i}_{\mu_i,k,-q}
    \right)^\dagger
  \right]
  +
  c_0\,\delta_{k,0}\,I .
  \label{eq:C_Gamma_two_body_with_lower}
\end{align}
The sum runs over symmetry-adapted one-body tensors with the same external tensor character $(k,q)$.  
The value of $\Delta N_i$ is fixed by the contraction channel.  
Since the one-body tensors have already been chosen to transform covariantly as in Eq.~\eqref{eq:C_Gamma_one_body_hat}, their contributions can be removed by redefining the two-body tensor.  
Away from the scalar identity ambiguity with $k = 0$, we define
\begin{align}
  \widehat O^{[2,\Delta N]}_{\mu,k,q}
  :=
  O^{[2,\Delta N]}_{\mu,k,q}
  +
  \sum_i
  \frac{c_i}{2}\,
  \widehat O^{[1,\Delta N_i],i}_{\mu_i,k,q},
  \qquad (k > 0).
  \label{eq:C_two_body_subtraction}
\end{align}
The factor $1/2$ appears because the lower-body contribution in Eq.~\eqref{eq:C_Gamma_two_body_with_lower} is shared between the operator and its Hermitian-conjugate partner.  
Using Eq.~\eqref{eq:C_Gamma_one_body_hat}, one then obtains
\begin{align}
  \Gamma
  \widehat O^{[2,\Delta N]}_{\mu,k,q}
  \Gamma^{-1}
  =
  (+1)
  (-1)^{k-q}
  \left(
    \widehat O^{[2,\Delta N]}_{\mu,k,-q}
  \right)^\dagger .
  \label{eq:C_Gamma_two_body_hat}
\end{align}

For the electric monopole $k=0$, the coefficients of $\sum_i c_i\,\widehat O^{[1,\Delta N_i]}_{0,0}$ are fixed by the condition of $\Gamma$-covariance.  
However, the identity operator $I$ has the same symmetry as the particle-number-conserving two-body electric monopole.  
Consequently, the coefficient of the identity component $c_0$ cannot be determined from $\Gamma$-covariance alone.  
To remove this remaining ambiguity, it is natural to impose an additional condition, such as tracelessness or orthogonality to all lower-body basis tensors.

The same idea gives the general recursive definition.  Suppose that all tensors with body-number smaller than $N_{\rm b}$ have already been redefined so as to transform covariantly under $\Gamma$.  For an $N_{\rm b}$-body tensor, we define
\begin{align}
  \widehat O^{[N_{\rm b},\Delta N]}_{\mu,k,q}
  =
  O^{[N_{\rm b},\Delta N]}_{\mu,k,q}
  +
  \text{(appropriate lower-body counterterms)}.
  \label{eq:C_hatO_definition}
\end{align}
The counterterms are chosen so as to cancel the lower-body contribution $\Delta^{(<N_{\rm b})}_{\mu,k,-q}$ in Eq.~\eqref{eq:C_on_general_O}.  The resulting symmetry-adapted tensor satisfies
\begin{align}
  \Gamma
  \widehat O^{[N_{\rm b},\Delta N]}_{\mu,k,q}
  \Gamma^{-1}
  =
  (-1)^{N_{\rm b}}
  (-1)^{k-q}
  \left(
    \widehat O^{[N_{\rm b},\Delta N]}_{\mu,k,-q}
  \right)^\dagger .
  \label{eq:C_on_hatO}
\end{align}

In view of Eq.~\eqref{eq:C_on_general_O}, one may formally define the following $\Gamma$-covariant operator:
\begin{align}
  \widehat O^{[N_{\rm b}, \Delta N]}_{\mu, k, q}
  =
  O^{[N_{\rm b}, \Delta N]}_{\mu, k, q}
  +
 (-1)^{N_{\rm b}}(-1)^{k-q}\,
  \Gamma \left(O^{[N_{\rm b}, \Delta N]}_{\mu, k, -q}\right)^\dagger
  \Gamma^{-1}.
\end{align}
Indeed, this definition satisfies
  \begin{align}
  \Gamma \widehat O^{[N_{\rm b}, \Delta N]}_{\mu, k, q} \Gamma^{-1}
  &=
  \Gamma O^{[N_{\rm b}, \Delta N]}_{\mu, k, q} \Gamma^{-1}
  +
 (-1)^{N_{\rm b}}(-1)^{k-q}\,
  \Gamma^2 O^{[N_{\rm b}, \Delta N]\dagger}_{\mu, k, -q}
  \Gamma^{-2}\nonumber\\
  &=
  \Gamma O^{[N_{\rm b}, \Delta N]}_{\mu, k, q} \Gamma^{-1}
  +
 (-1)^{N_{\rm b}}(-1)^{k-q}\,
  O^{[N_{\rm b}, \Delta N]\dagger}_{\mu, k, -q}\nonumber\\
  &=
  (-1)^{N_{\rm b}}(-1)^{k-q}
  \left(
  (-1)^{N_{\rm b}}(-1)^{k-q} \Gamma O^{[N_{\rm b}, \Delta N]}_{\mu, k, q} \Gamma^{-1}
  +\,
  O^{[N_{\rm b}, \Delta N]\dagger}_{\mu, k, -q}
  \right)\nonumber\\
  &=
  (-1)^{N_{\rm b}}(-1)^{k-q}
  \left(
  (-1)^{N_{\rm b}}(-1)^{k+q} \Gamma O^{[N_{\rm b}, \Delta N]}_{\mu, k, q} \Gamma^{-1}
  +\,
  O^{[N_{\rm b}, \Delta N]\dagger}_{\mu, k, -q}
  \right)\nonumber\\
  &=
  (-1)^{N_{\rm b}}(-1)^{k-q}
  \left(\widehat O^{[N_{\rm b}, \Delta N]}_{\mu, k, -q}\right)^\dagger,
\end{align}
where Eq.~\eqref{Gamma_2_O} has been used in the second equality.

Since the subtraction procedure involves only lower-body tensors with the same
tensor character, the time-reversal transformation property remains unchanged:
\begin{align}
  \mathcal T
  \widehat O^{[N_{\rm b},\Delta N]}_{\mu,k,q}
  \mathcal T^{-1}
  =
  (-1)^{k-q}
  \widehat O^{[N_{\rm b},\Delta N]}_{\mu,k,-q}.
  \label{eq:C_T_on_hatO}
\end{align}

\subsection{Hermitian many-body multipoles and symmetry labels}
\label{app:C_Hermitian_symmetry_labels}

Using the symmetry-adapted tensor $\widehat O$, we construct Hermitian many-body multipoles with definite $\mathcal T$ parity as
\begin{align}
  X^{[N_{\rm b},\Delta N]}_{\mu,k,q;+}
  &:=
  \widehat O^{[N_{\rm b},\Delta N]}_{\mu,k,q}
  +
  (-1)^{k-q}
  \left(
    \widehat O^{[N_{\rm b},\Delta N]}_{\mu,k,-q}
  \right)^\dagger ,
  \label{eq:C_def_Xplus}
  \\
  X^{[N_{\rm b},\Delta N]}_{\mu,k,q;-}
  &:=
  i\left[
  \widehat O^{[N_{\rm b},\Delta N]}_{\mu,k,q}
  -
  (-1)^{k-q}
  \left(
    \widehat O^{[N_{\rm b},\Delta N]}_{\mu,k,-q}
  \right)^\dagger
  \right].
  \label{eq:C_def_Xminus}
\end{align}
Here the daggered term has net particle-number change $-\Delta N$, because Hermitian conjugation exchanges creation and annihilation blocks.  
Both combinations satisfy the Hermiticity condition
\begin{align}
  \left(
    X^{[N_{\rm b},\Delta N]}_{\mu,k,q;\pm}
  \right)^\dagger
  =
  (-1)^{k-q}
  X^{[N_{\rm b},\Delta N]}_{\mu,k,-q;\pm}.
  \label{eq:C_X_Hermiticity}
\end{align}
The sign $\pm$ labels the time-reversal parity: $+$ denotes $\mathcal T$-even and $-$ denotes $\mathcal T$-odd. 
After assigning the conventional multipole name $X = Q,M,T, G$, this time-reversal parity is already encoded in the multipole symbol.

The symmetry properties of the Hermitian multipoles $X^{[N_{\rm b},\Delta N]}_{\mu,k,q;\pm}$ are summarized below. 

\subsubsection*{Rotation $R$}

Under rotations, they transform as rank-$k$ spherical tensors:
\begin{align}
  R
  X^{[N_{\rm b},\Delta N]}_{\mu,k,q;\pm}
  R^{-1}
  =
  \sum_{q'}
  D^{(k)}_{q'q}(R)\,
  X^{[N_{\rm b},\Delta N]}_{\mu,k,q';\pm}.
  \label{eq:C_X_summary_rotation}
\end{align}
\subsubsection*{Spatial inversion $\mathcal{P}$}
Their spatial inversion parity is
\begin{align}
  \mathcal P
  X^{[N_{\rm b},\Delta N]}_{\mu,k,q;\pm}
  \mathcal P^{-1}
  =
  \sigma_{\mathcal P}
  X^{[N_{\rm b},\Delta N]}_{\mu,k,q;\pm},
  \qquad
  \sigma_{\mathcal P}=\pm1 .
  \label{eq:C_X_summary_P}
\end{align}
Here $\sigma_{\mathcal P}$ is determined by the product of the one-particle inversion parities $\pi$ contained in the tensor.

\subsubsection*{Time reversal $\mathcal{T}$}

The two Hermitian combinations have definite time-reversal parity:
\begin{align}
  \mathcal T
  X^{[N_{\rm b},\Delta N]}_{\mu,k,q; \pm}
  \mathcal T^{-1}
  =
  \sigma_{\mathcal T}
  (-1)^{k-q}
  X^{[N_{\rm b},\Delta N]}_{\mu,k,-q; \pm},
  \qquad
  \sigma_{\mathcal T}=\pm1.
  \label{eq:C_X_summary_T}
\end{align}

\subsubsection*{Chiral transformation $\Gamma = \mathcal{T} \mathcal{C}_{A}$}

The $\Gamma$ parity follows from the symmetry-adapted transformation law of $\widehat O$:
\begin{align}
  \Gamma
  X^{[N_{\rm b},\Delta N]}_{\mu,k,q;\pm}
  \Gamma^{-1}
  =
  \sigma_{\Gamma}
  (-1)^{k-q}
  X^{[N_{\rm b},\Delta N]}_{\mu,k,-q;\pm},
  \qquad
  \sigma_{\Gamma}
  =
  (-1)^{N_{\rm b}}\sigma_{\mathcal T}.
  \label{eq:C_X_summary_Gamma}
\end{align}
Thus, the $\Gamma$ parity differs from the time-reversal parity only by the universal body-number factor $(-1)^{N_{\rm b}}$.

\subsubsection*{Particle-hole transformation $\mathcal C_{A}$}

Since $\Gamma=\mathcal T\mathcal C_A$, the particle-hole parity is
\begin{align}
  \mathcal C_A
  X^{[N_{\rm b},\Delta N]}_{\mu,k,q;\pm}
  \mathcal C_A^{-1}
  =
  \sigma_{\mathcal C_A}
  X^{[N_{\rm b},\Delta N]}_{\mu,k,q;\pm},
  \qquad
  \sigma_{\mathcal C_A}
  =
  \sigma_{\mathcal T}\sigma_{\Gamma}
  =
  (-1)^{N_{\rm b}}.
  \label{eq:C_X_summary_CA}
\end{align}
Thus the $\mathcal C_A$ label is determined solely by the parity of the body-number $N_{\rm b}$.

\subsubsection*{Gauge transformation $\mathcal{G}_{\pi/2}$}

Since we restrict ourselves to even-fermion-parity operators, $\Delta N\in 2\mathbb Z$.
The discrete gauge parity is fixed by the net particle-number change $\Delta N$:
\begin{align}
  \mathcal G_{\pi/2}
  X^{[N_{\rm b},\Delta N]}_{\mu,k,q;\sigma_{\mathcal T}}
  \mathcal G_{\pi/2}^{-1}
  =
  \sigma_{\mathcal G}
  X^{[N_{\rm b},\Delta N]}_{\mu,k,q;\sigma_{\mathcal T}},
  \label{eq:C_X_summary_G}
\end{align}
where
\begin{align}
  \sigma_{\mathcal{G}}
  :=
  e^{-i\pi \Delta N/2}
  =
  (-1)^{\Delta N/2}
  =
  \begin{cases}
    +1, & \Delta N = 0 \!\!\!\pmod 4,\\
    -1, & \Delta N = 2 \!\!\!\pmod 4 .
  \end{cases}
  \label{eq:G_parity_general_jp}
\end{align}
Thus, the $\mathcal G_{\pi/2}$ label is determined solely by $\Delta N\bmod 4$.

\subsubsection*{Summary of symmetry labels}

A symmetry-adapted Hermitian $N_{\rm b}$-body multipole $X^{[N_{\rm b},\Delta N]}_{\mu,k,q;\pm}$ is characterized by
\begin{align}
  \left(
    \sigma_{\mathcal P},
    \sigma_{\mathcal T},
    \sigma_{\mathcal C_A},
    \sigma_{\mathcal G}
  \right)
  =
  \left(
    \sigma_{\mathcal P},
    \pm 1,
    (-1)^{N_{\rm b}},
    (-1)^{\Delta N/2}
  \right),
  \label{eq:C_X_summary_labels_CA}
\end{align}
or equivalently by
\begin{align}
  \left(
    \sigma_{\mathcal P},
    \sigma_{\mathcal T},
    \sigma_{\Gamma},
    \sigma_{\mathcal G}
  \right)
  =
  \left(
    \sigma_{\mathcal P},
    \pm 1,
    (-1)^{N_{\rm b}} (\pm 1),
    (-1)^{\Delta N/2}
  \right).
  \label{eq:C_X_summary_labels_Gamma}
\end{align}
The conventional fourfold multipole classification is obtained from the spatial-inversion and time-reversal parities together with the tensor rank $k$. 
For a rank-$k$ multipole, the standard correspondence is
\begin{align} 
Q_{\mu,k,q}^{[N_{\rm b},\Delta N]} &: \quad (\sigma_{\mathcal P},\sigma_{\mathcal T}) = \left((-1)^k,+1\right), \notag\\ 
M_{\mu,k,q}^{[N_{\rm b},\Delta N]} &: \quad (\sigma_{\mathcal P},\sigma_{\mathcal T}) = \left((-1)^{k+1},-1\right), \notag\\ 
T_{\mu,k,q}^{[N_{\rm b},\Delta N]} &: \quad (\sigma_{\mathcal P},\sigma_{\mathcal T}) = \left((-1)^k,-1\right), \notag\\ 
G_{\mu,k,q}^{[N_{\rm b},\Delta N]} &: \quad (\sigma_{\mathcal P},\sigma_{\mathcal T}) = \left((-1)^{k+1},+1\right). 
\label{eq:C_conventional_QMTG} 
\end{align}
The additional labels 
\begin{align} 
\sigma_{\mathcal C_A}=(-1)^{N_{\rm b}}, \qquad \sigma_{\mathcal G}=(-1)^{\Delta N/2}, 
\label{eq:C_additional_labels_summary} 
\end{align} 
or equivalently
\begin{align} 
\sigma_{\Gamma}=(-1)^{N_{\rm b}}\sigma_{\mathcal T}, \qquad \sigma_{\mathcal G}=(-1)^{\Delta N/2}, 
\label{eq:C_Gamma_G_labels_summary} 
\end{align} 
refine this conventional classification in the many-body operator space. 
Thus each conventional multipole type $Q,M,T,G$ splits into four many-body sectors distinguished by the pair $(\sigma_{\mathcal C_A},\sigma_{\mathcal G})$, or equivalently by $(\sigma_{\Gamma},\sigma_{\mathcal G})$ once $\sigma_{\mathcal T}$ is fixed. 
This provides the compact origin of the sixteenfold classification employed in the main text.

\paragraph*{One-body sector ($N_{\rm b} = 1$)}
In this case,
$\sigma_{\Gamma}=-\sigma_T$ and
$\sigma_{C_A}=-1$.  Therefore,
\begin{align}
 c^\dagger c
  &:\quad
  (\sigma_{C_A}, \sigma_{\Gamma},\sigma_{\mathcal{G}})=(-1,-\sigma_T,+1),\\
  c^\dagger c^\dagger + cc
  &:\quad
  (\sigma_{C_A}, \sigma_{\Gamma},\sigma_{\mathcal{G}})=(-1,-\sigma_T,-1).
\end{align}

\paragraph*{Two-body sector ($N_{\rm b} = 2$)}
In this case,
$\sigma_{\Gamma}=+\sigma_T$ and
$\sigma_{C_A}=+1$.  Therefore,
\begin{align}
  c^\dagger c^\dagger c c
  &:\quad
  (\sigma_{C_A}, \sigma_{\Gamma},\sigma_{\mathcal{G}})=(+1,+\sigma_T,+1),\\
  c^\dagger c^\dagger c^\dagger c + \mathrm{H.c.}
  &:\quad
  (\sigma_{C_A}, \sigma_{\Gamma},\sigma_{\mathcal{G}})=(+1,+\sigma_T,-1).
\end{align}

 { 
  % \color{orange}

  % {\color{orange}
% \section{Explicit Form of the Canonical Transformations}
  \section{Operator Representation of the Canonical Transformations}
  \label{Representation_Canonical_Transformations}

In the previous section, we examined the compatibility of 
transformations acting on the space of multipole operators with even fermion parity.
In this section, we discuss how these transformations are implemented
on a finite-dimensional Fock space.
It should be emphasized that commutativity at the level of transformation
actions and commutativity of their implementing operators are, in general,
not identical.
For example, $\mathcal T\Gamma=\Gamma\mathcal T$ means that, as actions
on multipoles, the result is independent of the order of the two actions.
In terms of the Fock-space operators $U_{\mathcal T}$ and $U_\Gamma$
implementing these transformations, this statement reads
\begin{align}
  U_{\mathcal T}U_\Gamma O
  U_\Gamma^{-1}U_{\mathcal T}^{-1}
  =
  U_\Gamma U_{\mathcal T}O
  U_{\mathcal T}^{-1}U_\Gamma^{-1}.
\end{align}
However, this equality of adjoint actions does not necessarily imply
$U_{\mathcal T}U_\Gamma=U_\Gamma U_{\mathcal T}$.
In general, the two products may differ by a phase factor:
\begin{align}
  U_{\mathcal T}U_\Gamma
  =
  \lambda\,
  U_\Gamma U_{\mathcal T},
  \qquad
  |\lambda|=1.
\end{align}

We consider a single irreducible $j$ multiplet.
The total angular momentum operators on the Fock space are defined by
\begin{align}
  J_a
  =
  \sum_{m,m'=-j}^{j}
  c_m^\dagger
  (J_a)_{mm'}
  c_{m'},
  \qquad
  a=x,y,z.
\end{align}
Continuous rotations are then represented by
\begin{align}
  U_R(\bm\theta)
  =
  e^{-i\,\bm\theta\cdot\bm J}.
\end{align}

In the standard convention, time reversal is written as
\begin{align}
  U_{\mathcal T}
  =
  e^{-i\pi J_y}\mathcal K ,
\end{align}
where $\mathcal K$ denotes complex conjugation.
Its action on the creation and annihilation operators reproduces
Eq.~(\ref{eq:time}).

Next, we discuss spatial inversion.
For a single $j$ multiplet, spatial inversion is represented in terms of
the inversion eigenvalue $\pi_j=\pm1$ as
\begin{align}
  U_{\mathcal P}
  =
  \exp\!\left[
    -i\pi\frac{1-\pi_j}{2}\hat N_j
  \right],
  \qquad
  \hat N_j
  =
  \sum_{m=-j}^{j}n_m,
  \qquad
  n_m=c_m^\dagger c_m.
\end{align}
For $\pi_j=+1$, this gives $U_{\mathcal P}=1$, whereas for
$\pi_j=-1$,
\begin{align}
  U_{\mathcal P}
  =
  e^{-i\pi\hat N_j}
  =
  (-1)^{\hat N_j}.
\end{align}
This representation reproduces Eqs.~(\ref{eq:inversion_cd}),(\ref{eq:inversion_c}).

The particle-number gauge transformation by $\pi/2$ is represented as
\begin{align}
  U_{\mathcal G_{\pi/2}}
  =
  e^{-i\pi\hat N_j/2},
\end{align}
which reproduces Eq.~(\ref{g_pi_2}).

\subsection{Chiral and Particle--Hole Transformations}
To implement the particle--hole transformation on Fock space, we introduce
Majorana operators.  For each mode $m$, we define
\begin{align}
  \gamma_m
  &:=
  c_m+c_m^\dagger,
  \\
  \bar\gamma_m
  &:=
  -i(c_m-c_m^\dagger)
  =
  i(c_m^\dagger-c_m).
\end{align}
They satisfy
\begin{align}
  \label{majorana_relation}
  \gamma_m^\dagger&=\gamma_m,
  \qquad
  \bar\gamma_m^\dagger=\bar\gamma_m,
  \qquad
  \gamma_m^2=1,
  \\
  \label{majorana_relation_2}
  \{\gamma_m,\gamma_{m'}\}
  &=
  2\delta_{mm'},
  \qquad
  \{\bar\gamma_m,\bar\gamma_{m'}\}
  =
  2\delta_{mm'},
  \qquad
  \{\gamma_m,\bar\gamma_{m'}\}=0.
\end{align}

We fix the order of the product as $m=j,j-1,\dots,-j$ and define
\begin{align}
  \Omega_j
  &:=
  \prod_{m=j}^{-j}\gamma_m
  =
  \gamma_{m_1}\gamma_{m_2}\cdots\gamma_{m_{N_j}},
\end{align}
where
$
  m_1=j,
  m_2=j-1,
  \dots,
  m_{N_j}=-j
$.
The inverse of $\Omega_j$ is given by the product in the reverse order:
\begin{align}
  \Omega_j^{-1}
  =
  \gamma_{m_{N_j}}\cdots\gamma_{m_2}\gamma_{m_1}.
\end{align}
Using Eq.~(\ref{majorana_relation}), one finds
\begin{align}
  \Omega_j\Omega_j^{-1}
  =
  \gamma_{m_1}\cdots\gamma_{m_{N_j}}
  \gamma_{m_{N_j}}\cdots\gamma_{m_1}
  =
  1.
\end{align}
With $N_j:=2j+1$, Eqs.~(\ref{majorana_relation}) and
(\ref{majorana_relation_2}) give
\begin{align}
  \Omega_j^2
  =
  (-1)^{N_j(N_j-1)/2},
  \qquad
  \Omega_j^\dagger
  =
  \Omega_j^{-1}
  =
  (-1)^{N_j(N_j-1)/2}\Omega_j.
\end{align}
From the Majorana anticommutation relations, the adjoint action of
$\Omega_j$ is
\begin{align}
  \Omega_j\gamma_\mu\Omega_j^{-1}
  =
  (-1)^{N_j-1}\gamma_\mu,
  \qquad
  \Omega_j\bar\gamma_\mu\Omega_j^{-1}
  =
  (-1)^{N_j}\bar\gamma_\mu .
\end{align}
Equivalently, in terms of the fermion operators,
\begin{align}
  \Omega_j c_\mu^\dagger\Omega_j^{-1}
  =
  (-1)^{N_j-1}c_\mu,
  \qquad
  \Omega_j c_\mu\Omega_j^{-1}
  =
  (-1)^{N_j-1}c_\mu^\dagger .
\end{align}
Therefore, the particle--hole transformation can be represented on Fock
space as
\begin{align}
  U_{\mathcal C_A}
  =
  \Omega_j\mathcal K .
\end{align}
Its action on the creation and annihilation operators is
\begin{align}
  U_{\mathcal C_A}c_\mu^\dagger U_{\mathcal C_A}^{-1}
  =
  (-1)^{N_j-1}c_\mu,
  \qquad
  U_{\mathcal C_A}c_\mu U_{\mathcal C_A}^{-1}
  =
  (-1)^{N_j-1}c_\mu^\dagger.
\end{align}
The overall sign appearing here is visible on single-fermion operators
as a difference from $\mathcal G_\pi=(-1)^{\hat N_j}$, but it acts
trivially on the even-fermion operator space.
Thus, this representation is consistent with the discussion in the previous section on the even-fermion multipole space.

For the chiral transformation
$  \Gamma=\mathcal T\mathcal C_A $,
the representation on Fock space is
\begin{align}
  U_\Gamma
  =
  U_{\mathcal T}U_{\mathcal C_A}
  =
  e^{-i\pi J_y}\Omega_j.
\end{align}
Here we have used $\mathcal K\Omega_j=\Omega_j\mathcal K$.

Since $\Omega_j$ is invariant under the $\pi$ rotation about the
$y$ axis in the above convention, $e^{-i\pi J_y}$ and $\Omega_j$
commute:
\begin{align}
  \label{Omega_pi_y_rot}
  e^{-i\pi J_y} \Omega_j e^{i\pi J_y}
  &=
  \prod_{m=j}^{-j}
  (-1)^{j-m}\gamma_{-m}
  =
  (-1)^{\sum_{m}(j-m) + N_j(N_j-1)/2}
  \prod_{m=j}^{-j}\gamma_m
  =
  \Omega_j.
\end{align}
Here we used
$
  \sum_{m=-j}^{j}(j-m) = j(2j+1)
  =
  \frac{N_j(N_j-1)}{2}.
$
Therefore, 
\begin{align}
  U_\Gamma^2
  =
  e^{-i2\pi J_y}\Omega_j^2
  =
  (-1)^{2j\hat N_j}
  (-1)^{N_j(N_j-1)/2}.
\end{align}
Here we used the identity on Fock space
$
  e^{-i2\pi J_y}
  =
  (-1)^{2j\hat N_j}.
$
It is convenient to define
\begin{align}
  K_j
  &:=
  \frac{N_j(N_j-1)}{2},
  \qquad
  \epsilon_j
  :=
  (-1)^{K_j},
  \qquad
  g_j
  :=
  (-1)^{2j\hat N_j}.
\end{align}
Then
\begin{align}
  U_\Gamma^2=\epsilon_j g_j,
  \qquad
  U_\Gamma^\dagger
  =
  U_\Gamma^{-1}
  =
  \epsilon_j g_j U_\Gamma.
\end{align}

We now decompose $U_\Gamma$ into Hermitian components:
\begin{align}
  \mathbb U_+
  &:=
  \frac{1}{2}
  \left(
    U_\Gamma+U_\Gamma^\dagger
  \right)
  =
  \frac{1}{2}
  \left(
    1+\epsilon_j g_j
  \right)U_\Gamma,
  \\
  \mathbb U_-
  &:=
  \frac{i}{2}
  \left(
    U_\Gamma-U_\Gamma^\dagger
  \right)
  =
  \frac{i}{2}
  \left(
    1-\epsilon_j g_j
  \right)U_\Gamma.
\end{align}
These operators are Hermitian. Conversely, one has
\begin{align}
  U_\Gamma
  =
  \mathbb U_+ - i\mathbb U_-.
\end{align}

For integer $j$, one has $g_j=1$.
Therefore, only one of $\mathbb U_+$ and $\mathbb U_-$ is nonzero:
\begin{align}
  j=0:\mathbb U_+,\qquad
  j=1:\mathbb U_-,\qquad
  j=2:\mathbb U_+,\qquad
  j=3:\mathbb U_-.
\end{align}
For half-integer $j$, one has
$  g_j=(-1)^{\hat N_j}.$
Thus, the active Hermitian component depends on the fermion-parity
sector. For instance,
\begin{align}
  &j=\frac12:\quad g_j=+1
  \quad\Longrightarrow\quad
  \mathbb U_-,
\qquad g_j=-1
  \quad\Longrightarrow\quad
  \mathbb U_+,
  \\
  &j=\frac32:\quad g_j=+1
  \quad\Longrightarrow\quad
  \mathbb U_+,
\qquad g_j=-1
  \quad\Longrightarrow\quad
  \mathbb U_-.
\end{align}
We emphasize that restricting the action to the even-fermion operator space does not mean setting $F=+1$; the latter corresponds to restricting the Fock space itself to the even-particle-number sector.

For hybrid systems, the above results are modified only by
\begin{align}
  e^{-i2\pi J_y}
  =
  \prod_k (-1)^{2k\hat N_k},
  \qquad
  N_j
  \rightarrow
  N:=\sum_k N_k.
\end{align}
Hence,
\begin{align}
  U_\Gamma^2
  =
  \left[
    \prod_k (-1)^{2k\hat N_k}
  \right]
  (-1)^{N(N-1)/2}.
\end{align}
For spinless hybrid systems, the first factor is unity, so that the
active Hermitian component is determined solely by the total number of single-particle states $N$. For example, a spinless $s$-$p$ hybrid has $N=4$, and hence $\mathbb U_+$ is active. For spinful systems composed of half-integer multiplets, the first factor becomes the total fermion parity $(-1)^{\hat N}$. Thus, as in the single-multiplet case, the active component depends on the fermion-parity sector.

\subsubsection*{Multipole classification}

We first examine the transformation properties under rotations.
Since $ \Gamma J_a \Gamma^{-1}=J_a$, one obtains
\begin{align}
  U_\Gamma U_R(\bm\theta)U_\Gamma^{-1}
  &=
  \exp\!\left[
    -i\bm\theta\cdot\bm J
  \right]
  =
  U_R(\bm\theta).
\end{align}
Thus, $U_\Gamma$ transforms as a rotational scalar.

Next, we examine the time-reversal and spatial-inversion properties.
Since $\Omega_j$ is real, the action of time reversal on $\Omega_j$ is equivalent to conjugation by a $\pi$ rotation about the $y$ axis. The latter has already been evaluated in Eq.~\eqref{Omega_pi_y_rot}. Thus,
\begin{align}
  U_{\mathcal T}\Omega_jU_{\mathcal T}^{-1}
  =
  e^{-i\pi J_y}\Omega_j e^{+i\pi J_y}
  =
  \Omega_j .
\end{align}
Under spatial inversion, one has
\begin{align}
  U_{\mathcal P}\Omega_jU_{\mathcal P}^{-1}
  =
  \pi_j^{N_j}\Omega_j .
\end{align}
Since $e^{-i\pi J_y}$ is even under both time reversal and spatial
inversion, it follows that
\begin{align}
  U_{\mathcal T}U_\Gamma U_{\mathcal T}^{-1}
  &=
  U_\Gamma,
  \qquad
  U_{\mathcal P}U_\Gamma U_{\mathcal P}^{-1}
  =
  \pi_j^{N_j}U_\Gamma .
\end{align}

As a result, the Hermitian components of $U_\Gamma$ satisfy
\begin{align}
  U_{\mathcal T}\mathbb U_+U_{\mathcal T}^{-1}
  &=
  \mathbb U_+,
  \qquad
  U_{\mathcal T}\mathbb U_-U_{\mathcal T}^{-1}
  =
  -\mathbb U_-.
\end{align}
On the other hand,
\begin{align}
  U_{\mathcal P}\mathbb U_\pm U_{\mathcal P}^{-1}
  =
  \pi_j^{N_j}\mathbb U_\pm.
\end{align}
Hence, $\mathbb U_+$ has the transformation property
\begin{align}
  \pi_j^{N_j}=+1
  \,\text{: E~monopole type},
  \qquad
  \pi_j^{N_j}=-1
  \,\text{: ET~monopole type},
\end{align}
whereas $\mathbb U_-$ has the transformation property
\begin{align}
  \pi_j^{N_j}=+1
  \,\text{: MT~monopole type},
  \qquad
  \pi_j^{N_j}=-1
  \,\text{: M~monopole type}.
\end{align}

This classification refers to the transformation properties of the
Hermitian components of the Fock-space representative.
The representative itself need not be fermion-parity even: since
$\Omega_j$ is a product of $N_j$ Majorana operators, its fermion parity
is $(-1)^{N_j}$.
Thus, it is fermion-parity even for even $N_j$ and fermion-parity odd
for odd $N_j$.

\subsection{Examples}

\subsubsection*{Spinless $s$ orbital}

For a spinless $s$ orbital, one has $j=0$ and $N_j=1$. Since
$J_y=0$,
$
  e^{-i\pi J_y}=1,
$
and hence
\begin{align}
  U_\Gamma
  =
  e^{-i\pi J_y}\Omega
  =
  \Omega
  =
  \gamma_s
  =
  c_s+c_s^\dagger.
\end{align}
This is classified as E monopole.

\subsubsection*{Spinful $s$ orbital}

Since the eigenvalues of $J_y$ on the finite-dimensional Fock space are
$\{0,\pm1/2\}$, one obtains
\begin{align}
  e^{-i\pi J_y}
  =
  1-2iJ_y-4J_y^2.
\end{align}

Let 
\begin{align}
  N
  &:=
  n_\uparrow+n_\downarrow,
  \qquad
  U
  :=
  n_\uparrow n_\downarrow,
  \qquad
  P_{S=1/2}
  :=
  N-2U.
\end{align}
$P_{S=1/2}$ represents the projection operator for one-particle states.
Then
\begin{align}
  J_y^2
  =
  \frac{1}{4}P_{S=1/2}
  =
  \frac{1}{4}(N-2U).
\end{align}
% % 
We find
\begin{align}
  \Omega
  =
  \gamma_\uparrow\gamma_\downarrow
  =
  -i \eta_- + 2i J_y,
\end{align}
where $\eta_- =  - i(c_\uparrow^\dagger c_\downarrow^\dagger - c_\downarrow c_\uparrow)$
which is MT monopole type.
Therefore,
\begin{align}
  U_\Gamma
  &=
  e^{-i\pi J_y}\Omega
  =
  \left(
    1-2iJ_y-P_{S=1/2}
  \right)
  \left(  -i \eta_- + 2i J_y \right)
  =
  -i\eta_-+2iJ_y+4J_y^2-2iJ_y
  =
  P_{S=1/2}-i\eta_-.
\end{align}
Thus, $U_\Gamma$ acts as the identity on the one paritcle sector.
On the $N=0,2$ sector, the nontrivial part is carried by the MT monopole.

\subsubsection*{Spinless orbitals in a real basis}

For spinless orbitals, it is often more transparent to use a real
one-particle basis (cartesian basis) in which the $\pi$ rotation about the $y$ axis is diagonal.  We denote this basis by
$
  c_\alpha,\,
  \alpha=1,\dots,N,
$
and assume
\begin{align}
  R_y c_\alpha R_y^{-1}
  =
  s_\alpha c_\alpha,
  \qquad
  R_y c_\alpha^\dagger R_y^{-1}
  =
  s_\alpha c_\alpha^\dagger,
  \qquad
  s_\alpha=\pm1,
\end{align}
where
% \begin{align}
$ R_y:=e^{-i\pi J_y}.$
% \end{align}
The Fock-space representative is
\begin{align}
  R_y
  =
  \prod_{s_\alpha=-1}(-1)^{n_\alpha}.
\end{align}
In this real basis,
the chiral transfomation takes the compact form
\begin{align}
  \label{general_Omega}
  R_y\Omega
  &=
  \left[
    \prod_{s_\alpha=-1}(-1)^{n_\alpha}
  \right]
  \prod_{\alpha=1}^{N}
  \left(
    c_\alpha+c_\alpha^\dagger
  \right)=
  \prod_{\alpha=1}^{N}
  \left(
    c_\alpha+s_\alpha c_\alpha^\dagger
  \right).
\end{align}
Equivalently, for an orbital with $s_\alpha=-1$ the Majorana factor
$c_\alpha+c_\alpha^\dagger$ is replaced by
$c_\alpha-c_\alpha^\dagger=i\bar\gamma_\alpha$, whereas for
$s_\alpha=+1$ it remains $c_\alpha+c_\alpha^\dagger=\gamma_\alpha$.

\subsubsection*{Spinless $p$ orbital}

For a spinless $p$ orbital, we take the cartesian basis
% \begin{align}
$
  c_x,\, c_y,\, c_z.
% \end{align}
$
Under the $\pi$ rotation about the $y$ axis,
\begin{align}
  c_x\mapsto -c_x,
  \qquad
  c_y\mapsto c_y,
  \qquad
  c_z\mapsto -c_z.
\end{align}
Thus
% \begin{align}
$
  s_x=-1,
  \,
  s_y=+1,
  \,
  s_z=-1,
% \end{align}
$
and the general formula (\ref{general_Omega}) gives
\begin{align}
  U_\Gamma
  =
  i\bar\gamma_x\,\gamma_y\,i\bar\gamma_z
  =
  (c_x-c_x^\dagger)
  (c_y+c_y^\dagger)
  (c_z-c_z^\dagger).
\end{align}
Expanding this expression, one obtains
\begin{align}
  U_\Gamma
  =&\
  c_x c_y c_z
  - c_x c_y c_z^\dagger
  + c_x c_y^\dagger c_z
  - c_x c_y^\dagger c_z^\dagger
-
  c_x^\dagger c_y c_z
  + c_x^\dagger c_y c_z^\dagger
  - c_x^\dagger c_y^\dagger c_z
  + c_x^\dagger c_y^\dagger c_z^\dagger .
\end{align}
Let
\begin{align}
  A
  &:=
  c_x^\dagger c_y^\dagger c_z^\dagger,
  \\
  S
  &:=
  -c_x c_y^\dagger c_z^\dagger
  +
  c_y c_z^\dagger c_x^\dagger
  -
  c_z c_x^\dagger c_y^\dagger .
\end{align}
Then the above expression can be written as
\begin{align}
  U_\Gamma
  =
  (A+S)-(A+S)^\dagger .
\end{align}
Equivalently, defining
\begin{align}
  M_0
  &:=
  i(A-A^\dagger)
  =
  i
  \left(
    c_x^\dagger c_y^\dagger c_z^\dagger
    +
    c_x c_y c_z
  \right),
  \\
  M_0'
  &:=
  i(S-S^\dagger),
\end{align}
one obtains
\begin{align}
  U_\Gamma
  =
  -iM_0-iM_0' .
\end{align}
Therefore, in the spinless $p$ orbital, the Hermitian representative of
the chiral transformation is of M monopole type.

\subsubsection*{Spinless $s$-$p$ Hybrid System}

For a spinless $s$-$p$ hybrid system,
one obtains
\begin{align}
  U_\Gamma
  =
  \gamma_s\, i\bar\gamma_x\,\gamma_y\, i\bar\gamma_z
  =
  \gamma_s\,(-iM_0-iM_0')
\end{align}
Focusing on Hermiticity ($U_\Gamma = U_\Gamma^\dagger$) and symmetry, we find
\begin{align}
  G_0^{(sp)}
  &:=
  % \gamma_s(-iM_0),
  -i \gamma_s M_0,
  \\
  G_0^{\prime(sp)}
  &:=
  -i \gamma_sM_0'.
\end{align}
Since $\gamma_s$ and $M_0,M_0'$ are fermion-parity odd Hermitian
operators acting on different orbital sectors, these products are
Hermitian. They provide ET monopole type components of
the spinless $s$-$p$ hybrid chiral implementation. These components can
be further decomposed according to the particle-number gauge charge; in
particular, $G_0^{\prime(sp)}$ contains a particle-number-conserving
two-body electric-toroidal monopole component.

}

\section{Multipole operators used in the examples of Sec.~\ref{sec:example}}
\label{app:matrix}

In this Appendix, we present the explicit matrix representations of the multipole operators used in the examples discussed of Sec.~\ref{sec:example}.
All matrices are written in the ordered bases specified below.

\subsection{Spinless $s$-$p$ system in the $N=2$ sector}

For the spinless $s$-$p$ system, we use the following basis of the
$N=2$ sector:
\begin{align}
  \ket{p_{1}s},
  \quad
  \ket{p_{0}s},
  \quad
  \ket{p_{0}p_{1}},
  \quad
  \ket{p_{-1}s},
  \quad
  \ket{p_{-1}p_{1}},
  \quad
  \ket{p_{-1}p_{0}} .
\end{align}
In this basis, the chiral transformation is represented as
\begin{align}
  \Gamma
  =
  \begin{pmatrix}
    0 & 0 & -1 & 0 & 0 & 0\\
    0 & 0 & 0 & 0 & -1 & 0\\
    -1 & 0 & 0 & 0 & 0 & 0\\
    0 & 0 & 0 & 0 & 0 & -1\\
    0 & -1 & 0 & 0 & 0 & 0\\
    0 & 0 & 0 & -1 & 0 & 0
  \end{pmatrix}.
\end{align}
The multipole matrices used in Table~\ref{tab:exp} are listed below.  
The superscript $[1]$ denotes one-body multipoles, whereas $[2]$ denotes two-body multipoles.
\begin{align}
Q_0^{%(1)
[1]}= \left[\begin{matrix}- \frac{\sqrt{3}}{6} & 0 & 0 & 0 & 0 & 0\\0 & - \frac{\sqrt{3}}{6} & 0 & 0 & 0 & 0\\0 & 0 & \frac{\sqrt{3}}{6} & 0 & 0 & 0\\0 & 0 & 0 & - \frac{\sqrt{3}}{6} & 0 & 0\\0 & 0 & 0 & 0 & \frac{\sqrt{3}}{6} & 0\\0 & 0 & 0 & 0 & 0 & \frac{\sqrt{3}}{6}\end{matrix}\right],
\end{align}
\begin{align}
M_x^{%(1)
[1]}= \left[\begin{matrix}0 & \frac{1}{2} & 0 & 0 & 0 & 0\\\frac{1}{2} & 0 & 0 & \frac{1}{2} & 0 & 0\\0 & 0 & 0 & 0 & \frac{1}{2} & 0\\0 & \frac{1}{2} & 0 & 0 & 0 & 0\\0 & 0 & \frac{1}{2} & 0 & 0 & \frac{1}{2}\\0 & 0 & 0 & 0 & \frac{1}{2} & 0\end{matrix}\right],
\end{align}
\begin{align}
M_y^{%(1)
[1]}= \left[\begin{matrix}0 & \frac{i}{2} & 0 & 0 & 0 & 0\\- \frac{i}{2} & 0 & 0 & \frac{i}{2} & 0 & 0\\0 & 0 & 0 & 0 & \frac{i}{2} & 0\\0 & - \frac{i}{2} & 0 & 0 & 0 & 0\\0 & 0 & - \frac{i}{2} & 0 & 0 & \frac{i}{2}\\0 & 0 & 0 & 0 & - \frac{i}{2} & 0\end{matrix}\right],
\end{align}
\begin{align}
M_z^{%(1)
[1]}= \left[\begin{matrix}\frac{\sqrt{2}}{2} & 0 & 0 & 0 & 0 & 0\\0 & 0 & 0 & 0 & 0 & 0\\0 & 0 & \frac{\sqrt{2}}{2} & 0 & 0 & 0\\0 & 0 & 0 & - \frac{\sqrt{2}}{2} & 0 & 0\\0 & 0 & 0 & 0 & 0 & 0\\0 & 0 & 0 & 0 & 0 & - \frac{\sqrt{2}}{2}\end{matrix}\right],
\end{align}
\begin{align}
Q_{z^2}^{%(1)
[1]}= \left[\begin{matrix}\frac{\sqrt{6}}{6} & 0 & 0 & 0 & 0 & 0\\0 & - \frac{\sqrt{6}}{3} & 0 & 0 & 0 & 0\\0 & 0 & - \frac{\sqrt{6}}{6} & 0 & 0 & 0\\0 & 0 & 0 & \frac{\sqrt{6}}{6} & 0 & 0\\0 & 0 & 0 & 0 & \frac{\sqrt{6}}{3} & 0\\0 & 0 & 0 & 0 & 0 & - \frac{\sqrt{6}}{6}\end{matrix}\right],
\end{align}
\begin{align}
Q_{x^2-y^2}^{%(1)
[1]}= \left[\begin{matrix}0 & 0 & 0 & \frac{\sqrt{2}}{2} & 0 & 0\\0 & 0 & 0 & 0 & 0 & 0\\0 & 0 & 0 & 0 & 0 & - \frac{\sqrt{2}}{2}\\\frac{\sqrt{2}}{2} & 0 & 0 & 0 & 0 & 0\\0 & 0 & 0 & 0 & 0 & 0\\0 & 0 & - \frac{\sqrt{2}}{2} & 0 & 0 & 0\end{matrix}\right],
\end{align}
\begin{align}
Q_{xy}^{%(1)
[1]}= \left[\begin{matrix}0 & 0 & 0 & \frac{\sqrt{2} i}{2} & 0 & 0\\0 & 0 & 0 & 0 & 0 & 0\\0 & 0 & 0 & 0 & 0 & - \frac{\sqrt{2} i}{2}\\- \frac{\sqrt{2} i}{2} & 0 & 0 & 0 & 0 & 0\\0 & 0 & 0 & 0 & 0 & 0\\0 & 0 & \frac{\sqrt{2} i}{2} & 0 & 0 & 0\end{matrix}\right],
\end{align}
\begin{align}
Q_{yz}^{%(1)
[1]}= \left[\begin{matrix}0 & \frac{i}{2} & 0 & 0 & 0 & 0\\- \frac{i}{2} & 0 & 0 & - \frac{i}{2} & 0 & 0\\0 & 0 & 0 & 0 & - \frac{i}{2} & 0\\0 & \frac{i}{2} & 0 & 0 & 0 & 0\\0 & 0 & \frac{i}{2} & 0 & 0 & \frac{i}{2}\\0 & 0 & 0 & 0 & - \frac{i}{2} & 0\end{matrix}\right],
\end{align}
\begin{align}
Q_{zx}^{%(1)
[1]}= \left[\begin{matrix}0 & \frac{1}{2} & 0 & 0 & 0 & 0\\\frac{1}{2} & 0 & 0 & - \frac{1}{2} & 0 & 0\\0 & 0 & 0 & 0 & - \frac{1}{2} & 0\\0 & - \frac{1}{2} & 0 & 0 & 0 & 0\\0 & 0 & - \frac{1}{2} & 0 & 0 & \frac{1}{2}\\0 & 0 & 0 & 0 & \frac{1}{2} & 0\end{matrix}\right],
\end{align}
\begin{align}
T_x^{%(1)
[1]}= \left[\begin{matrix}0 & 0 & 0 & 0 & \frac{\sqrt{2}}{4} & 0\\0 & 0 & \frac{\sqrt{2}}{4} & 0 & 0 & \frac{\sqrt{2}}{4}\\0 & \frac{\sqrt{2}}{4} & 0 & 0 & 0 & 0\\0 & 0 & 0 & 0 & \frac{\sqrt{2}}{4} & 0\\\frac{\sqrt{2}}{4} & 0 & 0 & \frac{\sqrt{2}}{4} & 0 & 0\\0 & \frac{\sqrt{2}}{4} & 0 & 0 & 0 & 0\end{matrix}\right],
\end{align}
\begin{align}
T_y^{%(1)
[1]}= \left[\begin{matrix}0 & 0 & 0 & 0 & \frac{\sqrt{2} i}{4} & 0\\0 & 0 & - \frac{\sqrt{2} i}{4} & 0 & 0 & \frac{\sqrt{2} i}{4}\\0 & \frac{\sqrt{2} i}{4} & 0 & 0 & 0 & 0\\0 & 0 & 0 & 0 & - \frac{\sqrt{2} i}{4} & 0\\- \frac{\sqrt{2} i}{4} & 0 & 0 & \frac{\sqrt{2} i}{4} & 0 & 0\\0 & - \frac{\sqrt{2} i}{4} & 0 & 0 & 0 & 0\end{matrix}\right],
\end{align}
\begin{align}
T_z^{%(1)
[1]}= \left[\begin{matrix}0 & 0 & \frac{1}{2} & 0 & 0 & 0\\0 & 0 & 0 & 0 & 0 & 0\\\frac{1}{2} & 0 & 0 & 0 & 0 & 0\\0 & 0 & 0 & 0 & 0 & - \frac{1}{2}\\0 & 0 & 0 & 0 & 0 & 0\\0 & 0 & 0 & - \frac{1}{2} & 0 & 0\end{matrix}\right],
\end{align}
\begin{align}
Q_x^{%(1)
[1]}= \left[\begin{matrix}0 & 0 & 0 & 0 & - \frac{\sqrt{2} i}{4} & 0\\0 & 0 & - \frac{\sqrt{2} i}{4} & 0 & 0 & - \frac{\sqrt{2} i}{4}\\0 & \frac{\sqrt{2} i}{4} & 0 & 0 & 0 & 0\\0 & 0 & 0 & 0 & - \frac{\sqrt{2} i}{4} & 0\\\frac{\sqrt{2} i}{4} & 0 & 0 & \frac{\sqrt{2} i}{4} & 0 & 0\\0 & \frac{\sqrt{2} i}{4} & 0 & 0 & 0 & 0\end{matrix}\right],
\end{align}
\begin{align}
Q_y^{%(1)
[1]}= \left[\begin{matrix}0 & 0 & 0 & 0 & \frac{\sqrt{2}}{4} & 0\\0 & 0 & - \frac{\sqrt{2}}{4} & 0 & 0 & \frac{\sqrt{2}}{4}\\0 & - \frac{\sqrt{2}}{4} & 0 & 0 & 0 & 0\\0 & 0 & 0 & 0 & - \frac{\sqrt{2}}{4} & 0\\\frac{\sqrt{2}}{4} & 0 & 0 & - \frac{\sqrt{2}}{4} & 0 & 0\\0 & \frac{\sqrt{2}}{4} & 0 & 0 & 0 & 0\end{matrix}\right],
\end{align}
\begin{align}
Q_z^{%(1)
[1]}= \left[\begin{matrix}0 & 0 & - \frac{i}{2} & 0 & 0 & 0\\0 & 0 & 0 & 0 & 0 & 0\\\frac{i}{2} & 0 & 0 & 0 & 0 & 0\\0 & 0 & 0 & 0 & 0 & \frac{i}{2}\\0 & 0 & 0 & 0 & 0 & 0\\0 & 0 & 0 & - \frac{i}{2} & 0 & 0\end{matrix}\right],
\end{align}
\begin{align}
M_x^{%(2)
[2]}= \left[\begin{matrix}0 & \frac{1}{4} & 0 & 0 & 0 & 0\\\frac{1}{4} & 0 & 0 & \frac{1}{4} & 0 & 0\\0 & 0 & 0 & 0 & - \frac{1}{4} & 0\\0 & \frac{1}{4} & 0 & 0 & 0 & 0\\0 & 0 & - \frac{1}{4} & 0 & 0 & - \frac{1}{4}\\0 & 0 & 0 & 0 & - \frac{1}{4} & 0\end{matrix}\right],
\end{align}
\begin{align}
M_y^{%(2)
[2]}= \left[\begin{matrix}0 & \frac{i}{4} & 0 & 0 & 0 & 0\\- \frac{i}{4} & 0 & 0 & \frac{i}{4} & 0 & 0\\0 & 0 & 0 & 0 & - \frac{i}{4} & 0\\0 & - \frac{i}{4} & 0 & 0 & 0 & 0\\0 & 0 & \frac{i}{4} & 0 & 0 & - \frac{i}{4}\\0 & 0 & 0 & 0 & \frac{i}{4} & 0\end{matrix}\right],
\end{align}
\begin{align}
M_z^{%(2)
[2]}= \left[\begin{matrix}\frac{\sqrt{2}}{4} & 0 & 0 & 0 & 0 & 0\\0 & 0 & 0 & 0 & 0 & 0\\0 & 0 & - \frac{\sqrt{2}}{4} & 0 & 0 & 0\\0 & 0 & 0 & - \frac{\sqrt{2}}{4} & 0 & 0\\0 & 0 & 0 & 0 & 0 & 0\\0 & 0 & 0 & 0 & 0 & \frac{\sqrt{2}}{4}\end{matrix}\right],
\end{align}
\begin{align}
Q_{zx}^{%(2)
[2]}= \left[\begin{matrix}0 & - \frac{1}{4} & 0 & 0 & 0 & 0\\- \frac{1}{4} & 0 & 0 & \frac{1}{4} & 0 & 0\\0 & 0 & 0 & 0 & - \frac{1}{4} & 0\\0 & \frac{1}{4} & 0 & 0 & 0 & 0\\0 & 0 & - \frac{1}{4} & 0 & 0 & \frac{1}{4}\\0 & 0 & 0 & 0 & \frac{1}{4} & 0\end{matrix}\right],
\end{align}
\begin{align}
Q_{yz}^{%(2)
[2]}= \left[\begin{matrix}0 & - \frac{i}{4} & 0 & 0 & 0 & 0\\\frac{i}{4} & 0 & 0 & \frac{i}{4} & 0 & 0\\0 & 0 & 0 & 0 & - \frac{i}{4} & 0\\0 & - \frac{i}{4} & 0 & 0 & 0 & 0\\0 & 0 & \frac{i}{4} & 0 & 0 & \frac{i}{4}\\0 & 0 & 0 & 0 & - \frac{i}{4} & 0\end{matrix}\right],
\end{align}
\begin{align}
Q_{x^2-y^2}^{%(2)
[2]}= \left[\begin{matrix}0 & 0 & 0 & - \frac{\sqrt{2}}{4} & 0 & 0\\0 & 0 & 0 & 0 & 0 & 0\\0 & 0 & 0 & 0 & 0 & - \frac{\sqrt{2}}{4}\\- \frac{\sqrt{2}}{4} & 0 & 0 & 0 & 0 & 0\\0 & 0 & 0 & 0 & 0 & 0\\0 & 0 & - \frac{\sqrt{2}}{4} & 0 & 0 & 0\end{matrix}\right],
\end{align}
\begin{align}
Q_{xy}^{%(2)
[2]}= \left[\begin{matrix}0 & 0 & 0 & - \frac{\sqrt{2} i}{4} & 0 & 0\\0 & 0 & 0 & 0 & 0 & 0\\0 & 0 & 0 & 0 & 0 & - \frac{\sqrt{2} i}{4}\\\frac{\sqrt{2} i}{4} & 0 & 0 & 0 & 0 & 0\\0 & 0 & 0 & 0 & 0 & 0\\0 & 0 & \frac{\sqrt{2} i}{4} & 0 & 0 & 0\end{matrix}\right],
\end{align}
\begin{align}
Q_{z^2}^{%(2)
[2]}= \left[\begin{matrix}- \frac{\sqrt{6}}{12} & 0 & 0 & 0 & 0 & 0\\0 & \frac{\sqrt{6}}{6} & 0 & 0 & 0 & 0\\0 & 0 & - \frac{\sqrt{6}}{12} & 0 & 0 & 0\\0 & 0 & 0 & - \frac{\sqrt{6}}{12} & 0 & 0\\0 & 0 & 0 & 0 & \frac{\sqrt{6}}{6} & 0\\0 & 0 & 0 & 0 & 0 & - \frac{\sqrt{6}}{12}\end{matrix}\right],
\end{align}
\begin{align}
Q_{0}^{%(2)
[2]}= \left[\begin{matrix}- \frac{\sqrt{3}}{6} & 0 & 0 & 0 & 0 & 0\\0 & - \frac{\sqrt{3}}{6} & 0 & 0 & 0 & 0\\0 & 0 & - \frac{\sqrt{3}}{6} & 0 & 0 & 0\\0 & 0 & 0 & - \frac{\sqrt{3}}{6} & 0 & 0\\0 & 0 & 0 & 0 & - \frac{\sqrt{3}}{6} & 0\\0 & 0 & 0 & 0 & 0 & - \frac{\sqrt{3}}{6}\end{matrix}\right],
\end{align}
\begin{align}
G_0^{%(2)
[2]}= \left[\begin{matrix}0 & 0 & - \frac{\sqrt{3}}{6} & 0 & 0 & 0\\0 & 0 & 0 & 0 & - \frac{\sqrt{3}}{6} & 0\\- \frac{\sqrt{3}}{6} & 0 & 0 & 0 & 0 & 0\\0 & 0 & 0 & 0 & 0 & - \frac{\sqrt{3}}{6}\\0 & - \frac{\sqrt{3}}{6} & 0 & 0 & 0 & 0\\0 & 0 & 0 & - \frac{\sqrt{3}}{6} & 0 & 0\end{matrix}\right],
\end{align}
\begin{align}
M_0^{%(2)
[2]}= \left[\begin{matrix}0 & 0 & \frac{\sqrt{3} i}{6} & 0 & 0 & 0\\0 & 0 & 0 & 0 & \frac{\sqrt{3} i}{6} & 0\\- \frac{\sqrt{3} i}{6} & 0 & 0 & 0 & 0 & 0\\0 & 0 & 0 & 0 & 0 & \frac{\sqrt{3} i}{6}\\0 & - \frac{\sqrt{3} i}{6} & 0 & 0 & 0 & 0\\0 & 0 & 0 & - \frac{\sqrt{3} i}{6} & 0 & 0\end{matrix}\right],
\end{align}
\begin{align}
T_x^{%(2)
[2]}= \left[\begin{matrix}0 & 0 & 0 & 0 & 0 & 0\\0 & 0 & 0 & 0 & 0 & 0\\0 & 0 & 0 & 0 & 0 & 0\\0 & 0 & 0 & 0 & 0 & 0\\0 & 0 & 0 & 0 & 0 & 0\\0 & 0 & 0 & 0 & 0 & 0\end{matrix}\right],
\end{align}
\begin{align}
T_y^{%(2)
[2]}= \left[\begin{matrix}0 & 0 & 0 & 0 & 0 & 0\\0 & 0 & 0 & 0 & 0 & 0\\0 & 0 & 0 & 0 & 0 & 0\\0 & 0 & 0 & 0 & 0 & 0\\0 & 0 & 0 & 0 & 0 & 0\\0 & 0 & 0 & 0 & 0 & 0\end{matrix}\right],
\end{align}
\begin{align}
T_z^{%(2)
[2]}= \left[\begin{matrix}0 & 0 & 0 & 0 & 0 & 0\\0 & 0 & 0 & 0 & 0 & 0\\0 & 0 & 0 & 0 & 0 & 0\\0 & 0 & 0 & 0 & 0 & 0\\0 & 0 & 0 & 0 & 0 & 0\\0 & 0 & 0 & 0 & 0 & 0\end{matrix}\right],
\end{align}
\begin{align}
Q_x^{%(2)
[2]}= \left[\begin{matrix}0 & 0 & 0 & 0 & 0 & 0\\0 & 0 & 0 & 0 & 0 & 0\\0 & 0 & 0 & 0 & 0 & 0\\0 & 0 & 0 & 0 & 0 & 0\\0 & 0 & 0 & 0 & 0 & 0\\0 & 0 & 0 & 0 & 0 & 0\end{matrix}\right],
\end{align}
\begin{align}
Q_y^{%(2)
[2]}= \left[\begin{matrix}0 & 0 & 0 & 0 & 0 & 0\\0 & 0 & 0 & 0 & 0 & 0\\0 & 0 & 0 & 0 & 0 & 0\\0 & 0 & 0 & 0 & 0 & 0\\0 & 0 & 0 & 0 & 0 & 0\\0 & 0 & 0 & 0 & 0 & 0\end{matrix}\right],
\end{align}
\begin{align}
Q_z^{%(2)
[2]}= \left[\begin{matrix}0 & 0 & 0 & 0 & 0 & 0\\0 & 0 & 0 & 0 & 0 & 0\\0 & 0 & 0 & 0 & 0 & 0\\0 & 0 & 0 & 0 & 0 & 0\\0 & 0 & 0 & 0 & 0 & 0\\0 & 0 & 0 & 0 & 0 & 0\end{matrix}\right],
\end{align}
\begin{align}
M_{zx}^{%(2)
[2]}= \left[\begin{matrix}0 & 0 & 0 & 0 & \frac{i}{4} & 0\\0 & 0 & \frac{i}{4} & 0 & 0 & - \frac{i}{4}\\0 & - \frac{i}{4} & 0 & 0 & 0 & 0\\0 & 0 & 0 & 0 & - \frac{i}{4} & 0\\- \frac{i}{4} & 0 & 0 & \frac{i}{4} & 0 & 0\\0 & \frac{i}{4} & 0 & 0 & 0 & 0\end{matrix}\right],
\end{align}
\begin{align}
M_{yz}^{%(2)
[2]}= \left[\begin{matrix}0 & 0 & 0 & 0 & - \frac{1}{4} & 0\\0 & 0 & \frac{1}{4} & 0 & 0 & \frac{1}{4}\\0 & \frac{1}{4} & 0 & 0 & 0 & 0\\0 & 0 & 0 & 0 & - \frac{1}{4} & 0\\- \frac{1}{4} & 0 & 0 & - \frac{1}{4} & 0 & 0\\0 & \frac{1}{4} & 0 & 0 & 0 & 0\end{matrix}\right],
\end{align}
\begin{align}
M_{x^2-y^2}^{%(2)
[2]}= \left[\begin{matrix}0 & 0 & 0 & 0 & 0 & \frac{\sqrt{2} i}{4}\\0 & 0 & 0 & 0 & 0 & 0\\0 & 0 & 0 & - \frac{\sqrt{2} i}{4} & 0 & 0\\0 & 0 & \frac{\sqrt{2} i}{4} & 0 & 0 & 0\\0 & 0 & 0 & 0 & 0 & 0\\- \frac{\sqrt{2} i}{4} & 0 & 0 & 0 & 0 & 0\end{matrix}\right],
\end{align}
\begin{align}
M_{xy}^{%(2)
[2]}= \left[\begin{matrix}0 & 0 & 0 & 0 & 0 & - \frac{\sqrt{2}}{4}\\0 & 0 & 0 & 0 & 0 & 0\\0 & 0 & 0 & \frac{\sqrt{2}}{4} & 0 & 0\\0 & 0 & \frac{\sqrt{2}}{4} & 0 & 0 & 0\\0 & 0 & 0 & 0 & 0 & 0\\- \frac{\sqrt{2}}{4} & 0 & 0 & 0 & 0 & 0\end{matrix}\right],
\end{align}
\begin{align}
M_{z^2}^{%(2)
[2]}= \left[\begin{matrix}0 & 0 & \frac{\sqrt{6} i}{12} & 0 & 0 & 0\\0 & 0 & 0 & 0 & - \frac{\sqrt{6} i}{6} & 0\\- \frac{\sqrt{6} i}{12} & 0 & 0 & 0 & 0 & 0\\0 & 0 & 0 & 0 & 0 & \frac{\sqrt{6} i}{12}\\0 & \frac{\sqrt{6} i}{6} & 0 & 0 & 0 & 0\\0 & 0 & 0 & - \frac{\sqrt{6} i}{12} & 0 & 0\end{matrix}\right],
\end{align}
\begin{align}
G_{zx}^{%(2)
[2]}= \left[\begin{matrix}0 & 0 & 0 & 0 & - \frac{1}{4} & 0\\0 & 0 & - \frac{1}{4} & 0 & 0 & \frac{1}{4}\\0 & - \frac{1}{4} & 0 & 0 & 0 & 0\\0 & 0 & 0 & 0 & \frac{1}{4} & 0\\- \frac{1}{4} & 0 & 0 & \frac{1}{4} & 0 & 0\\0 & \frac{1}{4} & 0 & 0 & 0 & 0\end{matrix}\right],
\end{align}
\begin{align}
G_{yz}^{%(2)
[2]}= \left[\begin{matrix}0 & 0 & 0 & 0 & - \frac{i}{4} & 0\\0 & 0 & \frac{i}{4} & 0 & 0 & \frac{i}{4}\\0 & - \frac{i}{4} & 0 & 0 & 0 & 0\\0 & 0 & 0 & 0 & - \frac{i}{4} & 0\\\frac{i}{4} & 0 & 0 & \frac{i}{4} & 0 & 0\\0 & - \frac{i}{4} & 0 & 0 & 0 & 0\end{matrix}\right],
\end{align}
\begin{align}
G_{x^2-y^2}^{%(2)
[2]}= \left[\begin{matrix}0 & 0 & 0 & 0 & 0 & - \frac{\sqrt{2}}{4}\\0 & 0 & 0 & 0 & 0 & 0\\0 & 0 & 0 & - \frac{\sqrt{2}}{4} & 0 & 0\\0 & 0 & - \frac{\sqrt{2}}{4} & 0 & 0 & 0\\0 & 0 & 0 & 0 & 0 & 0\\- \frac{\sqrt{2}}{4} & 0 & 0 & 0 & 0 & 0\end{matrix}\right],
\end{align}
\begin{align}
G_{xy}^{%(2)
[2]}= \left[\begin{matrix}0 & 0 & 0 & 0 & 0 & - \frac{\sqrt{2} i}{4}\\0 & 0 & 0 & 0 & 0 & 0\\0 & 0 & 0 & - \frac{\sqrt{2} i}{4} & 0 & 0\\0 & 0 & \frac{\sqrt{2} i}{4} & 0 & 0 & 0\\0 & 0 & 0 & 0 & 0 & 0\\\frac{\sqrt{2} i}{4} & 0 & 0 & 0 & 0 & 0\end{matrix}\right],
\end{align}
\begin{align}
G_{z^2}^{%(2)
[2]}= \left[\begin{matrix}0 & 0 & - \frac{\sqrt{6}}{12} & 0 & 0 & 0\\0 & 0 & 0 & 0 & \frac{\sqrt{6}}{6} & 0\\- \frac{\sqrt{6}}{12} & 0 & 0 & 0 & 0 & 0\\0 & 0 & 0 & 0 & 0 & - \frac{\sqrt{6}}{12}\\0 & \frac{\sqrt{6}}{6} & 0 & 0 & 0 & 0\\0 & 0 & 0 & - \frac{\sqrt{6}}{12} & 0 & 0\end{matrix}\right],
\end{align}

\subsection{Four-site square cluster in the $N = 2$ sector}
For the four-site square cluster, we use the following ordering of the $N = 2$ states:
\begin{align}
 \ket{\mathrm{vac.}},
 \ket{0,x},
 \ket{0,y},
 \ket{0,xy},
 \ket{x,y},
 \ket{y,xy},
 \ket{x,xy},
 \ket{\mathrm{full}} .
\end{align}
Here $0,x,y,xy$ denote the symmetry-adapted one-particle cluster orbitals defined in Eq.~\eqref{eq:cluster_mode_def}.  
In this basis, the matrices of the multipole operators used in Table~\ref{tab:exp_H1_H5} are listed below.  
Normal one-body, normal two-body, and anomalous pairing-type multipoles are distinguished by the labels $[1,0]$, $[2,0]$, and $[1,2]$, respectively.

% # vec, 0, x, y, xy, x_2, y_2, xy_2, z'_2, x''_2, y''_2, full
% #
% # x_2   = |0,x>
% # y_2   = |0,y>
% # xy_2  = |0,xy>
% # z'_2  = |x,y>
% # x''_2 = |y,xy>
% # y''_2 = |x,xy>
% # -----------------------------
% \begin{align}
% Q_0^{(1)}= \left[\begin{matrix}- \frac{1}{2} & 0 & 0 & 0 & 0 & 0 & 0 & 0\\0 & \frac{1}{2} & 0 & 0 & 0 & 0 & 0 & 0\\0 & 0 & \frac{1}{2} & 0 & 0 & 0 & 0 & 0\\0 & 0 & 0 & \frac{1}{2} & 0 & 0 & 0 & 0\\0 & 0 & 0 & 0 & - \frac{1}{2} & 0 & 0 & 0\\0 & 0 & 0 & 0 & 0 & - \frac{1}{2} & 0 & 0\\0 & 0 & 0 & 0 & 0 & 0 & - \frac{1}{2} & 0\\0 & 0 & 0 & 0 & 0 & 0 & 0 & \frac{1}{2}\end{matrix}\right]
% \end{align}

\begin{align}
Q_0^{%(1)
[1,0]}= \left[\begin{matrix}- \frac{1}{2} & 0 & 0 & 0 & 0 & 0 & 0 & 0\\0 & \frac{1}{2} & 0 & 0 & 0 & 0 & 0 & 0\\0 & 0 & \frac{1}{2} & 0 & 0 & 0 & 0 & 0\\0 & 0 & 0 & \frac{1}{2} & 0 & 0 & 0 & 0\\0 & 0 & 0 & 0 & - \frac{1}{2} & 0 & 0 & 0\\0 & 0 & 0 & 0 & 0 & - \frac{1}{2} & 0 & 0\\0 & 0 & 0 & 0 & 0 & 0 & - \frac{1}{2} & 0\\0 & 0 & 0 & 0 & 0 & 0 & 0 & \frac{1}{2}\end{matrix}\right]
\end{align}
\begin{align}
Q_0^{'%(1)
[1,0]}= \left[\begin{matrix}-1 & 0 & 0 & 0 & 0 & 0 & 0 & 0\\0 & 0 & 0 & 0 & 0 & 0 & 0 & 0\\0 & 0 & 0 & 0 & 0 & 0 & 0 & 0\\0 & 0 & 0 & -1 & 0 & 0 & 0 & 0\\0 & 0 & 0 & 0 & 1 & 0 & 0 & 0\\0 & 0 & 0 & 0 & 0 & 0 & 0 & 0\\0 & 0 & 0 & 0 & 0 & 0 & 0 & 0\\0 & 0 & 0 & 0 & 0 & 0 & 0 & 1\end{matrix}\right]
\end{align}
\begin{align}
Q_{xy}^{%(1)
[1,0]}= \left[\begin{matrix}0 & 0 & 0 & 0 & 0 & 0 & 0 & 0\\0 & 0 & 0 & 0 & 0 & 0 & -1 & 0\\0 & 0 & 0 & 0 & 0 & -1 & 0 & 0\\0 & 0 & 0 & 0 & 0 & 0 & 0 & 0\\0 & 0 & 0 & 0 & 0 & 0 & 0 & 0\\0 & 0 & -1 & 0 & 0 & 0 & 0 & 0\\0 & -1 & 0 & 0 & 0 & 0 & 0 & 0\\0 & 0 & 0 & 0 & 0 & 0 & 0 & 0\end{matrix}\right]
\end{align}
\begin{align}
Q_v^{%(1)
[1,0]}= \left[\begin{matrix}0 & 0 & 0 & 0 & 0 & 0 & 0 & 0\\0 & 1 & 0 & 0 & 0 & 0 & 0 & 0\\0 & 0 & -1 & 0 & 0 & 0 & 0 & 0\\0 & 0 & 0 & 0 & 0 & 0 & 0 & 0\\0 & 0 & 0 & 0 & 0 & 0 & 0 & 0\\0 & 0 & 0 & 0 & 0 & -1 & 0 & 0\\0 & 0 & 0 & 0 & 0 & 0 & 1 & 0\\0 & 0 & 0 & 0 & 0 & 0 & 0 & 0\end{matrix}\right]
\end{align}
\begin{align}
Q_{xy}^{'%(1)
[1,0]}= \left[\begin{matrix}0 & 0 & 0 & 0 & 0 & 0 & 0 & 0\\0 & 0 & 1 & 0 & 0 & 0 & 0 & 0\\0 & 1 & 0 & 0 & 0 & 0 & 0 & 0\\0 & 0 & 0 & 0 & 0 & 0 & 0 & 0\\0 & 0 & 0 & 0 & 0 & 0 & 0 & 0\\0 & 0 & 0 & 0 & 0 & 0 & 1 & 0\\0 & 0 & 0 & 0 & 0 & 1 & 0 & 0\\0 & 0 & 0 & 0 & 0 & 0 & 0 & 0\end{matrix}\right]
\end{align}
\begin{align}
Q_x^{%(1)
[1,0]}= \left[\begin{matrix}0 & 0 & 0 & 0 & 0 & 0 & 0 & 0\\0 & 0 & 0 & 0 & 0 & 0 & 0 & 0\\0 & 0 & 0 & 0 & 1 & 0 & 0 & 0\\0 & 0 & 0 & 0 & 0 & 0 & 1 & 0\\0 & 0 & 1 & 0 & 0 & 0 & 0 & 0\\0 & 0 & 0 & 0 & 0 & 0 & 0 & 0\\0 & 0 & 0 & 1 & 0 & 0 & 0 & 0\\0 & 0 & 0 & 0 & 0 & 0 & 0 & 0\end{matrix}\right]
\end{align}
\begin{align}
Q_y^{%(1)
[1,0]}= \left[\begin{matrix}0 & 0 & 0 & 0 & 0 & 0 & 0 & 0\\0 & 0 & 0 & 0 & -1 & 0 & 0 & 0\\0 & 0 & 0 & 0 & 0 & 0 & 0 & 0\\0 & 0 & 0 & 0 & 0 & 1 & 0 & 0\\0 & -1 & 0 & 0 & 0 & 0 & 0 & 0\\0 & 0 & 0 & 1 & 0 & 0 & 0 & 0\\0 & 0 & 0 & 0 & 0 & 0 & 0 & 0\\0 & 0 & 0 & 0 & 0 & 0 & 0 & 0\end{matrix}\right]
\end{align}
\begin{align}
Q_x^{'%(1)
[1,0]}= \left[\begin{matrix}0 & 0 & 0 & 0 & 0 & 0 & 0 & 0\\0 & 0 & 0 & 0 & 0 & 0 & 0 & 0\\0 & 0 & 0 & 1 & 0 & 0 & 0 & 0\\0 & 0 & 1 & 0 & 0 & 0 & 0 & 0\\0 & 0 & 0 & 0 & 0 & 0 & 1 & 0\\0 & 0 & 0 & 0 & 0 & 0 & 0 & 0\\0 & 0 & 0 & 0 & 1 & 0 & 0 & 0\\0 & 0 & 0 & 0 & 0 & 0 & 0 & 0\end{matrix}\right]
\end{align}
\begin{align}
Q_y^{'%(1)
[1,0]}= \left[\begin{matrix}0 & 0 & 0 & 0 & 0 & 0 & 0 & 0\\0 & 0 & 0 & 1 & 0 & 0 & 0 & 0\\0 & 0 & 0 & 0 & 0 & 0 & 0 & 0\\0 & 1 & 0 & 0 & 0 & 0 & 0 & 0\\0 & 0 & 0 & 0 & 0 & -1 & 0 & 0\\0 & 0 & 0 & 0 & -1 & 0 & 0 & 0\\0 & 0 & 0 & 0 & 0 & 0 & 0 & 0\\0 & 0 & 0 & 0 & 0 & 0 & 0 & 0\end{matrix}\right]
\end{align}
\begin{align}
Q_0^{%(2)
[2,0]}= \left[\begin{matrix}0 & 0 & 0 & 0 & 0 & 0 & 0 & 0\\0 & \frac{1}{2} & 0 & 0 & 0 & 0 & 0 & 0\\0 & 0 & \frac{1}{2} & 0 & 0 & 0 & 0 & 0\\0 & 0 & 0 & -1 & 0 & 0 & 0 & 0\\0 & 0 & 0 & 0 & -1 & 0 & 0 & 0\\0 & 0 & 0 & 0 & 0 & \frac{1}{2} & 0 & 0\\0 & 0 & 0 & 0 & 0 & 0 & \frac{1}{2} & 0\\0 & 0 & 0 & 0 & 0 & 0 & 0 & 0\end{matrix}\right]
\end{align}
\begin{align}
Q_0^{'%(2)
[2,0]}= \left[\begin{matrix}3 & 0 & 0 & 0 & 0 & 0 & 0 & 0\\0 & -1 & 0 & 0 & 0 & 0 & 0 & 0\\0 & 0 & -1 & 0 & 0 & 0 & 0 & 0\\0 & 0 & 0 & -1 & 0 & 0 & 0 & 0\\0 & 0 & 0 & 0 & -1 & 0 & 0 & 0\\0 & 0 & 0 & 0 & 0 & -1 & 0 & 0\\0 & 0 & 0 & 0 & 0 & 0 & -1 & 0\\0 & 0 & 0 & 0 & 0 & 0 & 0 & 3\end{matrix}\right]
\end{align}
\begin{align}
Q_0^{''%(2)
[2,0]}= \left[\begin{matrix}0 & 0 & 0 & 0 & 0 & 0 & 0 & 0\\0 & 0 & 0 & 0 & 0 & 1 & 0 & 0\\0 & 0 & 0 & 0 & 0 & 0 & 1 & 0\\0 & 0 & 0 & 0 & 0 & 0 & 0 & 0\\0 & 0 & 0 & 0 & 0 & 0 & 0 & 0\\0 & 1 & 0 & 0 & 0 & 0 & 0 & 0\\0 & 0 & 1 & 0 & 0 & 0 & 0 & 0\\0 & 0 & 0 & 0 & 0 & 0 & 0 & 0\end{matrix}\right]
\end{align}
\begin{align}
Q_v^{%(2)
[2,0]}= \left[\begin{matrix}0 & 0 & 0 & 0 & 0 & 0 & 0 & 0\\0 & 0 & 0 & 0 & 0 & 0 & 0 & 0\\0 & 0 & 0 & 0 & 0 & 0 & 0 & 0\\0 & 0 & 0 & 0 & 1 & 0 & 0 & 0\\0 & 0 & 0 & 1 & 0 & 0 & 0 & 0\\0 & 0 & 0 & 0 & 0 & 0 & 0 & 0\\0 & 0 & 0 & 0 & 0 & 0 & 0 & 0\\0 & 0 & 0 & 0 & 0 & 0 & 0 & 0\end{matrix}\right]
\end{align}
\begin{align}
Q_v^{'%(2)
[2,0]}= \left[\begin{matrix}0 & 0 & 0 & 0 & 0 & 0 & 0 & 0\\0 & \frac{1}{2} & 0 & 0 & 0 & 0 & 0 & 0\\0 & 0 & - \frac{1}{2} & 0 & 0 & 0 & 0 & 0\\0 & 0 & 0 & 0 & 0 & 0 & 0 & 0\\0 & 0 & 0 & 0 & 0 & 0 & 0 & 0\\0 & 0 & 0 & 0 & 0 & \frac{1}{2} & 0 & 0\\0 & 0 & 0 & 0 & 0 & 0 & - \frac{1}{2} & 0\\0 & 0 & 0 & 0 & 0 & 0 & 0 & 0\end{matrix}\right]
\end{align}
\begin{align}
Q_v^{''%(2)
[2,0]}= \left[\begin{matrix}0 & 0 & 0 & 0 & 0 & 0 & 0 & 0\\0 & 0 & 0 & 0 & 0 & 1 & 0 & 0\\0 & 0 & 0 & 0 & 0 & 0 & -1 & 0\\0 & 0 & 0 & 0 & 0 & 0 & 0 & 0\\0 & 0 & 0 & 0 & 0 & 0 & 0 & 0\\0 & 1 & 0 & 0 & 0 & 0 & 0 & 0\\0 & 0 & -1 & 0 & 0 & 0 & 0 & 0\\0 & 0 & 0 & 0 & 0 & 0 & 0 & 0\end{matrix}\right]
\end{align}
\begin{align}
Q_{xy}^{%(2)
[2,0]}= \left[\begin{matrix}0 & 0 & 0 & 0 & 0 & 0 & 0 & 0\\0 & 0 & \frac{1}{2} & 0 & 0 & 0 & 0 & 0\\0 & \frac{1}{2} & 0 & 0 & 0 & 0 & 0 & 0\\0 & 0 & 0 & 0 & 0 & 0 & 0 & 0\\0 & 0 & 0 & 0 & 0 & 0 & 0 & 0\\0 & 0 & 0 & 0 & 0 & 0 & - \frac{1}{2} & 0\\0 & 0 & 0 & 0 & 0 & - \frac{1}{2} & 0 & 0\\0 & 0 & 0 & 0 & 0 & 0 & 0 & 0\end{matrix}\right]
\end{align}
\begin{align}
G_z^{%(2)
[2,0]}= \left[\begin{matrix}0 & 0 & 0 & 0 & 0 & 0 & 0 & 0\\0 & 0 & 0 & 0 & 0 & 0 & 1 & 0\\0 & 0 & 0 & 0 & 0 & -1 & 0 & 0\\0 & 0 & 0 & 0 & 0 & 0 & 0 & 0\\0 & 0 & 0 & 0 & 0 & 0 & 0 & 0\\0 & 0 & -1 & 0 & 0 & 0 & 0 & 0\\0 & 1 & 0 & 0 & 0 & 0 & 0 & 0\\0 & 0 & 0 & 0 & 0 & 0 & 0 & 0\end{matrix}\right]
\end{align}
\begin{align}
Q_x^{%(2)
[2,0]}= \left[\begin{matrix}0 & 0 & 0 & 0 & 0 & 0 & 0 & 0\\0 & 0 & 0 & 0 & 0 & 0 & 0 & 0\\0 & 0 & 0 & 0 & \frac{1}{2} & 0 & 0 & 0\\0 & 0 & 0 & 0 & 0 & 0 & - \frac{1}{2} & 0\\0 & 0 & \frac{1}{2} & 0 & 0 & 0 & 0 & 0\\0 & 0 & 0 & 0 & 0 & 0 & 0 & 0\\0 & 0 & 0 & - \frac{1}{2} & 0 & 0 & 0 & 0\\0 & 0 & 0 & 0 & 0 & 0 & 0 & 0\end{matrix}\right]
\end{align}
\begin{align}
Q_y^{%(2)
[2,0]}= \left[\begin{matrix}0 & 0 & 0 & 0 & 0 & 0 & 0 & 0\\0 & 0 & 0 & 0 & - \frac{1}{2} & 0 & 0 & 0\\0 & 0 & 0 & 0 & 0 & 0 & 0 & 0\\0 & 0 & 0 & 0 & 0 & - \frac{1}{2} & 0 & 0\\0 & - \frac{1}{2} & 0 & 0 & 0 & 0 & 0 & 0\\0 & 0 & 0 & - \frac{1}{2} & 0 & 0 & 0 & 0\\0 & 0 & 0 & 0 & 0 & 0 & 0 & 0\\0 & 0 & 0 & 0 & 0 & 0 & 0 & 0\end{matrix}\right]
\end{align}
\begin{align}
Q_x^{'%(2)
[2,0]}= \left[\begin{matrix}0 & 0 & 0 & 0 & 0 & 0 & 0 & 0\\0 & 0 & 0 & 0 & 0 & 0 & 0 & 0\\0 & 0 & 0 & \frac{1}{2} & 0 & 0 & 0 & 0\\0 & 0 & \frac{1}{2} & 0 & 0 & 0 & 0 & 0\\0 & 0 & 0 & 0 & 0 & 0 & - \frac{1}{2} & 0\\0 & 0 & 0 & 0 & 0 & 0 & 0 & 0\\0 & 0 & 0 & 0 & - \frac{1}{2} & 0 & 0 & 0\\0 & 0 & 0 & 0 & 0 & 0 & 0 & 0\end{matrix}\right]
\end{align}
\begin{align}
Q_y^{'%(2)
[2,0]}= \left[\begin{matrix}0 & 0 & 0 & 0 & 0 & 0 & 0 & 0\\0 & 0 & 0 & \frac{1}{2} & 0 & 0 & 0 & 0\\0 & 0 & 0 & 0 & 0 & 0 & 0 & 0\\0 & \frac{1}{2} & 0 & 0 & 0 & 0 & 0 & 0\\0 & 0 & 0 & 0 & 0 & \frac{1}{2} & 0 & 0\\0 & 0 & 0 & 0 & \frac{1}{2} & 0 & 0 & 0\\0 & 0 & 0 & 0 & 0 & 0 & 0 & 0\\0 & 0 & 0 & 0 & 0 & 0 & 0 & 0\end{matrix}\right]
\end{align}
\begin{align}
\tilde G_z^{%(1)
[1,2]}= \left[\begin{matrix}0 & 0 & 0 & 0 & 1 & 0 & 0 & 0\\0 & 0 & 0 & 0 & 0 & 0 & 0 & 0\\0 & 0 & 0 & 0 & 0 & 0 & 0 & 0\\0 & 0 & 0 & 0 & 0 & 0 & 0 & 1\\1 & 0 & 0 & 0 & 0 & 0 & 0 & 0\\0 & 0 & 0 & 0 & 0 & 0 & 0 & 0\\0 & 0 & 0 & 0 & 0 & 0 & 0 & 0\\0 & 0 & 0 & 1 & 0 & 0 & 0 & 0\end{matrix}\right]
\end{align}
\begin{align}
\tilde Q_{xy}^{%(1)
[1,2]}= \left[\begin{matrix}0 & 0 & 0 & 1 & 0 & 0 & 0 & 0\\0 & 0 & 0 & 0 & 0 & 0 & 0 & 0\\0 & 0 & 0 & 0 & 0 & 0 & 0 & 0\\1 & 0 & 0 & 0 & 0 & 0 & 0 & 0\\0 & 0 & 0 & 0 & 0 & 0 & 0 & 1\\0 & 0 & 0 & 0 & 0 & 0 & 0 & 0\\0 & 0 & 0 & 0 & 0 & 0 & 0 & 0\\0 & 0 & 0 & 0 & 1 & 0 & 0 & 0\end{matrix}\right]
\end{align}
\begin{align}
\tilde Q_x^{%(1)
[1,2]}= \left[\begin{matrix}0 & 1 & 0 & 0 & 0 & 0 & 0 & 0\\1 & 0 & 0 & 0 & 0 & 0 & 0 & 0\\0 & 0 & 0 & 0 & 0 & 0 & 0 & 0\\0 & 0 & 0 & 0 & 0 & 0 & 0 & 0\\0 & 0 & 0 & 0 & 0 & 0 & 0 & 0\\0 & 0 & 0 & 0 & 0 & 0 & 0 & 1\\0 & 0 & 0 & 0 & 0 & 0 & 0 & 0\\0 & 0 & 0 & 0 & 0 & 1 & 0 & 0\end{matrix}\right]
\end{align}
\begin{align}
\tilde Q_x^{'%(1)
[1,2]}= \left[\begin{matrix}0 & 0 & 0 & 0 & 0 & 1 & 0 & 0\\0 & 0 & 0 & 0 & 0 & 0 & 0 & 1\\0 & 0 & 0 & 0 & 0 & 0 & 0 & 0\\0 & 0 & 0 & 0 & 0 & 0 & 0 & 0\\0 & 0 & 0 & 0 & 0 & 0 & 0 & 0\\1 & 0 & 0 & 0 & 0 & 0 & 0 & 0\\0 & 0 & 0 & 0 & 0 & 0 & 0 & 0\\0 & 1 & 0 & 0 & 0 & 0 & 0 & 0\end{matrix}\right]
\end{align}
\begin{align}
\tilde Q_y^{%(1)
[1,2]}= \left[\begin{matrix}0 & 0 & 1 & 0 & 0 & 0 & 0 & 0\\0 & 0 & 0 & 0 & 0 & 0 & 0 & 0\\1 & 0 & 0 & 0 & 0 & 0 & 0 & 0\\0 & 0 & 0 & 0 & 0 & 0 & 0 & 0\\0 & 0 & 0 & 0 & 0 & 0 & 0 & 0\\0 & 0 & 0 & 0 & 0 & 0 & 0 & 0\\0 & 0 & 0 & 0 & 0 & 0 & 0 & -1\\0 & 0 & 0 & 0 & 0 & 0 & -1 & 0\end{matrix}\right]
\end{align}
\begin{align}
\tilde Q_y^{'%(1)
[1,2]}= \left[\begin{matrix}0 & 0 & 0 & 0 & 0 & 0 & 1 & 0\\0 & 0 & 0 & 0 & 0 & 0 & 0 & 0\\0 & 0 & 0 & 0 & 0 & 0 & 0 & -1\\0 & 0 & 0 & 0 & 0 & 0 & 0 & 0\\0 & 0 & 0 & 0 & 0 & 0 & 0 & 0\\0 & 0 & 0 & 0 & 0 & 0 & 0 & 0\\1 & 0 & 0 & 0 & 0 & 0 & 0 & 0\\0 & 0 & -1 & 0 & 0 & 0 & 0 & 0\end{matrix}\right]
\end{align}


% Provide your bibliography here. You have two options:

%%% FIRST OPTION
% Write your entries here directly, following the example below, including:
% Author(s), Title, Journal Ref. with year in parentheses at the end, followed by the DOI number.

%\begin{thebibliography}{99}
%\bibitem{1931_Bethe_ZP_71} H. A. Bethe, {\it Zur Theorie der Metalle. i. Eigenwerte und Eigenfunktionen der linearen Atomkette}, Zeit. f{\"u}r Phys. {\bf 71}, 205 (1931), \doi{10.1007\%2FBF01341708}.
%\bibitem{arXiv:1108.2700} P. Ginsparg, {\it It was twenty years ago today... }, \url{http://arxiv.org/abs/1108.2700}.
%\end{thebibliography}

% Use your bibtex library, formatted by the SciPost style file.

% \bibliography{Multipole_theory.bib, exp.bib}%%一番最後で良い
\bibliography{ref.bib}%%一番最後で良い

%%%%%%%%%% END TODO: BIBLIOGRAPHY

\end{document}